\documentclass[12pt]{article}

\usepackage[dvipdfmx]{graphicx}
\usepackage[dvipdfmx]{color}
\usepackage{epsfig}
\usepackage{color}
\usepackage{amssymb,amsmath}
\usepackage{graphicx}
\usepackage{here}
\usepackage{comment}
\newcommand{\ba}{\begin{align}}
\newcommand{\ea}{\end{align}}

 \makeatletter
\def\alt{\mathrel{\mathpalette\gl@align<}}
\def\agt{\mathrel{\mathpalette\gl@align>}}
\def\gl@align#1#2{\lower.6ex\vbox{\baselineskip\z@skip\lineskip\z@
\ialign{$\m@th#1\hfil##\hfil$\crcr#2\crcr\sim\crcr}}} \makeatother

\begin{document}
\begin{flushright}
\end{flushright}
\vspace*{1.0cm}

\begin{center}
\baselineskip 20pt 
{\Large\bf 
Charged Lepton Flavor Violating processes 
\\ in Neutrinophilic Higgs+Seesaw model
}
\vspace{1cm}

{\large 
Naoyuki Haba, \ Tsuneharu Omija \ and \ Toshifumi Yamada
} \vspace{.5cm}

{\baselineskip 20pt \it
Graduate School of Science and Engineering, Shimane University, Matsue 690-8504, Japan
}

\vspace{.5cm}

\vspace{1.5cm} {\bf Abstract} \end{center}
We investigate charged lepton flavor violating (CLFV) processes in the ``neutrinophilic Higgs+seesaw model'',
 in which right-handed neutrinos couple only with an extra Higgs field which develops a tiny vacuum expectation value and the right-handed neutrinos also have Majorana mass.
The model realizes a seesaw mechanism around TeV scale without extremely small Dirac Yukawa couplings.
A phenomenological feature of the model is CLFV processes induced by loop diagrams of the charged scalar particles and heavy neutrinos.
Therefore, first we constrain the model's parameter space from the search for $\mu\to e\gamma$. 
Next, we predict the branching ratios of other CLFV processes including the $\mu\to3e$, $\mu+{\rm Al}\to e+{\rm Al}$, $\mu+{\rm Ti}\to e+{\rm Ti}$, $Z\to e\mu$,
$Z\to e\tau$, $Z\to \mu\tau$,  $h\to e\tau$ and $h\to\mu\tau$ processes, and discuss their detectability in future experiments.

\thispagestyle{empty}

\newpage

\setcounter{footnote}{0}
\baselineskip 18pt
%

\section{Introduction}

The origin of the smallness of the neutrino mass is one of the prime open questions in particle physics.
One candidate solution to the above mystery is the neutrinophilic Two Higgs Doublet Model~\cite{Gabriel:2006ns},
 where there is an extra Higgs doublet （called ``neutrinophilic Higgs'') 
 that couples to the lepton doublets and right-handed neutrinos while the coupling of the Standard Model (SM)
 Higgs doublet to right-handed neutrinos is forbidden by a $Z_2$ symmetry, and
 the smallness of the vacuum expectation value (VEV) of the neutrinophilic Higgs explains the smallness of the neutrino mass.
In the original proposal~\cite{Gabriel:2006ns}, Majorana mass for right-handed neutrinos is absent and the neutrinos are purely Dirac particles.
However, the $Z_2$ symmetry that forbids the coupling of the SM Higgs and right-handed neutrinos does not exclude the possibility that 
 right-handed neutrinos have a Majorana mass term.
If Majorana mass for right-handed neutrinos is introduced, the model becomes a low-scale realization of the seesaw mechanism~\cite{seesaw1}-\cite{seesaw4},
 where the smallness of the neutrino mass is accounted for by the seesaw mechanism in addition to the tininess of the neutrinophilic Higgs VEV.
We call the new model ``neutrinophilic higgs+seesaw model'', for the obvious reason.

Important experimental signatures of the neutrinophilic Higgs+seesaw model are (i) the presence of new scalar particles $H^{\pm}, H$ and $A$
 originating dominantly from the neutrinophilic Higgs field, and (ii) charged lepton flavor violating (CLFV) processes, such as $\mu\to e\gamma$,
 mediated by a loop of the charged scalar $H^{\pm}$ and a heavy neutrino.
In this paper, we investigate CLFV processes in the neutrinophilic Higgs+seesaw model in detail.
First, we constrain the parameter space of the neutrinophilic Higgs+seesaw model from current experimental bounds on CLFV processes,
 the most stringent bound coming from the $\mu\to e\gamma$ decay.
Next, we predict branching ratios (or conversion rates) of various CLFV processes and discuss whether it is possible to detect these processes in the future.

Previously, Ref.~\cite{Bertuzzo:2015ada} has studied CLFV processes in the neutrinophilic Two Higgs Doublet Model,
 but in that work, the Majorana mass term is not considered and the neutrinos are purely Dirac particles.
Our work extends it by introducing Majorana mass for right-handed neutrinos.
Also, Ref.~\cite{Toma:2013zsa} has studied CLFV processes in a model with similar phenomenological features,
 but CLFV decays of the SM-like Higgs particle and $Z$ boson are not included, unlike in our paper.

The rest of the paper is organized as follows. In Section~2, we describe the neutrinophilic Higgs+seesaw model. 
In Section~3, we give the formulas for the branching ratios of the CLFV process.
In Section~4, we present our numerical results, which include the current constraints on the neutrinophilic Higgs+seesaw model
 and predictions for various CLFV processes.
Finally, we summarize our results in Section~5.
\\

\section{Neutrinophilic Higgs+Seesaw Model}

The model contains two Higgs doublet fields, $H_1$ and $H_2$, left-handed leptons, $\ell_L^\alpha$, right-handed charged leptons, $e_R^\alpha$, and right-handed neutrinos, $\nu_R^i$, 
 where $\alpha=e,\mu,\tau$ is the flavor index for charged leptons and $i=1,2,3$ is another flavor index.
It also contains quarks, $q_L^k,\,u_R^k$ and $\,d_R^k$, but they play no role in this study.
The fields are charged under the SM $SU(3)_C\times SU(2)_L\times U(1)_Y$ gauge group and a $Z_2$ symmetry
 as given in Table~\ref{fields}  (bold numbers indicate the representations).
\begin{table}[H]
\begin{center}
  \caption{The field content and charge assignments.}
  \begin{tabular}{|c||c|c|c|c|} \hline
    Field & $SU(3)_C$ & $SU(2)_L$ & $U(1)_Y$ & $Z_2$ \\ \hline
    $H_1$  & {\bf 1} & {\bf 2}          & $-1/2$      & + \\
    $H_2$  & {\bf 1} & {\bf 2}          & $-1/2$      &  $-$ \\ \hline
    $\ell_L^\alpha$  & {\bf 1} & \bf{2}          & $-1/2$      & + \\
    $e_R^\alpha$      & {\bf 1} & \bf{1}          & $-1$      & + \\
    $\nu_R^i$            & {\bf 1} & \bf{1}          & 0      & $-$ \\ \hline
    $q_L^k$  & {\bf 3} & {\bf 2}          & $1/6$      & $+$ \\
    $u_R^k$   & {\bf 3} &{\bf 1}          & $2/3$      & $+$ \\
    $d_R^k$   & {\bf 3} &{\bf 1}          & $-1/3$      & $+$ \\ \hline
    \end{tabular}
  \label{fields} 
  \end{center}
\end{table}
Note that the above $Z_2$ charge assignment allows Majorana mass for right-handed neutrinos,
 while it forbids the Yukawa couplings of SM fermions with $H_2$ and the Yukawa coupling of right-handed neutrinos with $H_1$.

We assume that the $Z_2$ symmetry is softly broken in the scalar potential.
The most general scalar potential and Yukawa couplings are then
\begin{align} 
-{\cal L} &= m_1^2 \, H_1^\dagger H_1 + m_2^2 \, H_2^\dagger H_2 - m_3^2 \, (H_1^\dagger H_2 + H_2^\dagger H_1)
\nonumber \\
&+ \frac{\lambda_1}{2}(H_1^\dagger H_1)^2 + \frac{\lambda_2}{2}(H_2^\dagger H_2)^2 + \lambda_3(H_1^\dagger H_1)(H_2^\dagger H_2)
+ \lambda_4(H_1^\dagger H_2)(H_2^\dagger H_1) + \lambda_5(H_1^\dagger H_2)^2 + \lambda_5^*(H_2^\dagger H_1)^2
\nonumber \\
&+ (Y_e)_{\alpha\beta} \, \ell_L^{\alpha\dagger} \, \epsilon_g H_1^* \, e_R^\beta 
+ (Y_D)_{\alpha i} \, \ell_L^{\alpha\dagger} \, H_2 \, \nu_R^i + \frac{1}{2}M_{N_i} \, \nu_R^{iT} \epsilon_s \nu_R^i + {\rm H.c.},
\label{lagrangian}
\end{align}
 where $\epsilon_g$ denotes the antisymmetric tensor acting on $SU(2)_L$ indices and $\epsilon_s$ denotes that acting on spinor indices.
Here, we have taken the flavor basis in which the Majorana mass for right-handed neutrinos is diagonal,
 and we have made $m_3^2$ real positive by a phase redefinition.
Note that only the $m_3^2$ term explicitly breaks the $Z_2$ symmetry,
 and so the limit with $m_3^2/m_1^2\to0$ and $m_3^2/m_2^2\to0$ can be taken naturally at the quantum level.
We also assume 
\begin{align} 
m_1^2 &< 0, \ \ \ \ \ m_2^2 >0,
\label{m1m2}
\end{align}
 so that $H_1$ develops a VEV around the scale $\sqrt{\vert m_1^2\vert}$, and then $H_2$ gains a VEV through the term 
 $m_3^2 (H_1^\dagger H_2 + H_2^\dagger H_1)$.
Consequently, the VEV of $H_2$ is proportional to $m_3^2$ and is controlled by the explicit breaking of the $Z_2$ symmetry,
 and therefore the VEV of $H_2$ can naturally take a small value.
Additionally, we assume that $\lambda_5$ is suppressed as $|\lambda_5|\ll 8\pi^2 m_\nu |M_k| /|(Y_D)_{\alpha i}|^2v^2$ and 
 $|\lambda_5|\ll 8\pi^2 m_\nu m_2^2/|(Y_D)_{\alpha i}|^2v^2|M_k|$, where $m_\nu$ denotes the mass scale of active neutrinos and $v\simeq246$~GeV.
It follows that, unlike the model of Ref.~\cite{Ma:2006km}, the one-loop correction to the neutrino mass involving $\lambda_5$ is much smaller than the tree-level mass.
The above suppression of $\lambda_5$ is realized naturally by promoting the $Z_2$ symmetry to a global $U(1)$ symmetry under which $H_2$ is charged by $+1$,
 $\nu_R^i$ is charged by $-1$ and the other fields have no charge and which is broken only softly.

In the neutrinophilic Higgs+seesaw model, we take the limit with $m_3^2/m_2^2\to0$.
Then, writing the Higgs VEVs as $\langle H_1^0 \rangle = v_1/\sqrt{2}$ and $\langle H_2^0 \rangle = v_2/\sqrt{2}$, we find
\begin{align} 
v_1 &\simeq \sqrt{-2m_1^2/\lambda_1}, \\
v_2 &\simeq \frac{2m_3^2}{m_2^2+(\lambda_3+\lambda_4)v_1^2/2}v_1,
\label{vevs}
\end{align}
 and hence $v_2\ll v_1$.
The physical particles are
 the lighter CP (Charge conjugation Parity) -even scalar, $h$, which is identified with the observed 125~GeV scalar particle,
 the heavier CP-even scalar, $H$, the $CP$-odd scalar, $A$, and the charged scalar, $H^\pm$.
The masses of $A$ and $H^\pm$ are given by
\begin{align} 
m_A^2 &= \frac{m_3^2}{\sin\beta\cos\beta}, \ \ \ \ \
m_{H^\pm}^2 = \frac{m_3^2}{\sin\beta\cos\beta} - \frac{\lambda_4}{2}v^2
\end{align}
 and the masses of $h$ and $H$ are given, in the limit with $v_1\gg v_2$, by
\begin{align} 
m_h^2 &\simeq \lambda_1 v_1^2, \ \ \ \ \
m_H^2 \simeq m_A^2.
\end{align}
 where $\tan\beta \equiv v_1/v_2$.
In terms of $h,H,A,H^\pm$ and would-be Nambu-Goldstone modes $G^0$ and $G^\pm$, the Higgs fields are decomposed as
\begin{align} 
H_1 &=   \begin{pmatrix} 
      \frac{1}{\sqrt{2}}\left(\sin\beta \ v+\cos\alpha \ h+\sin\alpha \ H -i \sin\beta \ G^0 -i\cos\beta \ A\right) \\
      -\sin\beta \ G^- - \cos\beta \ H^- \\      
   \end{pmatrix},
   \nonumber \\
H_2 &=   \begin{pmatrix} 
      \frac{1}{\sqrt{2}}\left(\cos\beta \ v-\sin\alpha \ h+\cos\alpha \ H -i \cos\beta \ G^0 +i\sin\beta \ A\right) \\
       -\cos\beta \ G^- +\sin\beta \ H^- \\
   \end{pmatrix},
\end{align}
 where $\alpha$ is the mixing angle of the CP-even scalars. 
$\alpha$ satisfies
\begin{align} 
0&>\alpha>-\frac{\pi}{2}, \ \ \ \tan2\alpha \simeq -\frac{2}{\tan\beta}
\end{align}
 in the limit with $\tan\beta \gg 1$, and hence $\alpha\simeq0$ in the neutrinophilic Higgs+seesaw model.

The interaction of the charged scalar $H^\pm$ is the dominant source of CLFV processes and is particularly important.
The three-point interaction term for $H^+ H^- h$ is given by
 \begin{align} 
-{\cal L} &\supset v \lambda_3 \, h H^+H^-
\end{align}
 in the limit with $\tan\beta\gg1$.
The Yukawa interaction terms of $H^\pm$ are
\begin{align} 
-{\cal L} &\supset 
-(Y_e)_{\alpha\beta} \ \nu_L^{\alpha\dagger} e_R^\beta \ \cos\beta \ H^+
+ (Y_D)_{\alpha i} \ e_L^{\alpha\dagger} \nu_R^i \ \sin\beta \ H^- +{\rm H.c.}
\nonumber
\end{align}

We turn our attention to the lepton mass.
The Dirac and Majorana mass terms are given by
\begin{align} 
-{\cal L} &\supset 
\frac{v_1}{\sqrt{2}}(Y_e)_{\alpha\beta} \ e_L^{\alpha\dagger} e_R^\beta + \frac{v_2}{\sqrt{2}}(Y_D)_{\alpha i} \ \nu_L^{\alpha\dagger} \nu_R^i
+\frac{1}{2}M_{N_i} \ \nu_R^{iT} \epsilon_s \nu_R^i+{\rm H.c.}
\end{align}
Then, the mass matrix for neutrinos is obtained as
\begin{align} 
-{\cal L}  &\supset \frac{1}{2}\begin{pmatrix} 
      \nu_L^{\beta T} & \nu_R^{j\dagger}\epsilon_s^T
   \end{pmatrix}
    \epsilon_s
  \begin{pmatrix} 
      O & -\frac{v_2}{\sqrt{2}}(Y_D^*)_{\beta i} \\
      -\frac{v_2}{\sqrt{2}}(Y_D^*)_{\alpha j} & \delta_{ij} \, M_{N_i}^* \\
   \end{pmatrix}
    \begin{pmatrix} 
      \nu_L^\alpha \\
      \epsilon_s\, \nu_R^{i*}  \\
    \end{pmatrix}+{\rm H.c.}
\end{align}
The above mass matrix is diagonalized by a unitary matrix, $U$, as
\begin{align} 
\begin{pmatrix} 
      O & -\frac{v_2}{\sqrt{2}}(Y_D^*)_{\beta i} \\
      -\frac{v_2}{\sqrt{2}} (Y_D^*)_{\alpha j} & \delta_{ij} \, M_{N_i}^* \\
   \end{pmatrix}
= U^* \, {\rm diag}\left(m_{\nu_1}, \, m_{\nu_2}, \, m_{\nu_3}, \, m_{\nu_4}, \, m_{\nu_5}, \, m_{\nu_6} \right) \, U^\dagger,
\end{align}
 where $m_{\nu_1}, \, m_{\nu_2}$ and $\, m_{\nu_3}$ correspond to the tiny active neutrino masses, and $m_{\nu_4}, \, m_{\nu_5}$ and $\, m_{\nu_6}$
 to the masses of heavy neutrinos.
We assume $v_2\ll M_{N_j}$.
The unitary matrix $U$ is then approximated by
\begin{align} 
U &\simeq    \begin{pmatrix} 
      U_{PMNS} & O \\
      O & I_3 \\
   \end{pmatrix}
   ,
\end{align}
 where $U_{PMNS}$ denotes the PMNS (Pontecorvo-Maki-Nakagawa-Sakata) mixing matrix~\cite{mns,p} and $I_3$ denotes the 3-dimensional identity matrix,
 and we obtain the following seesaw formula:
\begin{align} 
-\frac{v_2^2}{2} (Y_D^*)_{\beta i}  (Y_D^*)_{\alpha i} \frac{1}{M_{N_i}}
&\simeq \left[ \, U_{PMNS} \begin{pmatrix} 
      m_{\nu_1} & 0 & 0 \\
      0 & m_{\nu_2} & 0 \\
      0 & 0 &m_{\nu_3} \\
   \end{pmatrix} U_{PMNS} \, \right]_{\alpha \beta \, {\rm component}}.
 \label{seesaw}
\end{align}
Inverting the relation Eq.~(\ref{seesaw}), one can express the neutrino Dirac Yukawa coupling $Y_D$ as 
\begin{align} 
Y_D &= i\frac{\sqrt{2}}{v_2} U_{PMNS}    \begin{pmatrix} 
      \sqrt{m_{\nu_1}} & 0 & 0 \\
      0 & \sqrt{m_{\nu_2}} & 0 \\
      0 & 0 & \sqrt{m_{\nu_3}} \\
   \end{pmatrix}
   R_{3\times3}
   \begin{pmatrix} 
      \sqrt{M_{N_1}} & 0 & 0 \\
      0 & \sqrt{M_{N_2}} & 0 \\
      0 & 0 & \sqrt{M_{N_3}} \\
   \end{pmatrix}   
   \label{yukawa}
\end{align}
 where $R_{3\times3}$ is an arbitrary complex-valued $3\times3$ rotation matrix~\cite{ci}.
The masses of heavy neutrinos are approximated as
\begin{align}
m_{\nu_4}\simeq M_{N1}^*, \ \ \ m_{\nu_5}\simeq M_{N2}^*, \ \ \ m_{\nu_6}\simeq M_{N3}^*,
\end{align}
 and the mass eigenstates belonging to $m_{\nu_4}, \, m_{\nu_5}$ and $\, m_{\nu_6}$ are mostly given by
 the right-handed neutrinos, namely, we find
\begin{align} 
\nu_4\simeq\epsilon_s\, \nu_R^{1*}, \ \ \ \nu_5\simeq\epsilon_s\, \nu_R^{2*}, \ \ \ \nu_6\simeq\epsilon_s\, \nu_R^{3*}.
\end{align}
\\

We comment on the constraints from electroweak precision tests.
The constraint from the Peskin-Takeuchi $T$-parameter~\cite{Peskin:1990zt,Peskin:1991sw} can be avoided by taking the coupling constants $\lambda_4,\lambda_5$ close to 0 so that the charged scalar $H^\pm$ and the heavy neutral scalars $H,A$ are nearly mass-degenerate (note $\beta\simeq\pi/2$, $\alpha\simeq0$). 
Taking $\lambda_4=\lambda_5=0$ does not affect the CLFV processes we discuss in the ensuing sections.
When $m_H^2\simeq m_A^2\simeq m_{H^\pm}^2$, the Peskin-Takeuchi $S$-parameter is explicitly calculated as
\begin{align}
S &\simeq \frac{1}{6\pi}(2s_W^4-2s_W^2+1)
\nonumber\\
&\times
\left\{-\frac{8}{3}+\frac{8m_{H^\pm}^2}{m_Z^2}+\frac{1}{m_Z^2}\left(1-\frac{4m_{H^\pm}^2}{m_Z^2}\right)
\sqrt{4m_Z^2m_{H^\pm}^2-m_Z^4}\arctan\frac{\sqrt{4m_Z^2m_{H^\pm}^2-m_Z^4}}{2m_{H^\pm}^2-m_Z^2}\right\}.
\end{align}
When $m_H^2\simeq m_A^2\simeq m_{H^\pm}^2=(300$~GeV$)^2$, which will be the benchmark value of our numerical analysis,
 we get $S\simeq-0.0003$. This is consistent with the current experimental bound~\cite{Tanabashi:2018oca}.
\\

Finally, we comment on new physics contributions to the electron electric dipole moment.
The two-loop Barr-Zee diagrams that involve $Y_D$ and contribute to the electron dipole moments are highly suppressed 
 by the coupling of $H^+$ to quarks (proportional to $\cos\beta$) and also by the mixing of a heavy neutrino with active flavor (proportional to $Y_D v/M_{Ni}$).
On the other hand, the one-loop diagrams contributing to the electron dipole moments are proportional to $\sum_{i=1}^3 (Y_D)_{ei}\,f_i\,(Y_D^\dagger)_{ie}$ where $f_i$ are real constants depending on the heavy neutrino masses. The quantity $\sum_{i=1}^3 (Y_D)_{ei}\,f_i\,(Y_D^\dagger)_{ie}$ is always real and hence no electric dipole moment arises in the model.
\\

\section{Branching Ratios of Charged Lepton Flavor Violating Processes}
\label{branchingratios}

The limits with $m_\beta/m_\alpha \to 0$ and $m_\alpha/M_Z \to 0$ are taken throughout this section.
We only consider the dominant contribution coming from one-loop diagrams of the charged scalar $H^\pm$ and heavy neutrinos $\nu_4,\nu_5,\nu_6$.

\subsection{$e_\alpha \to e_\beta \gamma$}

	\begin{figure}[H]
		\begin{center}
			\includegraphics[width=100mm]{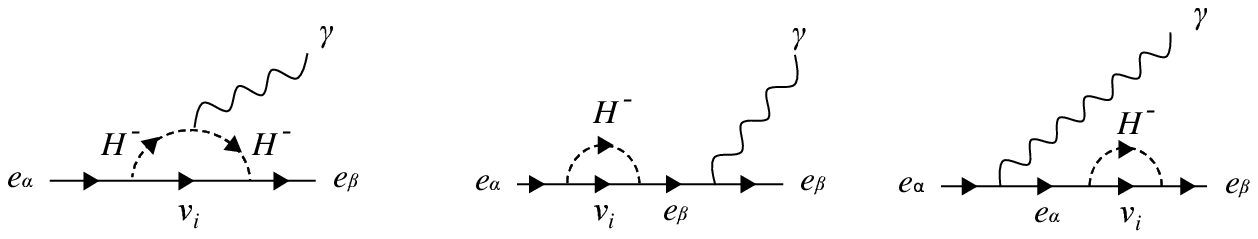}
			\end{center}
		\caption{Feynman diagrams contributing to $e_{\alpha}\to e_{\beta}\gamma$ at the one-loop level.}
		\label{etoegammapic}
	\end{figure}
CLFV decays of a charged lepton into a charged lepton and a photon, $e_\alpha \to e_\beta \gamma$, arise from the following dipole term,
 induced by loop diagrams of the charged scalar $H^\pm$ and heavy neutrinos $\nu_4,\nu_5,\nu_6$ in Fig.~\ref{etoegammapic}:
\begin{align} 
{\cal L}_{eff} &= \frac{1}{2}e\, A_D^{\beta\alpha}\, m_\alpha\bar{e}_\beta \sigma_{\mu\nu}e_\alpha F^{\mu\nu},
\\
A_D^{\beta\alpha} &= \frac{1}{16\pi^2}\frac{1}{2M_{H^\pm}^2} \sum_{i=1}^3 \, (Y_D)_{\beta i}F_2(r_i)(Y_D^\dagger)_{i\alpha}, \ \ \ \ \ 
r_i \equiv\frac{M_{N_i}^2}{M_{H^\pm}^2}, \label{AD}
\\
F_2(x) &= \frac{1-6x+3x^2+2x^3-6x^2\log x}{6(1-x)^4}. \label{F2}
\end{align}
The branching ratio is given by
\begin{align} 
Br(e_\alpha \to e_\beta \gamma) &= 48\pi^3\frac{\alpha}{G_F^2}\vert A_D^{\beta\alpha}\vert^2 Br(e_\alpha \to e_\beta \nu_\alpha \bar{\nu}_\beta).
\label{br}
\end{align}
\\

\subsection{$e_\alpha \to e_\beta \bar{e}_\beta e_\beta$}

	\begin{figure}[H]
		\begin{center}
			\includegraphics[width=100mm]{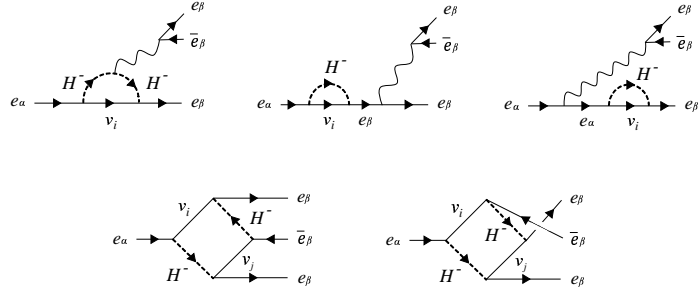}
			\end{center}
		\caption{Feynman diagrams contributing to $e_\alpha \to e_\beta \bar{e}_\beta e_\beta$ at the one-loop level.   In the upper row are $\gamma$-penguin diagrams or $Z$-penguin diagrams. However, we neglect the $Z$-penguin diagrams in Eq. (\ref{br3e}) because they are suppressed by $m_\alpha m_\beta/M_Z^2$.
	In the lower row are Box diagrams. }
	\label{etoeeepic}
	\end{figure}
CLFV decays of a charged lepton into three charged leptons, $e_\alpha \to e_\beta \bar{e}_\beta e_\beta$, arise from
 the following dipole, non-dipole and box-induced terms, induced by loop diagrams of the charged scalar and heavy neutrinos in Fig.~\ref{etoeeepic}:
\begin{align} 
{\cal L}_{eff} &= \frac{1}{2}e\, A_D^{\beta\alpha}\, m_\alpha\bar{e}_\beta \sigma_{\mu\nu}P_R e_\alpha F^{\mu\nu}
+e\, A_{ND}^{\beta\alpha}\, q^2 \, \bar{e}_\beta \gamma_\mu P_L e_\alpha A^\mu
+e^2\, B^{\beta\alpha}\, (\bar{e}_\beta \gamma_\mu P_L e_\beta)(\bar{e}_\beta \gamma^\mu P_L e_\alpha),
\label{eff2}\\
A_{ND}^{\beta\alpha} &= \frac{1}{16\pi^2}\frac{1}{6M_{H^\pm}^2} \sum_{i=1}^3 \, (Y_D)_{\beta i}G_2(r_i)(Y_D^\dagger)_{i\alpha},
\\
e^2B^{\beta\alpha} &= \frac{1}{16\pi^2}\frac{1}{4M_{H^\pm}^2} \sum_{i,j=1}^3 \, \left\{
\frac{1}{2}(Y_D)_{\beta i}(Y_D^\dagger)_{i\alpha}(Y_D)_{\beta j}(Y_D^\dagger)_{j\beta}D_1(r_i,r_j)\right.
\nonumber \\
&\left.+(Y_D^*)_{\beta i}(Y_D^\dagger)_{i\alpha}(Y_D)_{\beta j}(Y_D^T)_{j\beta}\sqrt{r_i r_j}D_2(r_i,r_j) \right\},
\\
G_2(x) &= \frac{2-9x+18x^2-11x^3+6x^3\log x}{6(1-x)^4},
\label{g2}\\
D_1(x,y)&=-\frac{x^2\log x}{(1-x)^2(x-y)}-\frac{1}{(1-x)(1-y)}-\frac{y^2\log y}{(1-y)^2(y-x)}, \label{D1}
\\
D_2(x,y)&=-\frac{x\log x}{(1-x)^2(x-y)}-\frac{1}{(1-x)(1-y)}-\frac{y\log y}{(1-y)^2(y-x)}. \label{D2}
\end{align}
The branching ratio is given by
\begin{align} 
Br(e_\alpha \to e_\beta \bar{e}_\beta e_\beta) &= 6\pi^2\frac{\alpha^2}{G_F^2}
\left\{ \ \vert A_D^{\beta\alpha}\vert^2\left(\frac{16}{3}\log\left(\frac{m_\alpha}{m_\beta}\right)-\frac{22}{3}\right)+\vert A_{ND}^{\beta\alpha}\vert^2+\frac{1}{6}\vert B^{\beta\alpha}\vert^2 \right.
\nonumber\\
&\left. +\left(-2A_D^{\beta\alpha} A_{ND}^{\beta\alpha*}+\frac{1}{3}A_{ND}^{\beta\alpha}B^{\beta\alpha*}-\frac{2}{3}A_D^{\beta\alpha} B^{\beta\alpha*}+{\rm H.c.}\right) \ \right\}
Br(e_\alpha \to e_\beta \nu_\alpha \bar{\nu}_\beta). \label{br3e}
\end{align}
Here, the contribution from the $Z$-penguin diagram is neglected because it is suppressed by $m_\alpha m_\beta/M_Z^2$ compared to 
 the contribution from the photon-penguin diagram.
\\

\subsection{$\mu N \to e N$}

\begin{figure}[H]
		\begin{center}
			\includegraphics[width=100mm]{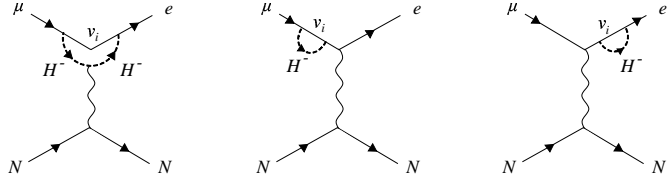}
			\end{center}
		\caption{Feynman diagrams contributing to $\mu N \to e N$ at the one-loop level.   They are $\gamma$-penguin diagrams or $Z$-penguin diagrams. However, we neglect the $Z$-penguin diagrams in Eq. (\ref{cr}) because they are suppressed by $m_\alpha m_\beta/M_Z^2$. In addition, the Higgs-penguin diagram is neglected because the up and down quark Yukawa couplings are tiny. }
		\label{munenpic}
	\end{figure}

The $\mu\to e$ conversion processes in a muonic atom arise from the dipole term $A_D$ and the non-dipole term $A_{ND}$.
We show Feynman diagrams contributing to $\mu N \to e N$ in Fig.~\ref{munenpic}.
The conversion rate divided by the muon capture rate, $CR(\mu\to e)$, reads
\begin{align} 
CR(\mu\to e) &= \frac{1}{\Gamma_{\rm capture}}\frac{p_e E_e m_\mu^3 \alpha^3 G_F^2}{8\pi^2 Z}Z_{eff}^4F_p^2
\left\vert (Z+N)g_{LV}^{(0)}+(Z-N)g_{LV}^{(1)} \right\vert^2,
\label{cr}\\
g_{LV}^{(0)} &= \frac{1}{2}\sum_{q=u,d}\left(g_{LV}^{(q)}G_V^{(q,p)}+g_{LV}^{(q)}G_V^{(q,n)}\right),
\ \ \ g_{LV}^{(1)} = \frac{1}{2}\sum_{q=u,d}\left(g_{LV}^{(q)}G_V^{(q,p)}-g_{LV}^{(q)}G_V^{(q,n)}\right),
\\
g_{LV}^{(q)} &= \frac{\sqrt{2}}{G_F}e^2 Q_q (A_{ND}^{\mu e}-A_D^{\mu e}) ,
\end{align}
 where $p_e$ and $E_e$ are the momentum and energy of the final state electron, and $Z$ and $N$ are the number of protons and neutrons, respectively.
 $Z_{eff}$ is the effective atomic charge, $F_p$ is the nuclear matrix element, and $g_{LV}^{(0)},g_{LV}^{(1)}$ are effective charges.
$\Gamma_{\rm capture}$ denotes the muon capture rate, and $ Q_q{\rm \ is \ the \ electric \ charge \ of \ quark} \ q$.
Here, the contribution from the $Z$-penguin diagram is again neglected, and that from the Higgs-penguin diagram is neglected because the up and down quark Yukawa couplings are tiny.
Also, since $\cos\beta\simeq0$, box diagrams involving two quarks and two leptons do not contribute.
\\

\subsection{$Z \to \bar{e}_\alpha e_\beta$}

\begin{figure}[H]
		\begin{center}
			\includegraphics[width=100mm]{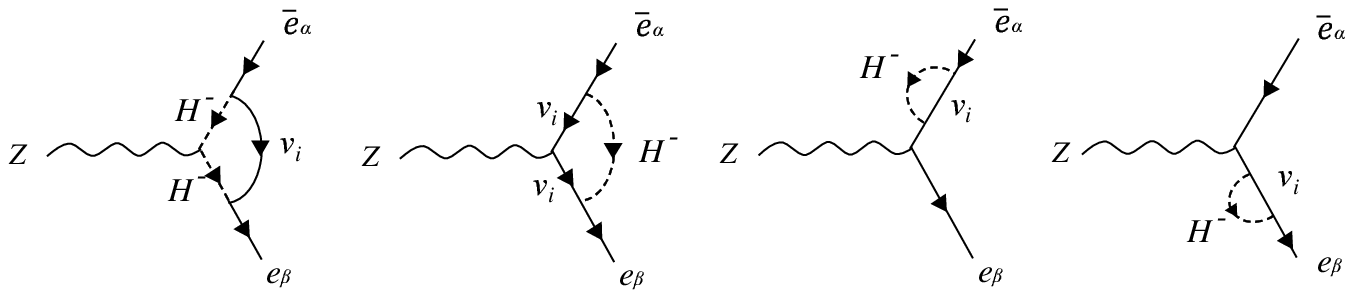}
			\end{center}
		\caption{Feynman diagrams contributing to $Z \to \bar{e}_\alpha e_\beta$ at the one-loop level.}
		\label{zeepic}
	\end{figure}

CLFV decays of a Z boson arise from the non-dipole term $A_{ND}$.
We show Feynman diagrams contributing to $Z \to \bar{e}_\alpha e_\beta$ in Fig.~\ref{zeepic}.
In the leading order of $M_Z^2/M_{H^\pm}^2$, the effective Lagrangian contributing to $Z\to \bar{e}_\alpha e_\beta$ decay is given by
\begin{align} 
{\cal L}_{eff} &= -A_{ND}^{\beta\alpha}
\left(-\frac{1}{2}+\sin^2\theta_W\right)g_Z \ \bar{e}_\beta\gamma_\mu P_L e_\alpha \ Z^\mu.
\end{align}
The branching ratio for $Z\to \bar{e}_\alpha e_\beta$ is
\begin{align} 
Br(Z\to \bar{e}_\alpha e_\beta) &= Br(Z\to \bar{e}_\alpha e_\alpha)\frac{g_{eL}^2}{g_{eL}^2+g_{eR}^2}
M_Z^4\left\vert A_{ND}^{\beta\alpha}\right\vert^2,
\\
(\ g_{eL}&=-\frac{1}{2}+\sin^2\theta_W, \ \ \ g_{eR}=\sin^2\theta_W \ ).
\nonumber
\end{align}
\\

\subsection{$h \to \bar{e}_\alpha e_\beta$}

\begin{figure}[H]
		\begin{center}
			\includegraphics[width=100mm]{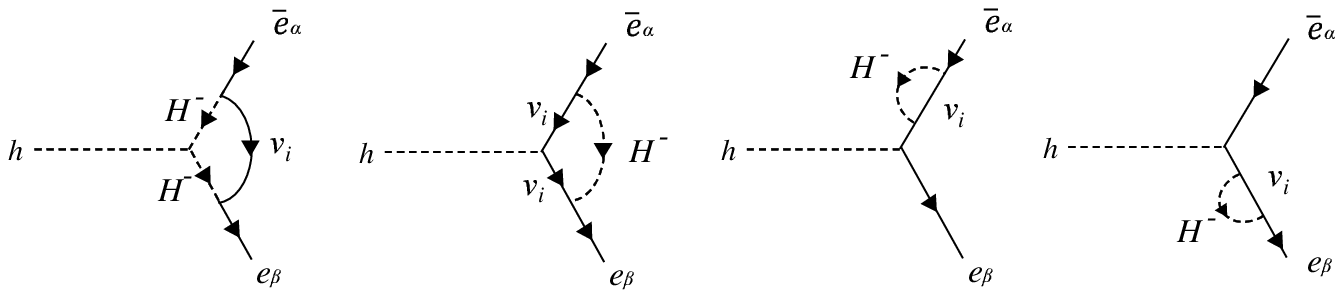}
			\end{center}
		\caption{Feynman diagrams contributing to $h \to \bar{e}_\alpha e_\beta$ at the one-loop level.}
		\label{heepic}
	\end{figure}

We show Feynman diagrams contributing to $h \to \bar{e}_\alpha e_\beta$ in Fig.~\ref{heepic}.
In the leading order of $m_h^2/M_{H^\pm}^2$, the effective Lagrangian contributing to $h\to \bar{e}_\alpha e_\beta$ decay is given by
\begin{align} 
{\cal L}_{eff} &= \frac{1}{16\pi^2} \frac{\lambda_3v \, m_\alpha}{M_{H^\pm}^2}\sum_{i=1}^3(Y_D)_{\beta i}G_H(r_i)(Y_D^\dagger)_{i\alpha}
\ \bar{e}_\beta P_R e_\alpha \ h,
\\
G_H(x) &= \frac{1-4x+3x^2-2x^2\log x}{4(1-x)^3},
\end{align}
 where $\lambda_3$ is the scalar quartic coupling that appears in Eq.~(\ref{lagrangian}).
$G_H$ is a novel function different from $F_2$ in $A_D$ or $G_2$ in $A_{ND}$.
The branching ratio for $h\to \bar{e}_\alpha e_\beta$ is 
\begin{align} 
Br(h\to \bar{e}_\alpha e_\beta) &= Br(h\to \bar{e}_\alpha e_\alpha)\frac{1}{2}
\left\vert \frac{1}{16\pi^2}\frac{1}{M_{H^\pm}^2}\frac{\lambda_3 v^2}{\sqrt{2}}\sum_{i=1}^3(Y_D)_{\beta i}G_H(r_i)(Y_D^\dagger)_{i\alpha} \right\vert^2.
\label{brh}
\end{align}
\\

\section{Numerical Study}

We investigate CLFV processes in the neutrinophilic Higgs+seesaw model,
 based on the branching ratio formulas in Sect.~\ref{branchingratios}.
First, we use current experimental upper limits on CLFV branching ratios to constrain the parameters of the model.
Next, under the above constraint, we predict the branching ratios of various CLFV processes
 including $\mu\to3e$, $\mu+{\rm Al}\to e+{\rm Al}$, $\mu+{\rm Ti}\to e+{\rm Ti}$, $Z\to e\mu$, $Z\to e\tau$, $Z\to \mu\tau$, $h\to \mu\tau$ and $h\to e\tau$,
 and assess their detectability in the future.

\subsection{Assumptions on the Model Parameters}
\label{assumptions}

The branching ratio formulas of CLFV processes
 depend on the neutrino Dirac Yukawa matrix Eq.~(\ref{yukawa}),
 the charged scalar mass $m_{H^{\pm}}$ and the right-handed neutrino Majorana masses $M_{N_1}, M_{N_2} $ and $M_{N_3}$.
The neutrino Dirac Yukawa matrix depends on 
 $v_2$, $M_{N_1}$, $M_{N_2}$, $M_{N_3}$, $m_{\nu_1}$, $m_{\nu_2}$, $m_{\nu_3}$ and $U_{PMNS}$ as well as a complex-valued $3\times3$ rotation matrix $R_{\rm 3 \times 3 }$.
There are too many parameters and it is not easy to gain physical insight to the phenomenology of the model.
Therefore, we reduce the number of parameters by considering the following situation.

For the charged scalar mass $m_{H^\pm}$, 
 the most phenomenologically interesting situation is when the charged scalar particle is detectable at the LHC.
Hence, we assume
\begin{align}
m_{H^\pm}=0.3 ~{\rm TeV}.
\end{align}

For the tiny active neutrino masses $m_{\nu_1}, ~m_{\nu_2} ~{\rm and}~ m_{\nu_3}$,
 we consider both Normal Hierarchy (NH) and Inverse Hierarchy (IH) cases, while focusing on the case where the lightest neutrino mass is 0;
 namely, we assume
\begin{align}
m_{\nu_1} = 0 ~({\rm NH}),~m_{\nu_3} = 0 ~({\rm IH})
\end{align}
The values of $m_{\nu_2}$ and $m_{\nu_3}$ ($m_{\nu_1}$ and $m_{\nu_2}$) in the NH (IH) case
 are obtained from the mass differences measured in neutrino oscillation experiments.
In this paper, we employ the central values of the mass differences given in NuFIT~4.1~\cite{Esteban:2018azc,nufit}.

For the parameters of $U_{PMNS}$, we employ the central values of the three mixing angles in NuFIT~4.1~\cite{Esteban:2018azc,nufit}.
As benchmark values of the Dirac phase $\delta$, we take
 the 3$\sigma$ bounds and central value in the NuFIT~4.1 result~\cite{Esteban:2018azc,nufit} as
\begin{align}
\delta&=144^\circ,~221^\circ,~357^\circ ~({\rm NH})\\
\delta&=205^\circ,~282^\circ,~348^\circ ~({\rm IH})
\end{align}
We set the Majorana phase to be 0.

For the Majorana masses of right-handed neutrinos, we assume them to be degenerate as
\begin{align}
 M_{N_1}=M_{N_2}=M_{N_3}=M_N, \label{a2}
\end{align} 
 where $M_N$ is taken as real positive by a phase redefinition.
We have found numerically that the branching ratios of CLFV processes do not change significantly
 even when the Majorana masses are hierarchical as $ M_{N_1}=0.1M_N$, $M_{N_2}=M_{N_3}=M_N$ or $M_{N_1}=M_{N_2}=0.1M_N$ and $M_{N_3}=M_N$.

For the neutrinophilic Higgs VEV $v_2$, we take it to be proportional to $\sqrt{M_N}$ as
\begin{align}
v_2=1\times10^{-6}\times&\sqrt{\frac{M_N}{{\rm TeV}}} ~ {\rm TeV}~({\rm NH}),\label{v2-nh}\\
v_2=2\times10^{-6}\times&\sqrt{\frac{M_N}{{\rm TeV}}} ~ {\rm TeV}~({\rm IH}).
\label{v2-ih}
\end{align}
These values of $v_2$ ensure $\left|Y_D\right|\sim0.05$ in each hierarchy,
 where we have defined $\left|Y_D\right|$ as the minimum absolute value of the Yukawa matrix components when 
 Im$\theta_1=$Im$\theta_2=$Im$\theta_3=0$.
Note that the motivation for the neutrinophilic Higgs+seesaw model is to realize low-scale seesaw without taking very small values for the neutrino Dirac Yukawa coupling,
 hence it is essential to have $\left|Y_D\right|$ not much smaller than 1.

For $R_{3\times3}$, we parametrize it in terms of three complex rotation angles $\theta_j ={\rm Re}\theta_j+\mathrm{i}{\rm Im}\theta_j$ $(j=1,2,3)$ as
\begin{align}
R_{3\times3}
  =
  \left(
    \begin{array}{ccc}
      1 & 0 & 0 \\
      0 & \cos{\theta_1} & -\sin{\theta_1} \\
      0 & \sin{\theta_1} & \cos{\theta_1}
    \end{array}
  \right)
  \left(
    \begin{array}{ccc}
      \cos{\theta_2} & 0 & -\sin{\theta_2} \\
      0 & 1 & 0 \\
      \sin{\theta_2} & 0 & \cos{\theta_2}
    \end{array}
  \right)
  \left(
    \begin{array}{ccc}
      \cos{\theta_3} & -\sin{\theta_3} & 0 \\
      \sin{\theta_3} & \cos{\theta_3} & 0 \\
      0 & 0 & 1
    \end{array}
  \right)
  \label{r33}
\end{align}
For the sake of simplifying the analysis,
 we vary each $\theta_j$ separately while fixing the other complex angles at zero. 
When we vary each $\theta_j$, its real part Re$\theta_j$ does not affect the branching ratios of CLFV processes, which is understood as follows:
Let us focus on the case where we vary $\theta_1$ while fixing $\theta_2=\theta_3=0$. $R_{3\times3}$ can then be decomposed as
\begin{align}
R_{3\times3}
  =
  \left(
    \begin{array}{ccc}
      1 & 0 & 0 \\
      0 & \cos(\mathrm{i}{\rm Im}\theta_1) & -\sin(\mathrm{i}{\rm Im}\theta_1) \\
      0 & \sin(\mathrm{i}{\rm Im}\theta_1) & \cos(\mathrm{i}{\rm Im}\theta_1)
    \end{array}
  \right)
  \left(
    \begin{array}{ccc}
      1 & 0 & 0 \\
      0 & \cos({\rm Re}\theta_1) & -\sin({\rm Re}\theta_1) \\
      0 & \sin({\rm Re}\theta_1) & \cos({\rm Re}\theta_1)
    \end{array}
  \right).
  \label{matrixdecom}
\end{align}
Since we are assuming that the Majorana masses are degenerate as Eq.~(\ref{a2}), the matrix with ${\rm Re}\theta_1$ in Eq.~(\ref{matrixdecom}) cancels
 in the combination $\sum_{i=1}^3(Y_D)_{\beta i}f(r_i)(Y_D^\dagger)_{i\alpha}$ where $f$ is any function and $r_i=M_{N_i}^2/M_{H^\pm}^2$.
Therefore, we only regard the imaginary parts Im$\theta_1$, Im$\theta_2$ and Im$\theta_3$ as the parameters of $R_{3\times3}$. 
The larger the absolute value of Im$\theta_j$ is, the larger $Y_D$ becomes. 
Thus, to maintain perturbativity, we restrict the range as $-2<$ Im$\theta_j<2$.

The above are our assumptions on the model parameters. 
Consequently, for one CLFV process such as $\mu\to3e$, we show 18 plots on ($M_N$, Im$\theta_j$)-parameter space 
[$3~(\delta)\times3~({\rm Im}\theta)\times2~({\rm NH,~IH})=18$].
\\

\subsection{Constraints on the Neutrinophilic Higgs+Seesaw Model from Charged Lepton Flavor Violating Processes}

The CLFV processes experimentally searched for are $e_\alpha \to e_\beta \gamma,~e_\alpha \to 3e_\beta,~ \mu N \to e N,~Z \to \bar{e}_\alpha e_\beta$ and $h \to \bar{e}_\alpha e_\beta$.
For each process, the upper limit on the branching ratio (or conversion rate) is obtained by experiments and it constrains the model parameter space.
At present, the strongest constraint comes from the upper limit on the $\mu\to e\gamma$ branching ratio,
 $Br(\mu\to e\gamma)<4.2\times10^{-13}$~\cite{meg}, in the entire parameter space.
Therefore, in the study of the current experimental constraints, we can concentrate on the $\mu\to e\gamma$ process
 while neglecting bounds from other CLFV processes~\cite{babar}-\cite{conversionAu}.

The constraint on the $(M_{N}, {\rm Im}\theta_j)$-parameter space from the bound $Br(\mu\to e\gamma)<4.2\times10^{-13}$
  is displayed by the blue solid line in every figure,
  for both NH and IH, for $m_{H^\pm}=0.3$ TeV, $v_2=1~(2)\times10^{-6}\times\sqrt{\frac{M}{{\rm TeV}}}$ TeV in NH (IH),
  and for the benchmark values of the Dirac phase $\delta$.
Additionally, we show the constraint when $v_2$ is multiplied by $1/3$ and thus $Y_D$ is uniformly multiplied by 3,
 by the dashed blue line.

We observe that the constraint tends to be weaker for smaller $M_N$ and $|$Im$\theta_j|$.
This is because $Y_D$ is proportional to $\sqrt{M_N}$ and $R_{3\times3}$ (see Eq. (\ref{yukawa})),
 and so $Br(\mu\to e\gamma)$ is suppressed for small $M_N$ and $|$Im$\theta_j|$.
\\

\subsection{Prediction on Charged Lepton Flavor Violating Processes}

\subsubsection{$ \mu \to 3e$}

\begin{figure}[H]
  \centering
  \thispagestyle{empty}
    \begin{tabular}{c}
 
 
      \begin{minipage}{0.33\hsize}
        \centering
          \includegraphics[keepaspectratio, scale=0.45, angle=0]
                          {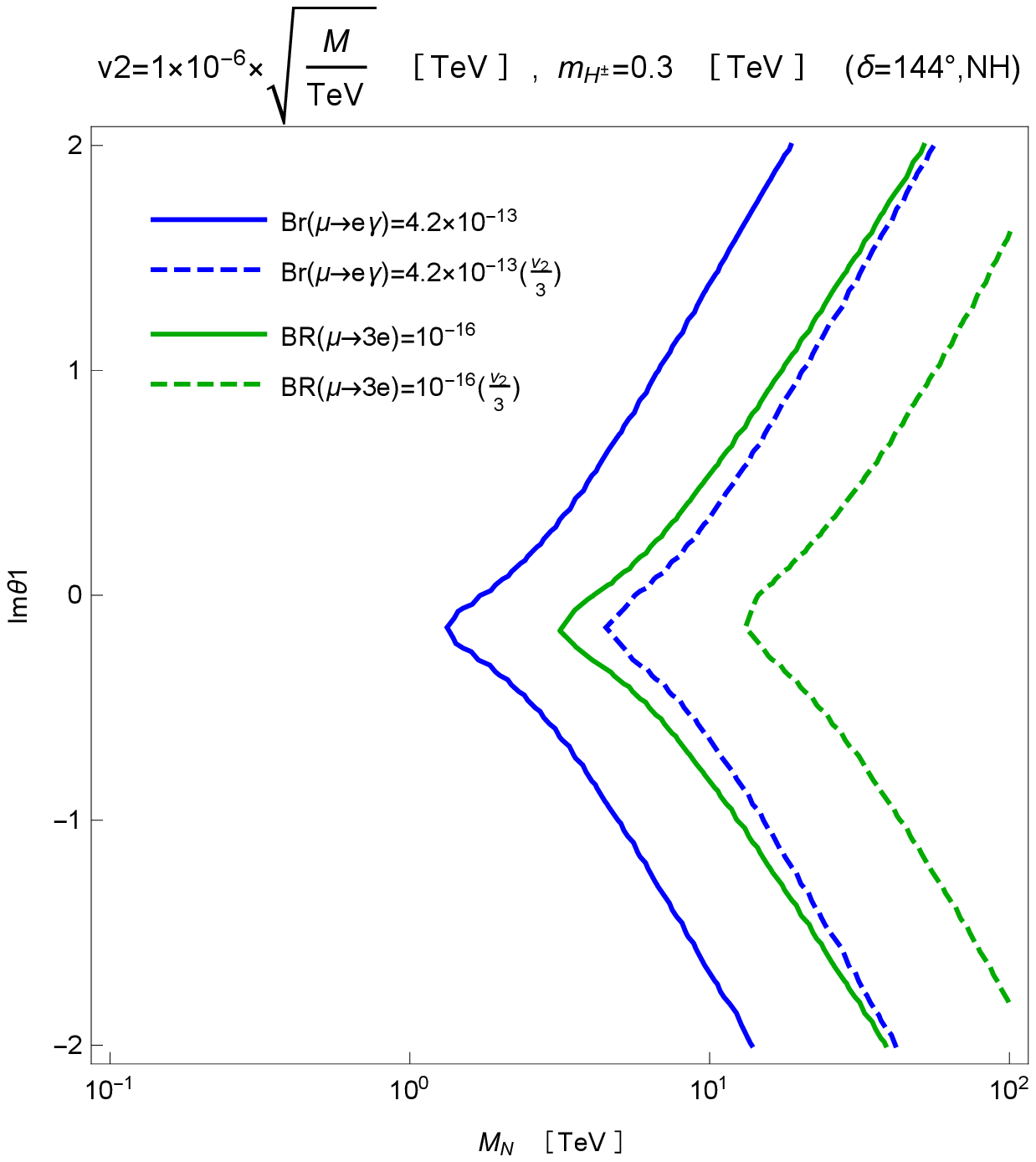}
      \end{minipage}

 
      \begin{minipage}{0.33\hsize}
        \centering
          \includegraphics[keepaspectratio, scale=0.44, angle=0]
                          {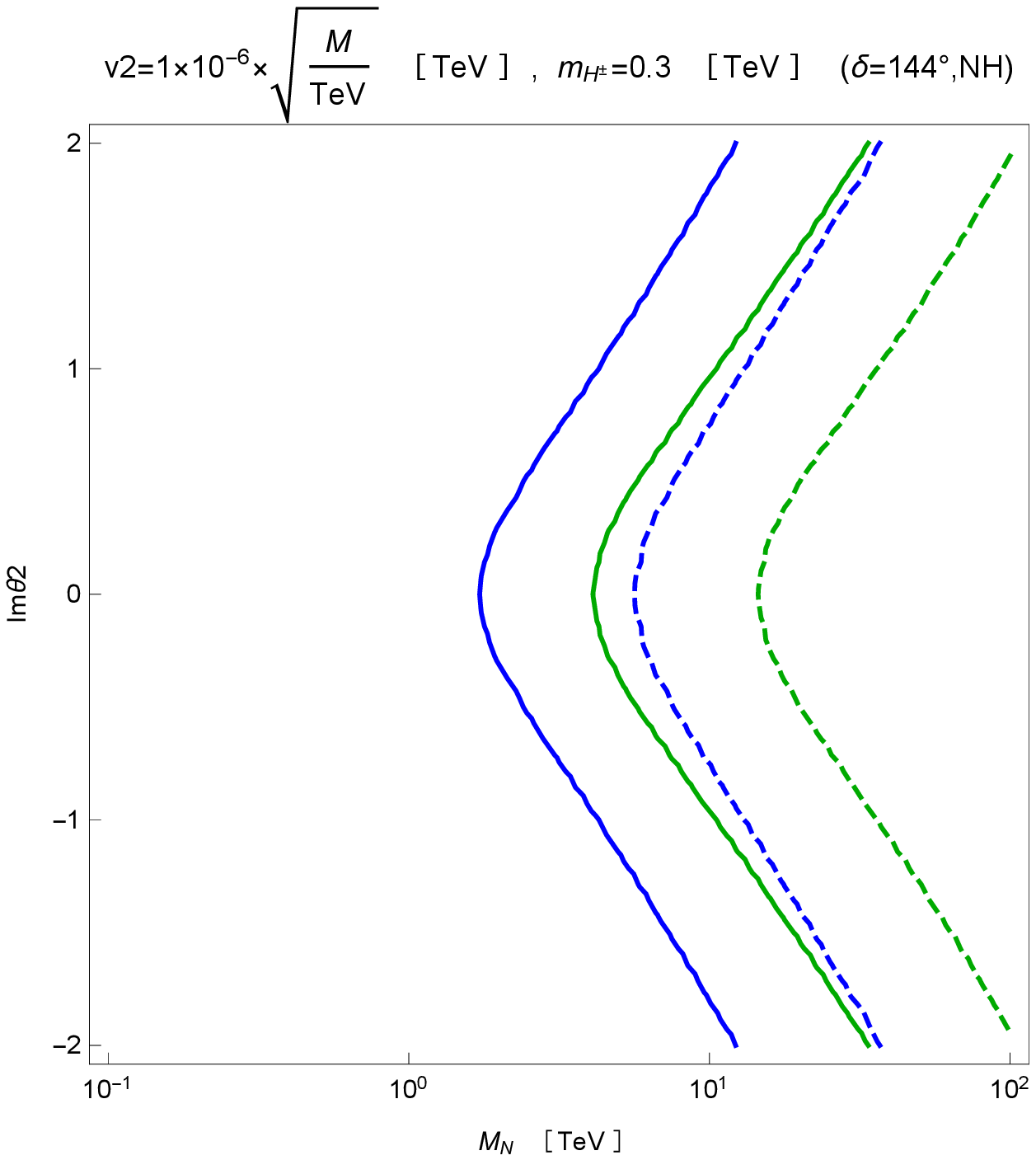}

      \end{minipage}
 
 
      \begin{minipage}{0.33\hsize}
        \centering
          \includegraphics[keepaspectratio, scale=0.44, angle=0]
                          {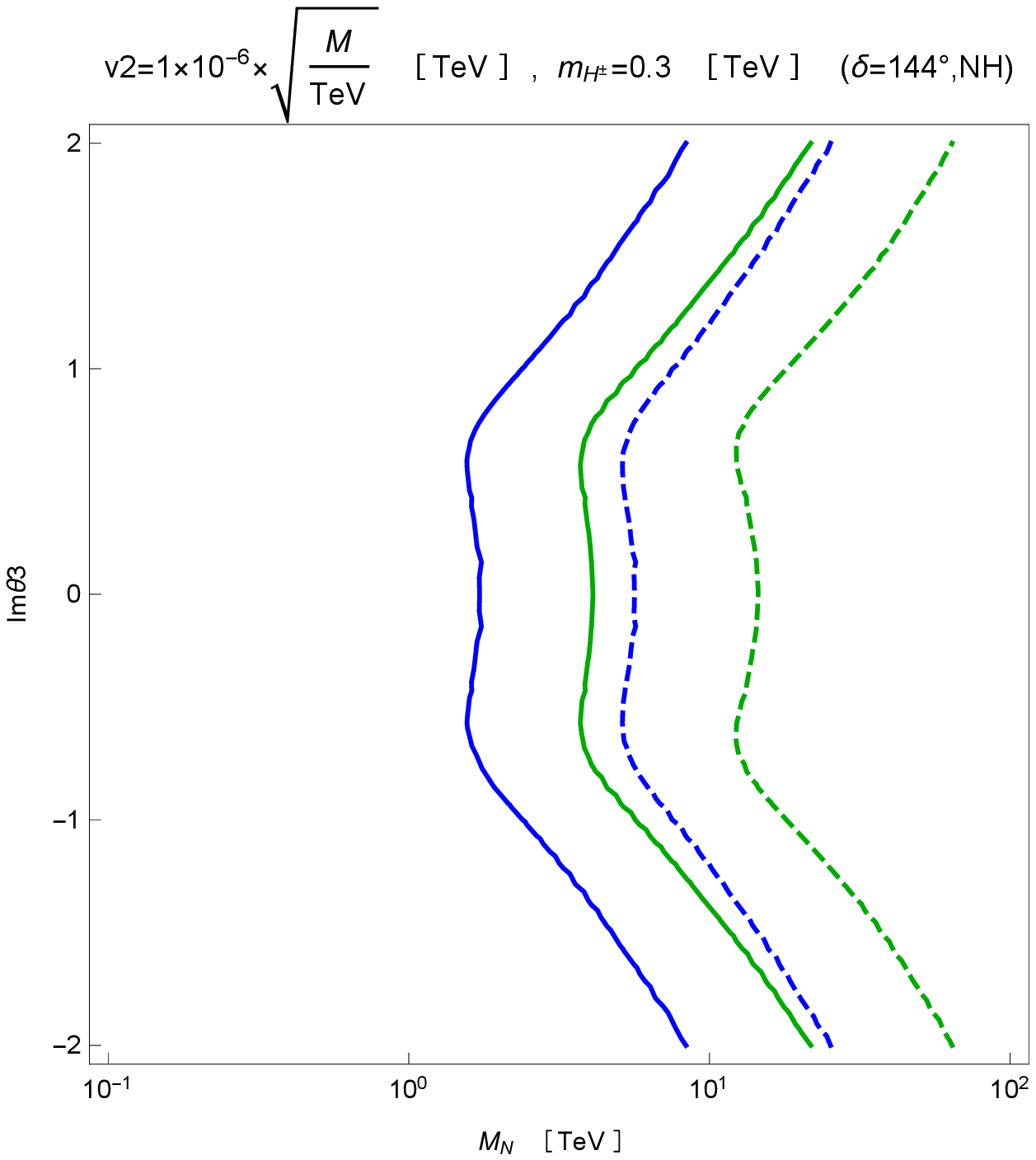}
      \end{minipage} \\
      \\
 
      \begin{minipage}{0.33\hsize}
        \centering
          \includegraphics[keepaspectratio, scale=0.44, angle=0]
                          {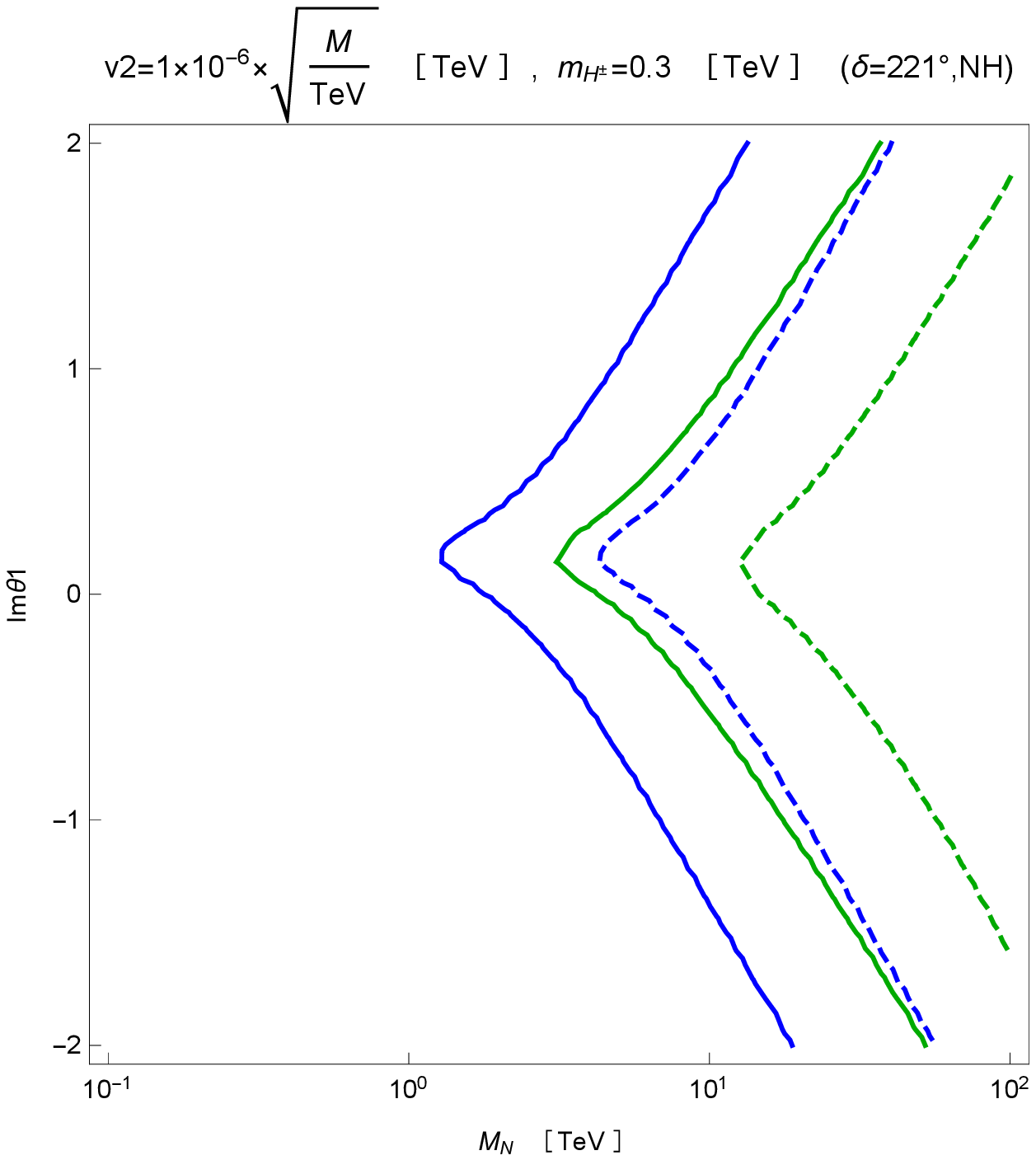}
      \end{minipage}

 
      \begin{minipage}{0.33\hsize}
        \centering
          \includegraphics[keepaspectratio, scale=0.44, angle=0]
                          {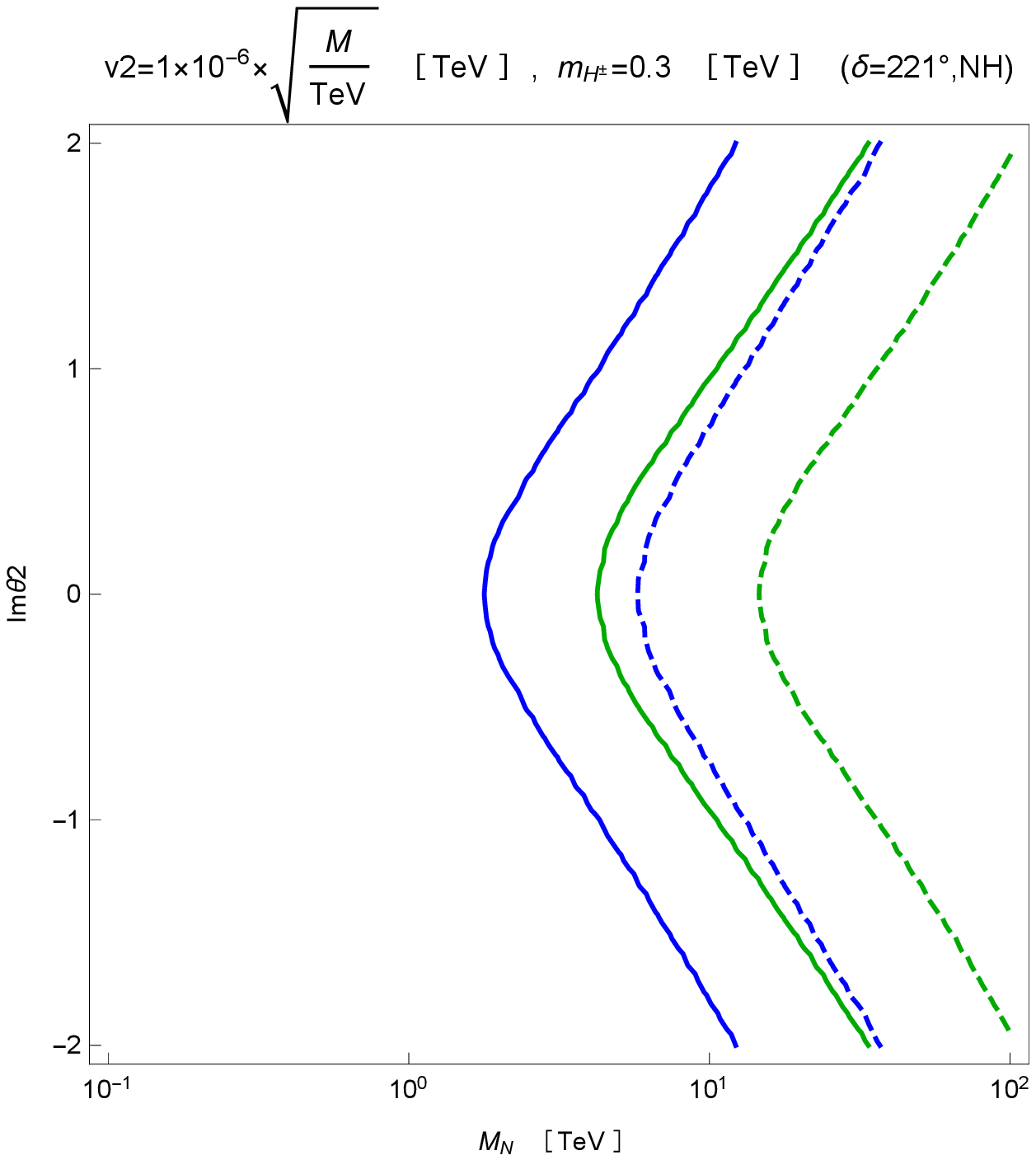}
      \end{minipage}
 
 
      \begin{minipage}{0.33\hsize}
        \centering
          \includegraphics[keepaspectratio, scale=0.44, angle=0]
                          {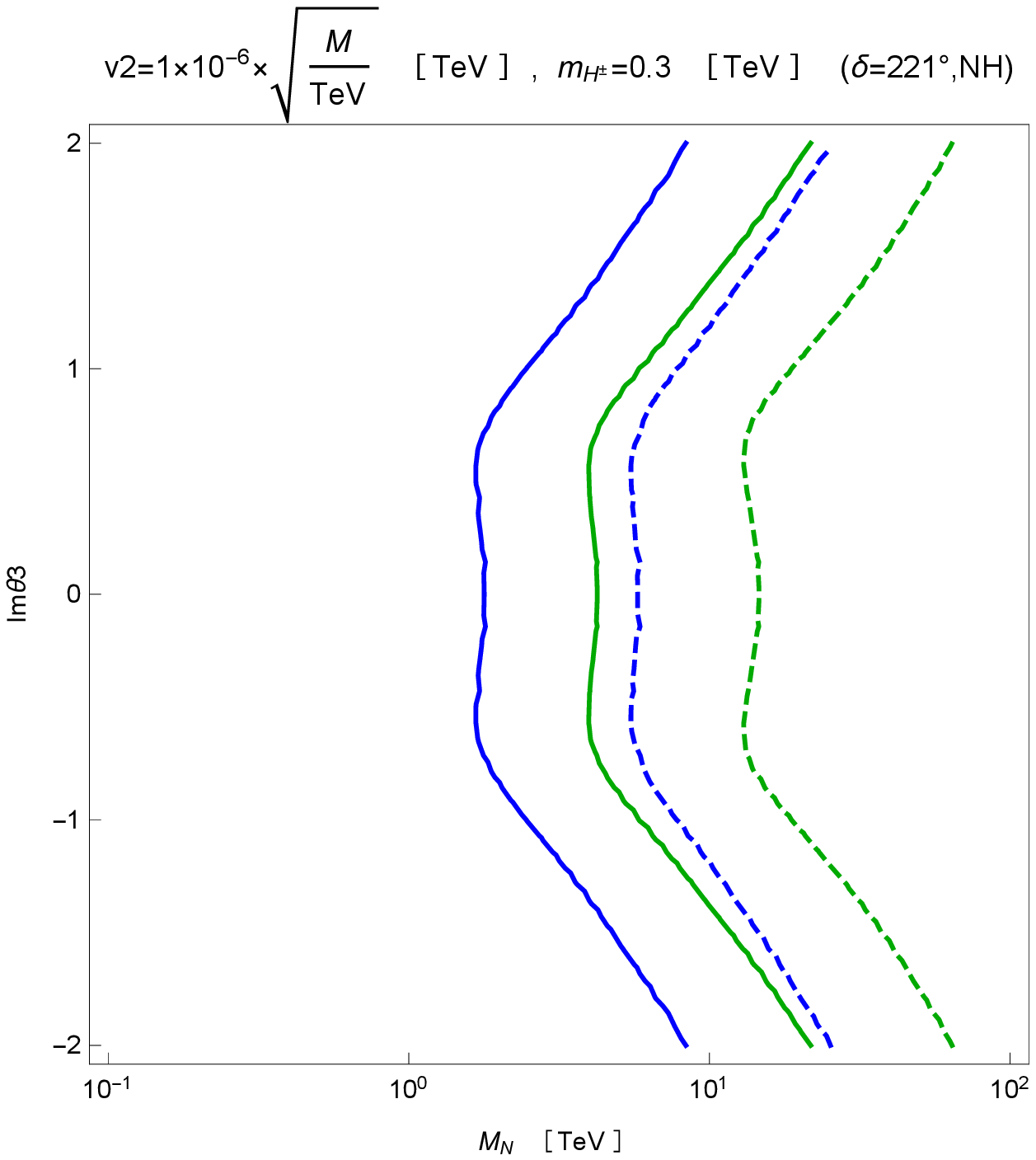}
      \end{minipage} \\ 
      \\
 
      \begin{minipage}{0.33\hsize}
        \centering
          \includegraphics[keepaspectratio, scale=0.44, angle=0]
                          {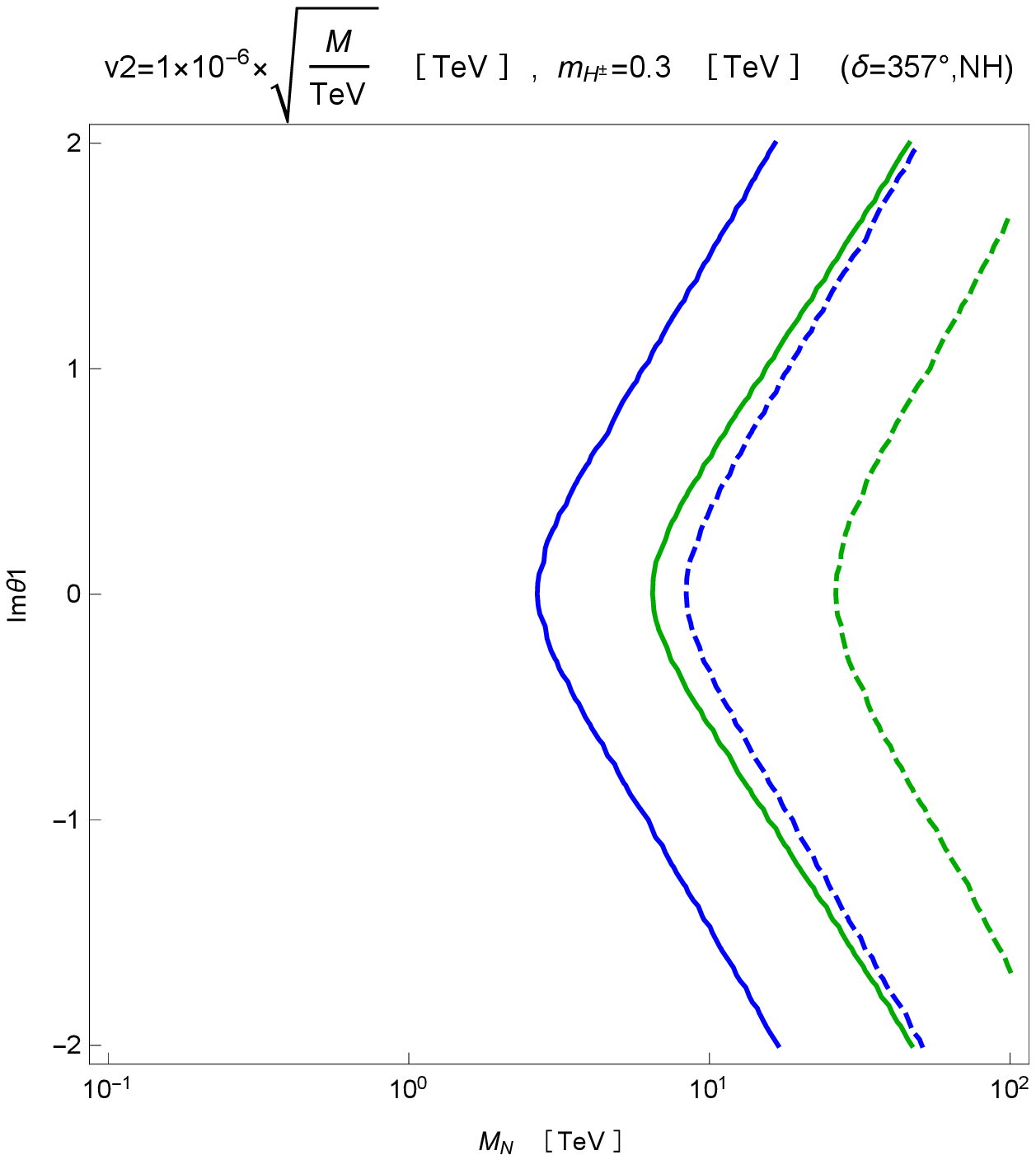}
      \end{minipage}

 
      \begin{minipage}{0.33\hsize}
        \centering
          \includegraphics[keepaspectratio, scale=0.44, angle=0]
                          {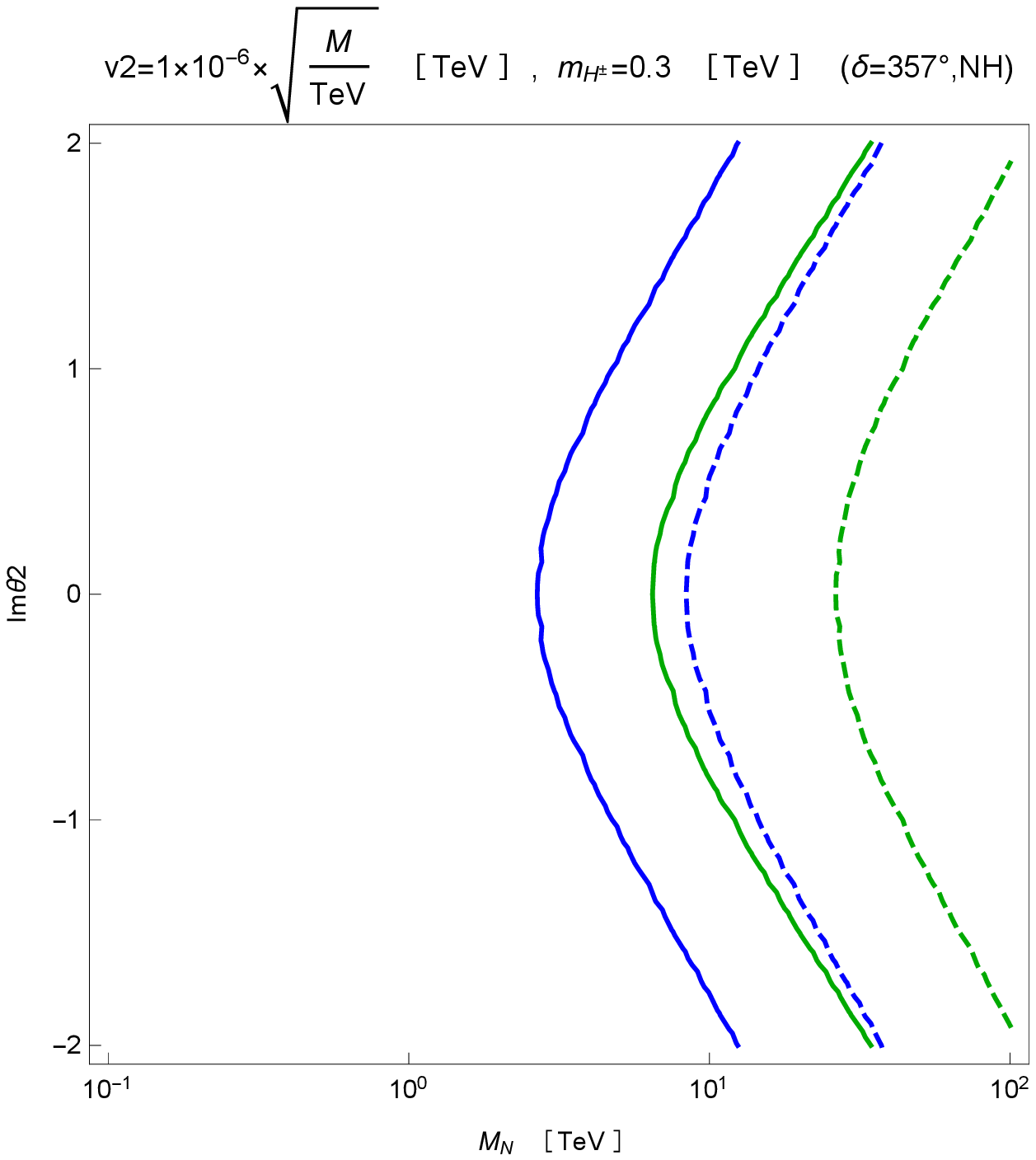}
      \end{minipage}
 
 
      \begin{minipage}{0.33\hsize}
        \centering
          \includegraphics[keepaspectratio, scale=0.44, angle=0]
                          {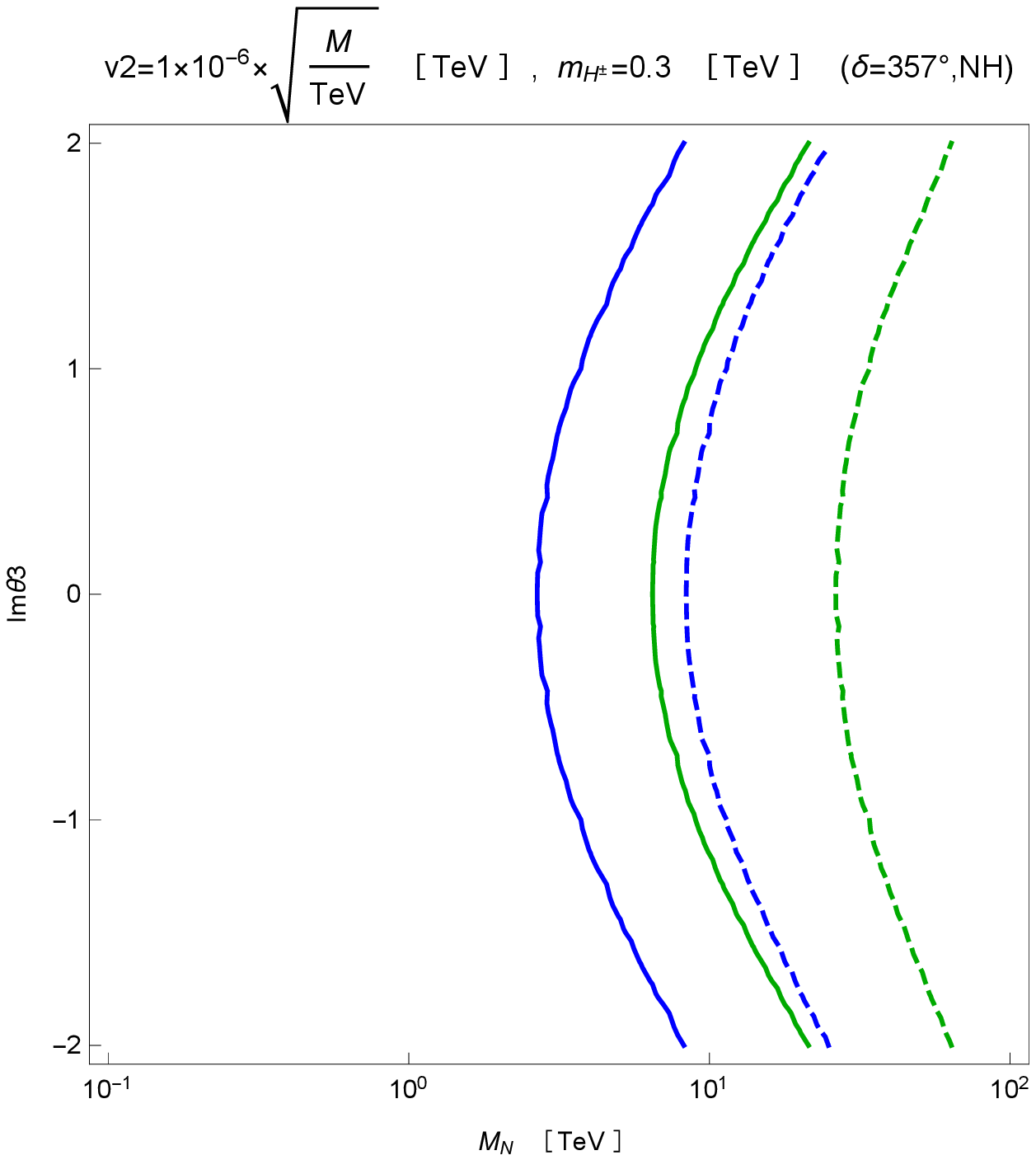}
      \end{minipage} 

    \end{tabular}
     \caption{\footnotesize Prediction for $Br(\mu\to3e)$, along with the values of $Br(\mu\to e\gamma)$. 
     The neutrino mass hierarchy is Normal Hierarchy, and we fix $m_{H^\pm}=0.3$ TeV. We take $\delta=144^\circ,~221^\circ~{\rm and}~357^\circ$ in the first, second and third rows. In the first column, we vary Im$\theta_1\neq0$ while fixing Im$\theta_2$=Im$\theta_3=0$. 
In the second column, we vary Im$\theta_2\neq0$ while fixing Im$\theta_1$=Im$\theta_3=0$. In the third column, we vary Im$\theta_3\neq0$ while fixing Im$\theta_1$=Im$\theta_2=0$.
The blue solid line corresponds to $Br(\mu\to e\gamma)=4.2\times10^{-13}$ for $v_2$ in Eq.~(\ref{v2-nh}), 
 and the region on the left of the blue solid line is excluded by the search for $Br(\mu\to e\gamma)$.
The green solid line corresponds to $Br(\mu\to3e)=10^{-16}$, the future sensitivity, for $v_2$ in Eq.~(\ref{v2-nh}).
The blue dashed line corresponds to $Br(\mu\to e\gamma)=4.2\times10^{-13}$ and the green dashed line corresponds to $Br(\mu\to3e)=10^{-16}$
 in the case when $v_2$ is multiplied by $1/3$ and thus $Y_D$ is uniformly multiplied by 3.
}
 \label{figpreeN}
\end{figure}

\begin{figure}[H]
  \centering
    \begin{tabular}{c}
 
 
      \begin{minipage}{0.33\hsize}
        \centering
          \includegraphics[keepaspectratio, scale=0.45, angle=0]
                          {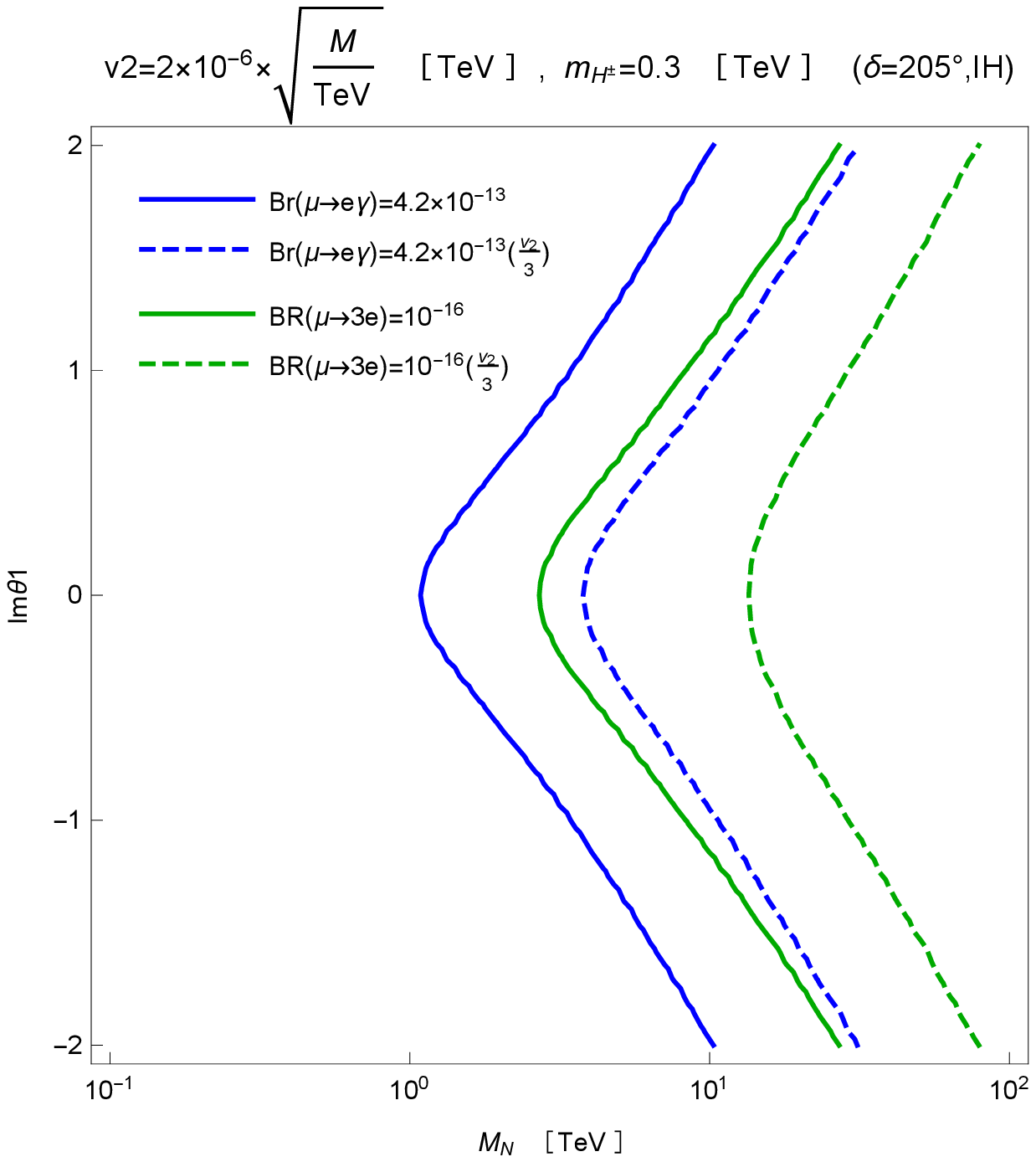}
      \end{minipage}

 
      \begin{minipage}{0.33\hsize}
        \centering
          \includegraphics[keepaspectratio, scale=0.44, angle=0]
                          {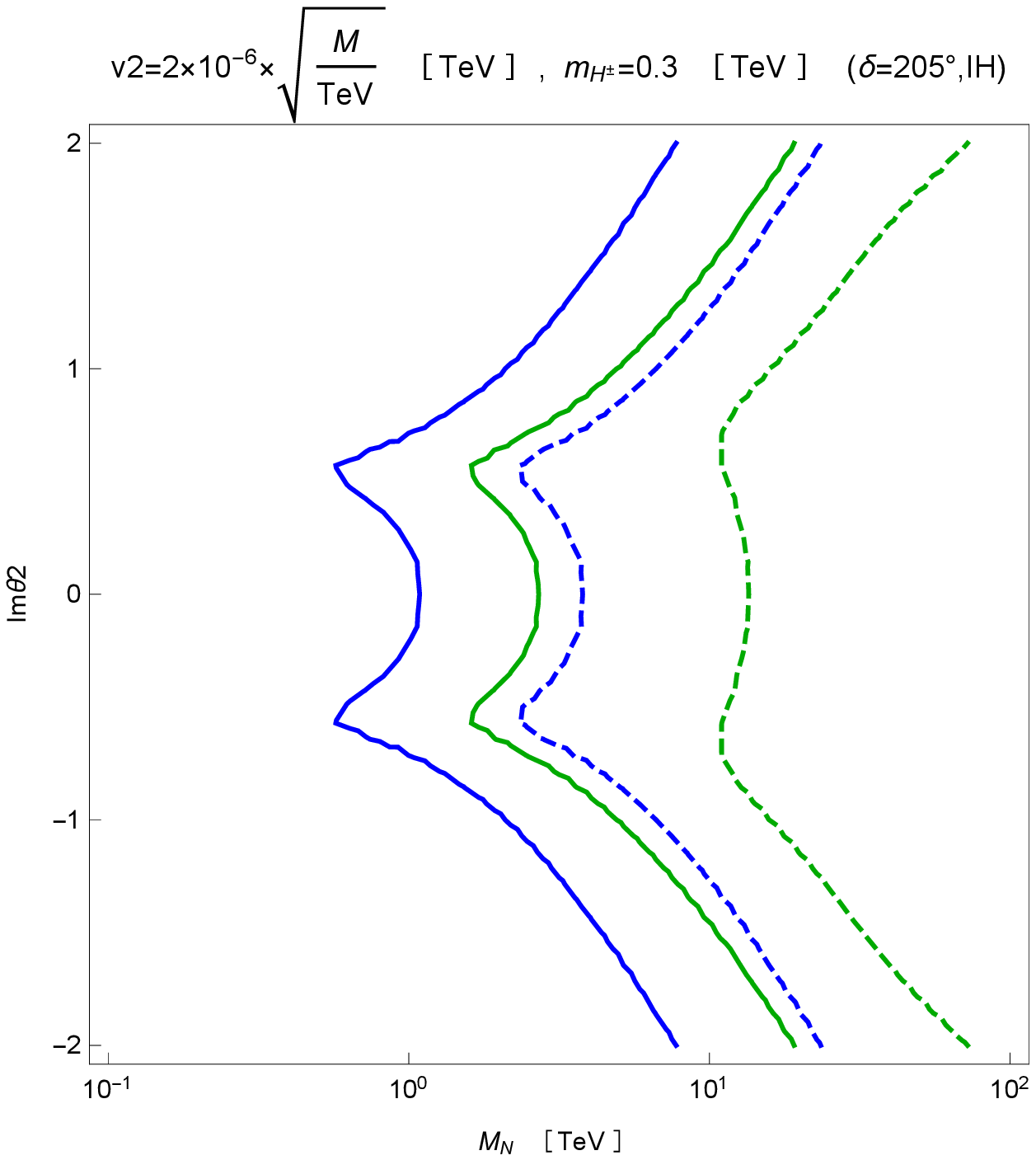}

      \end{minipage}
 
 
      \begin{minipage}{0.33\hsize}
        \centering
          \includegraphics[keepaspectratio, scale=0.44, angle=0]
                          {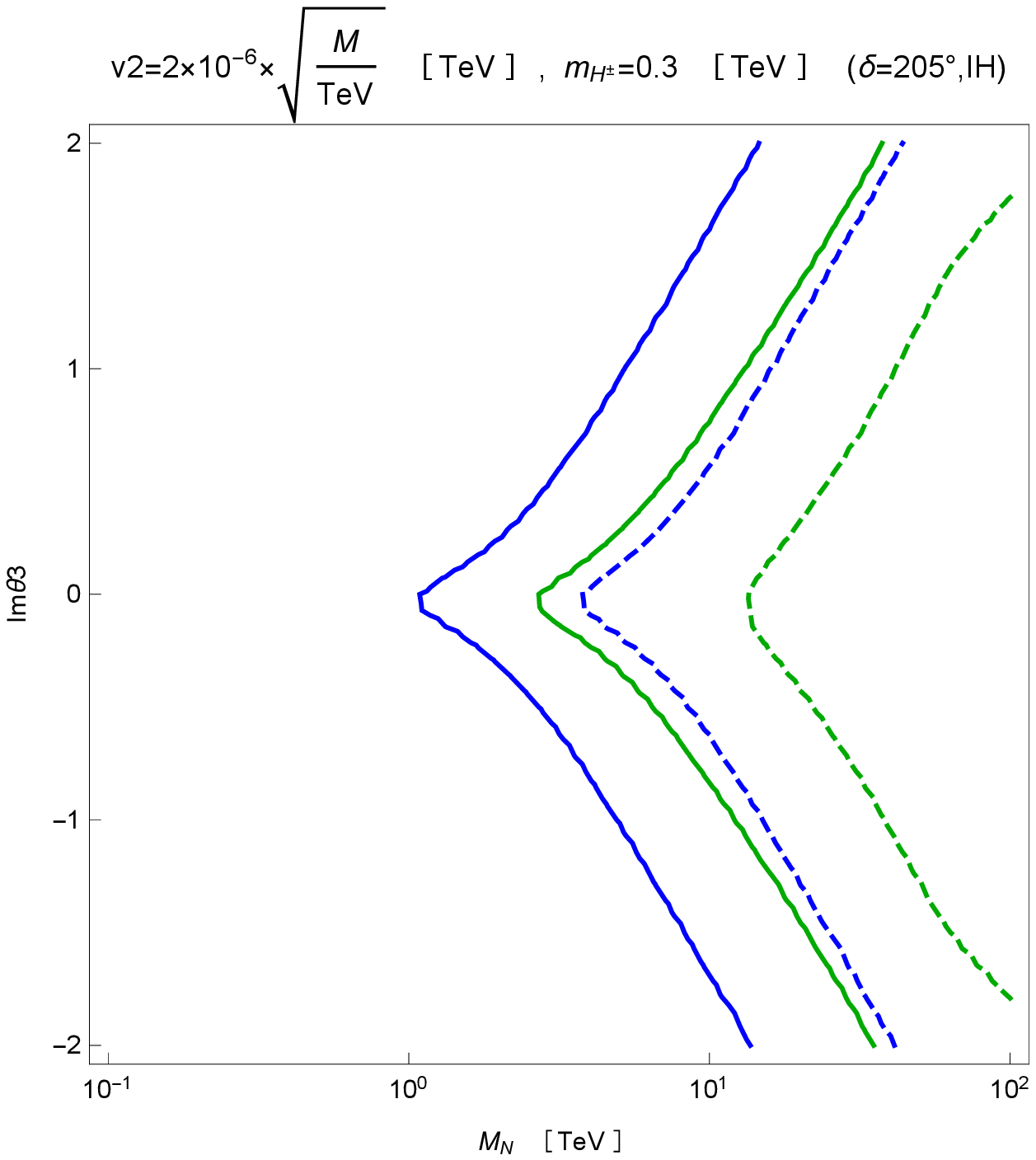}
      \end{minipage} \\
      \\
 
      \begin{minipage}{0.33\hsize}
        \centering
          \includegraphics[keepaspectratio, scale=0.44, angle=0]
                          {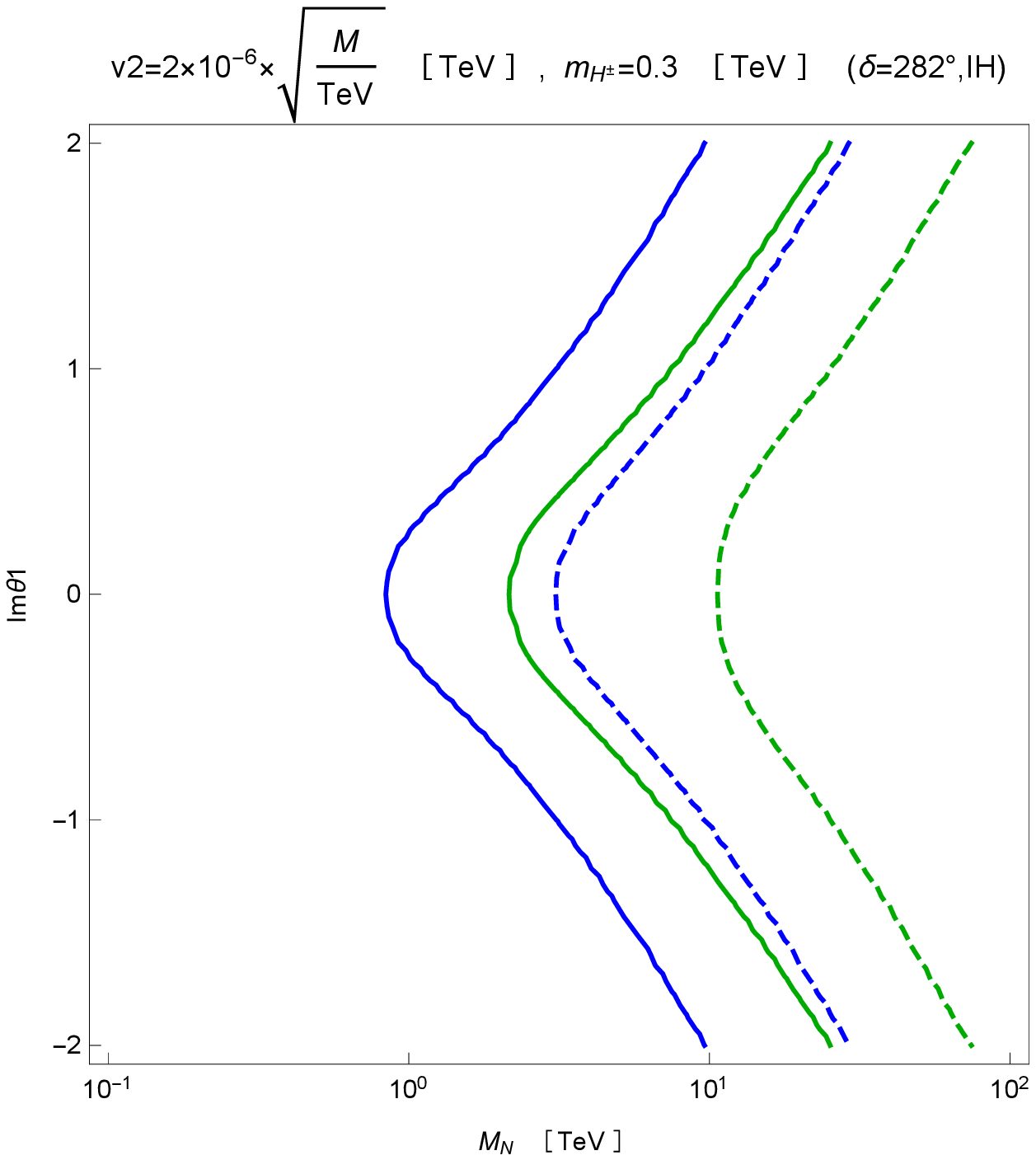}
      \end{minipage}

 
      \begin{minipage}{0.33\hsize}
        \centering
          \includegraphics[keepaspectratio, scale=0.44, angle=0]
                          {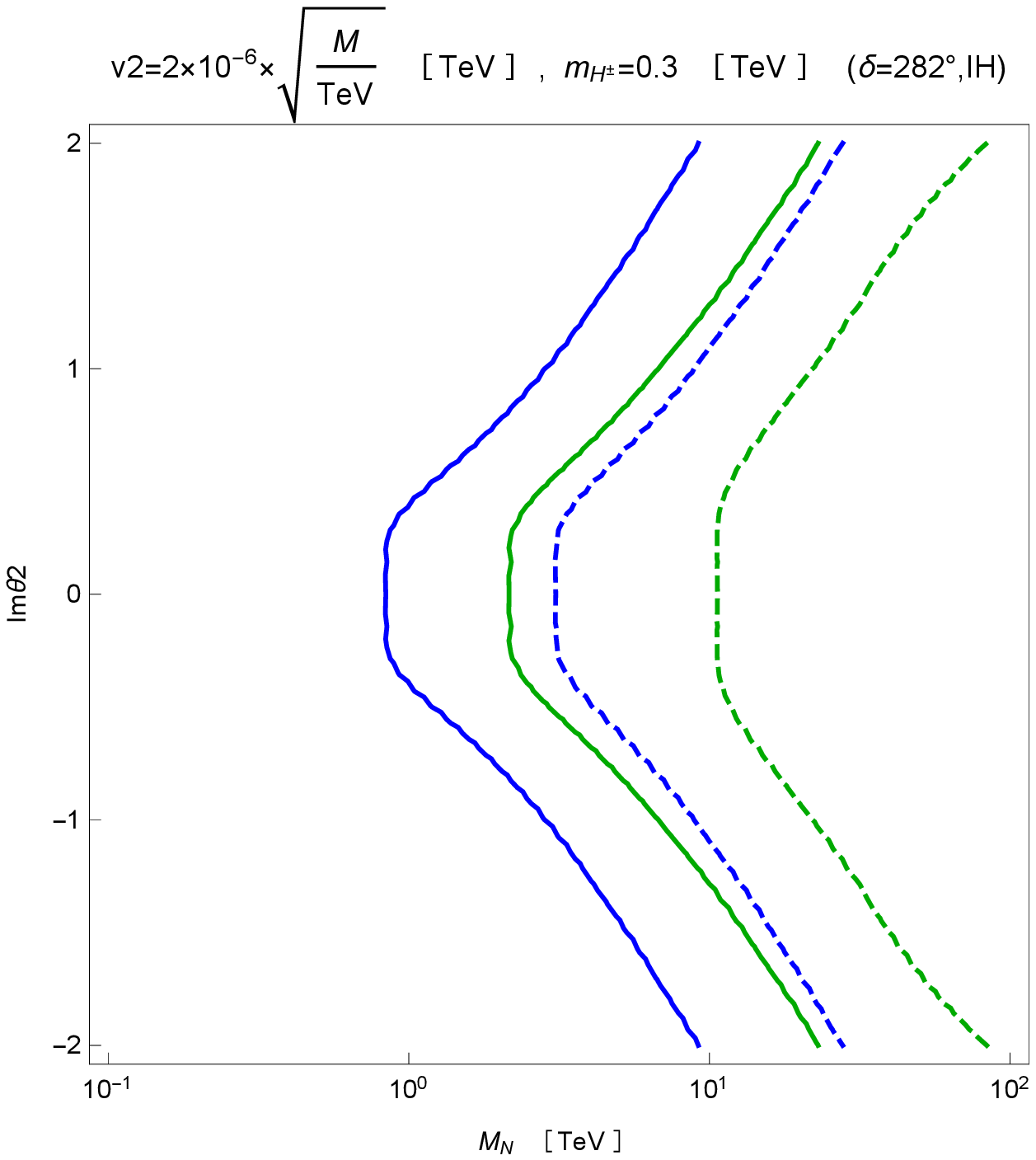}
      \end{minipage}
 
 
      \begin{minipage}{0.33\hsize}
        \centering
          \includegraphics[keepaspectratio, scale=0.44, angle=0]
                          {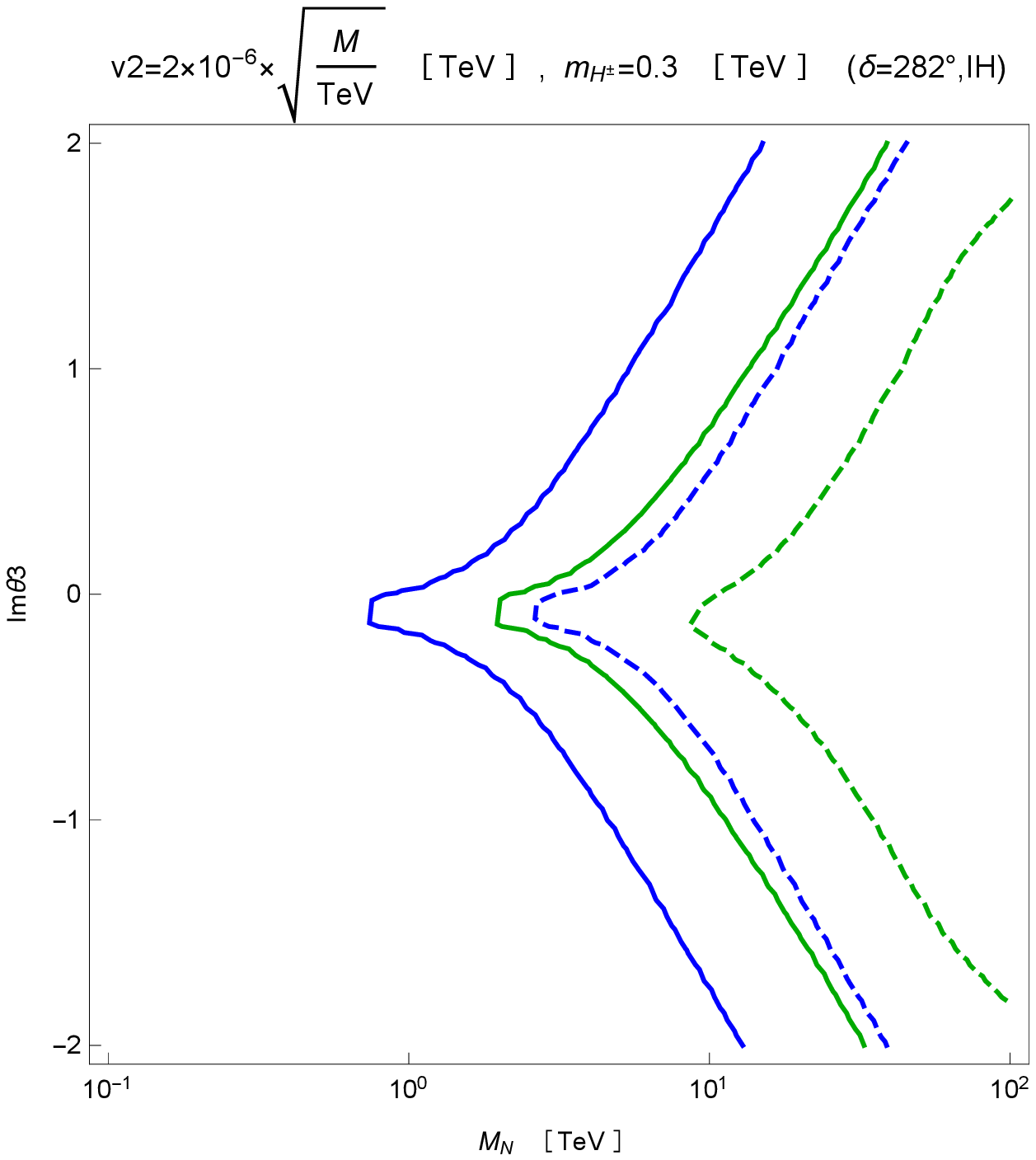}
      \end{minipage} \\ 
      \\
 
      \begin{minipage}{0.33\hsize}
        \centering
          \includegraphics[keepaspectratio, scale=0.44, angle=0]
                          {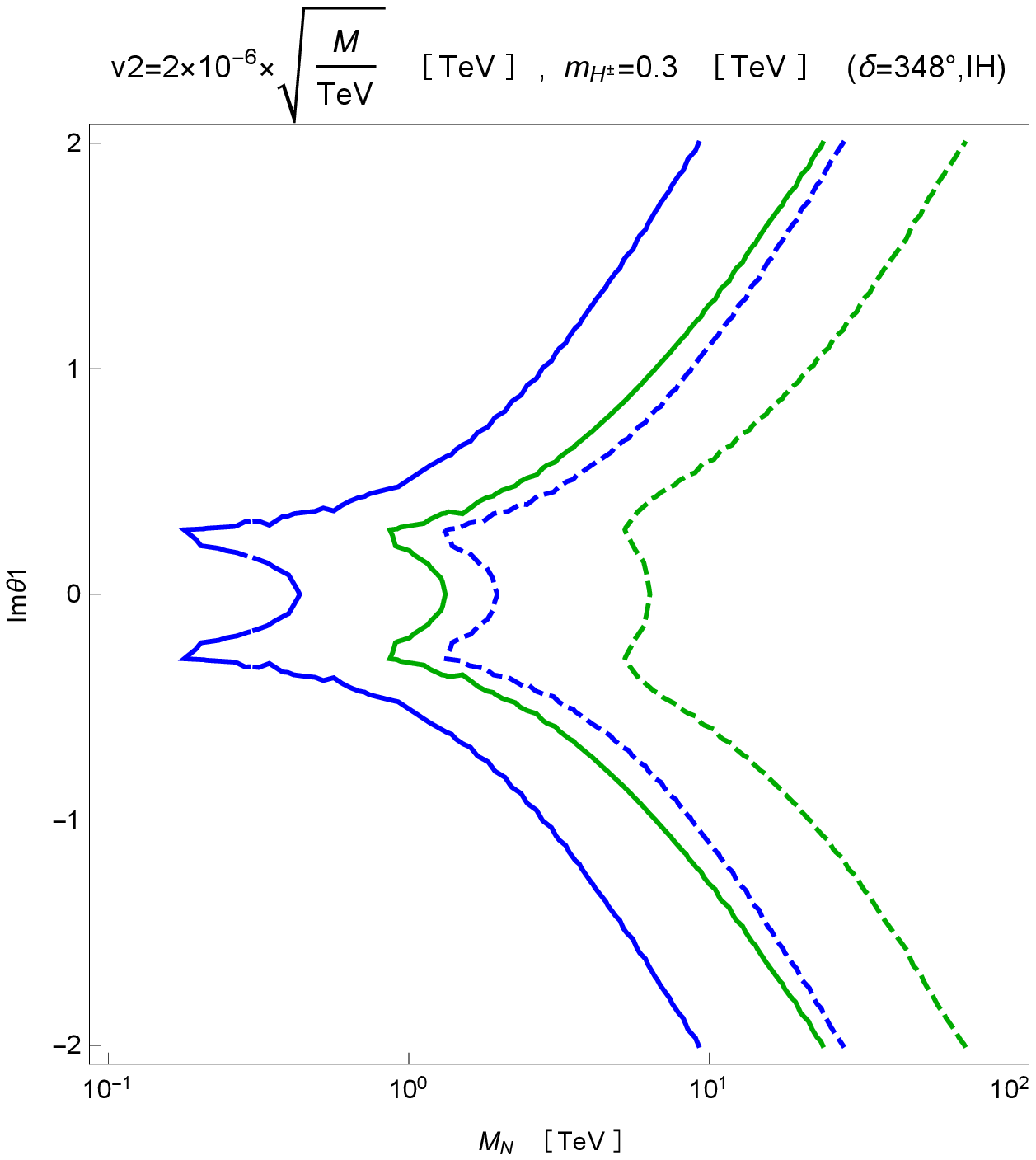}
      \end{minipage}

 
      \begin{minipage}{0.33\hsize}
        \centering
          \includegraphics[keepaspectratio, scale=0.44, angle=0]
                          {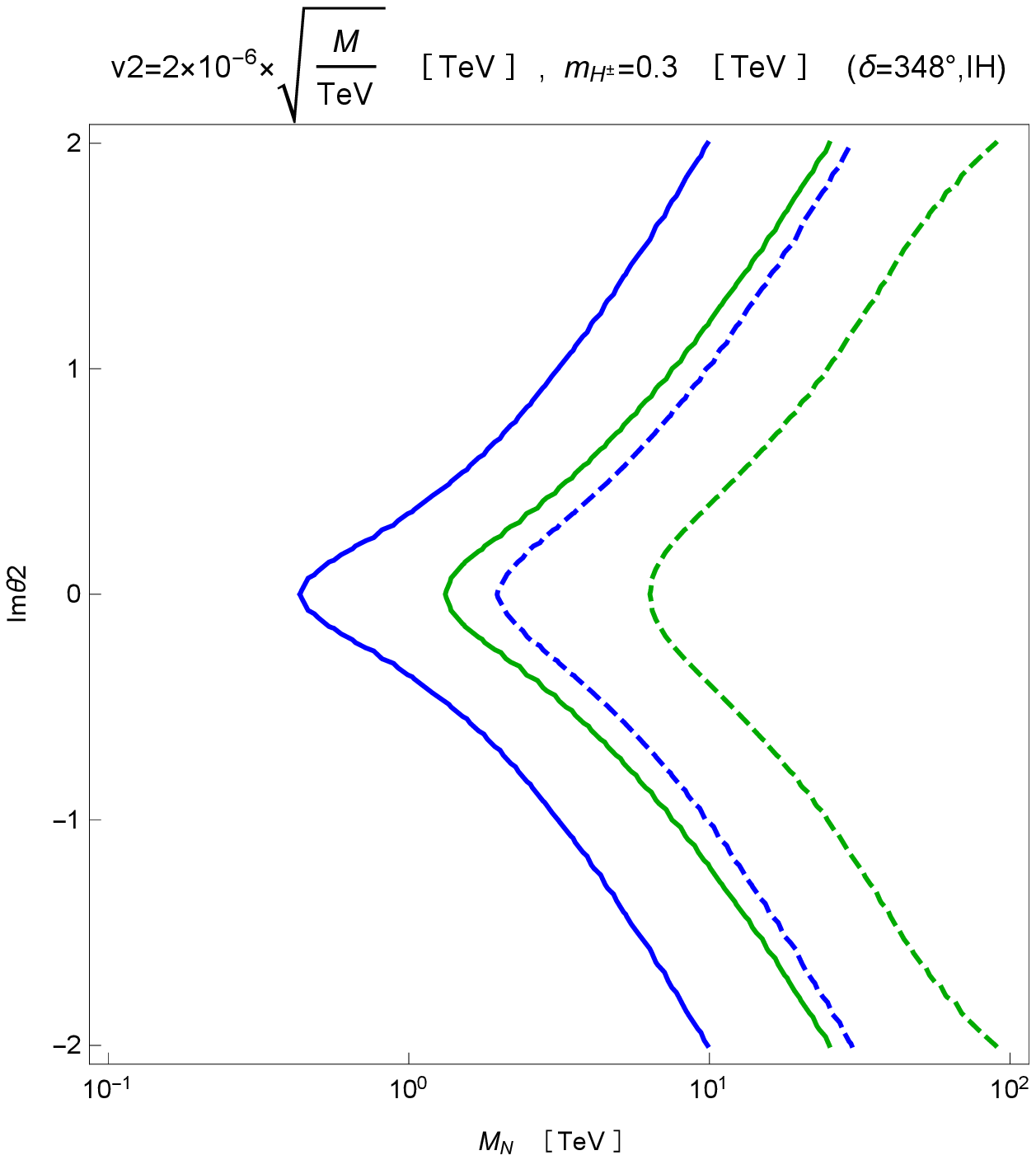}
      \end{minipage}
 
 
      \begin{minipage}{0.33\hsize}
        \centering
          \includegraphics[keepaspectratio, scale=0.44, angle=0]
                          {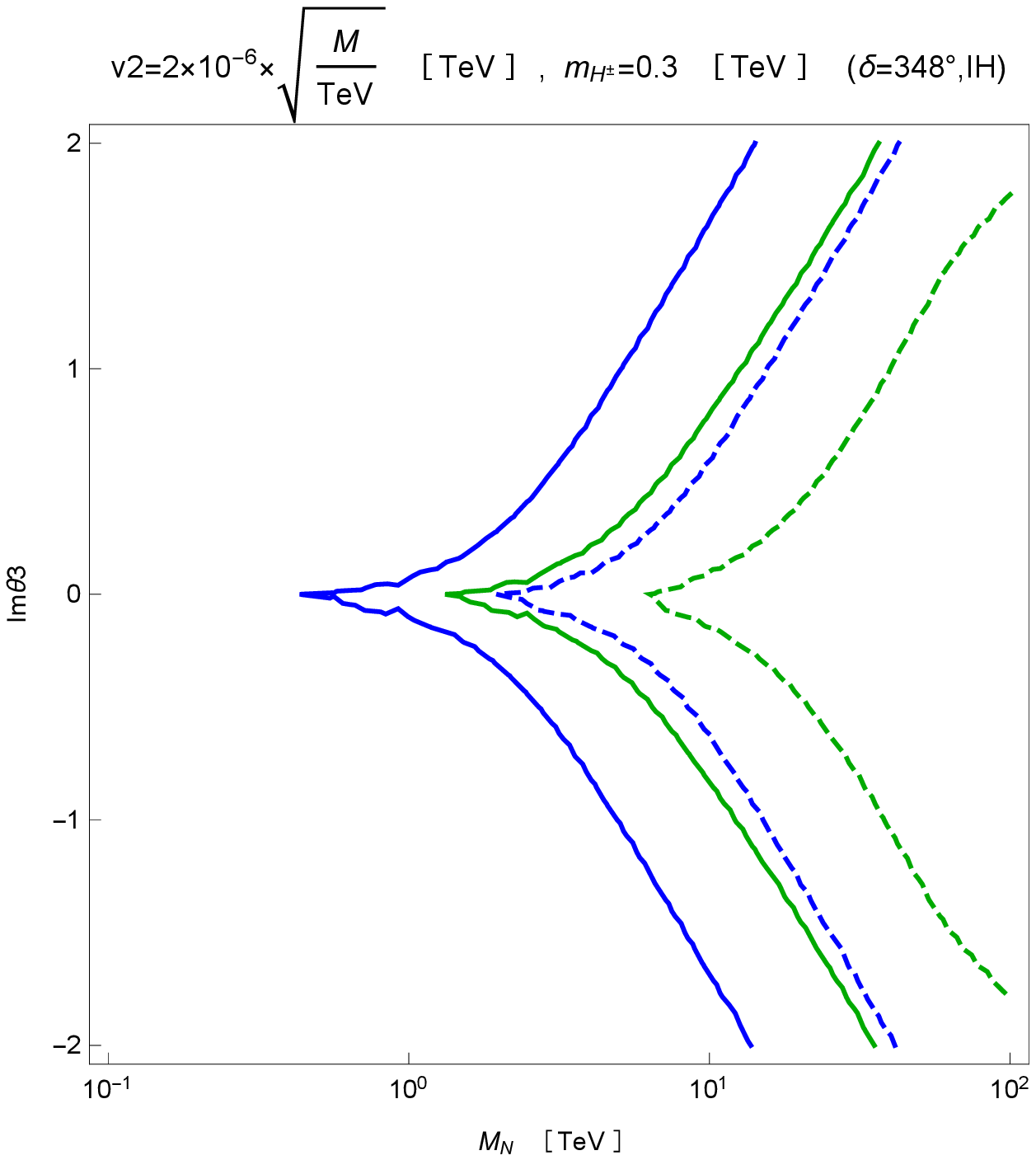}
      \end{minipage} 

    \end{tabular}
     \caption{\footnotesize Same as figure~\ref{figpreeN} except that the neutrino mass hierarchy is Inverted Hierarchy and $v_2$ is given in Eq.~(\ref{v2-ih}).}
 \label{figpreeI}
\end{figure}

Among the $e_\alpha \to e_\beta \bar{e}_\beta e_\beta$ processes, the future sensitivity for the $\mu\to 3e$ decay reaches $Br(\mu\to 3e)=10^{-16}$~\cite{Blondel:2013ia} and so there is a large chance that this mode is detected even when the model satisfies the current experimental bound on $Br(\mu\to e\gamma)$.
Therefore, we show in Fig.~\ref{figpreeN} (Normal Hierarchy) and Fig.~\ref{figpreeI} (Inverse Hierarchy) the prediction on $Br(\mu\to 3e)$,
 along with the value of $Br(\mu\to e\gamma)$.

In Fig.~\ref{figpreeN}, the blue solid line agrees with $Br(\mu\to e\gamma)=4.2\times10^{-13}$ for NH and $v_2$ in Eq.~(\ref{v2-nh}),
 and the region to the left of the blue solid line is excluded by the search for $\mu\to e\gamma$. 
The green solid line agrees with $Br(\mu\to 3e)=10^{-16}$, the future sensitivity.
Therefore, in the region between the blue solid line and the green solid line, the $\mu\to 3e$ process can be detected in the future.
Figure~\ref{figpreeI} is the corresponding figure for IH and $v_2$ in Eq.~(\ref{v2-ih}).

In the same figures, the blue and green dashed lines are contours of $Br(\mu\to e\gamma)=4.2\times10^{-13}$ and $Br(\mu\to 3e)=10^{-16}$
 in the case when $v_2$ is multiplied by $1/3$ and thus $Y_D$ is uniformly multiplied by 3 according to Eq.~(\ref{yukawa}).
Since the dipole and non-dipole terms $A_D,A_{ND}$ are proportional to $Y_D^2$ whereas the box-induced term $B$ is proportional to $Y_D^4$,
 reducing $v_2$ affects $Br(\mu\to 3e)$ and $Br(\mu\to e\gamma)$ differently.
However, such an effect is not clearly seen in the figures, as the region between the blue and green dashed lines
 has a similar size to that between the blue and green solid lines.

 \newpage
\newpage

 \newpage
\newpage

\subsubsection{$\mu$-$e$ Conversions}

The processes whose sensitivity will be improved in the future are the $\mu +{\rm Al}\to e+{\rm Al}$ and $\mu +{\rm Ti}\to e+{\rm Ti}$ processes. 
The future sensitivity for $CR(\mu +{\rm Al}\to e+{\rm Al})$ is $2\times10^{-17}$~\cite{Kuno:2013mha}, and that for $CR(\mu +{\rm Ti}\to e+{\rm Ti})$ is $10^{-18}$~\cite{Bertuzzo:2015ada}.
Therefore, we study whether the $\mu +{\rm Al}\to e+{\rm Al}$ and $\mu +{\rm Ti}\to e+{\rm Ti}$ processes can be detected in the future.
In the numerical calculation of the conversion rates, we employ the values of $Z_{eff}$, $F_p$, $\Gamma_{\rm capture}$ from Ref.~\cite{conversion}.

We comment that a peculiar property of the conversion rates $CR(\mu N\to e N)$ is that they are zero if $M_N=m_{H^\pm}$,
because $A_{ND}=A_D$ at $\frac{M_{N_i}^2}{M_{H^\pm}^2}=1$.
Therefore, the plots of $CR(\mu +{\rm Al}\to e+{\rm Al})$ and $CR(\mu +{\rm Ti}\to e+{\rm Ti})$ show a different behavior from other processes
 around the region $M_N\simeq m_{H^\pm}=0.3$~TeV.
However, this region is excluded by the $\mu\to e\gamma$ search and so such a behavior is unimportant.

\begin{figure}[H]
  \centering
  \thispagestyle{empty}
    \begin{tabular}{c}
 
 
      \begin{minipage}{0.33\hsize}
        \centering
          \includegraphics[keepaspectratio, scale=0.45, angle=0]
                          {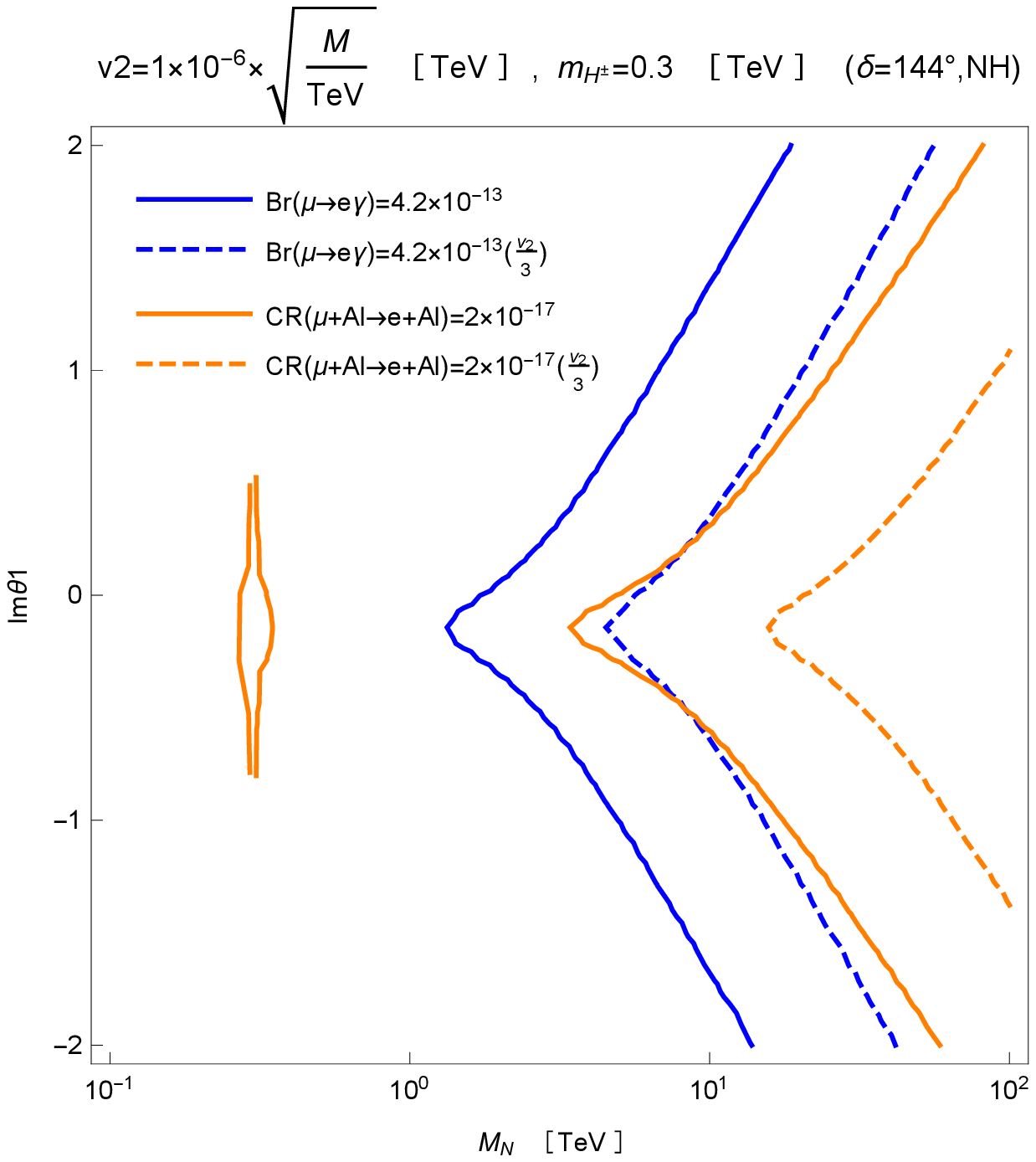}
      \end{minipage}

 
      \begin{minipage}{0.33\hsize}
        \centering
          \includegraphics[keepaspectratio, scale=0.44, angle=0]
                          {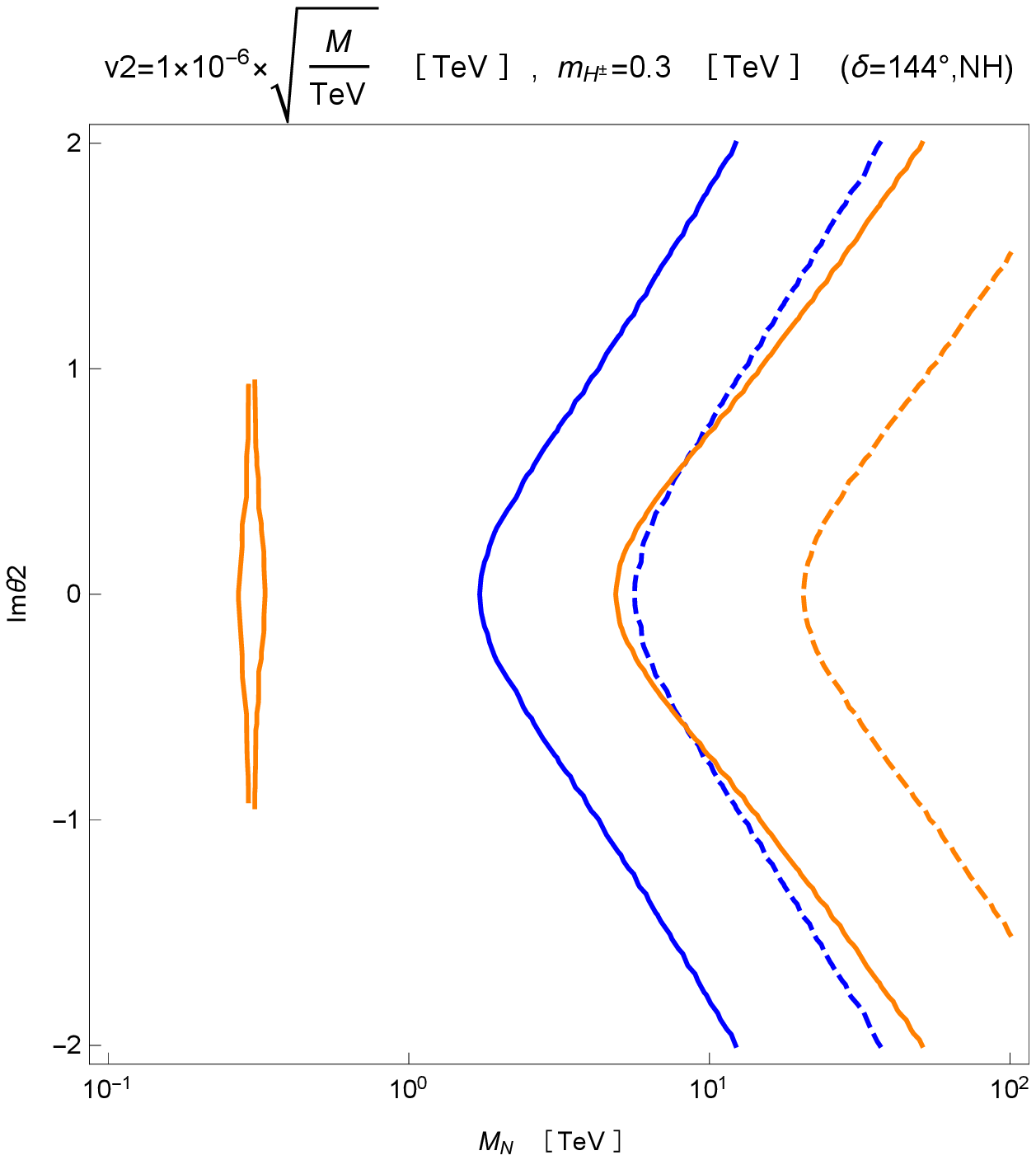}

      \end{minipage}
 
 
      \begin{minipage}{0.33\hsize}
        \centering
          \includegraphics[keepaspectratio, scale=0.44, angle=0]
                          {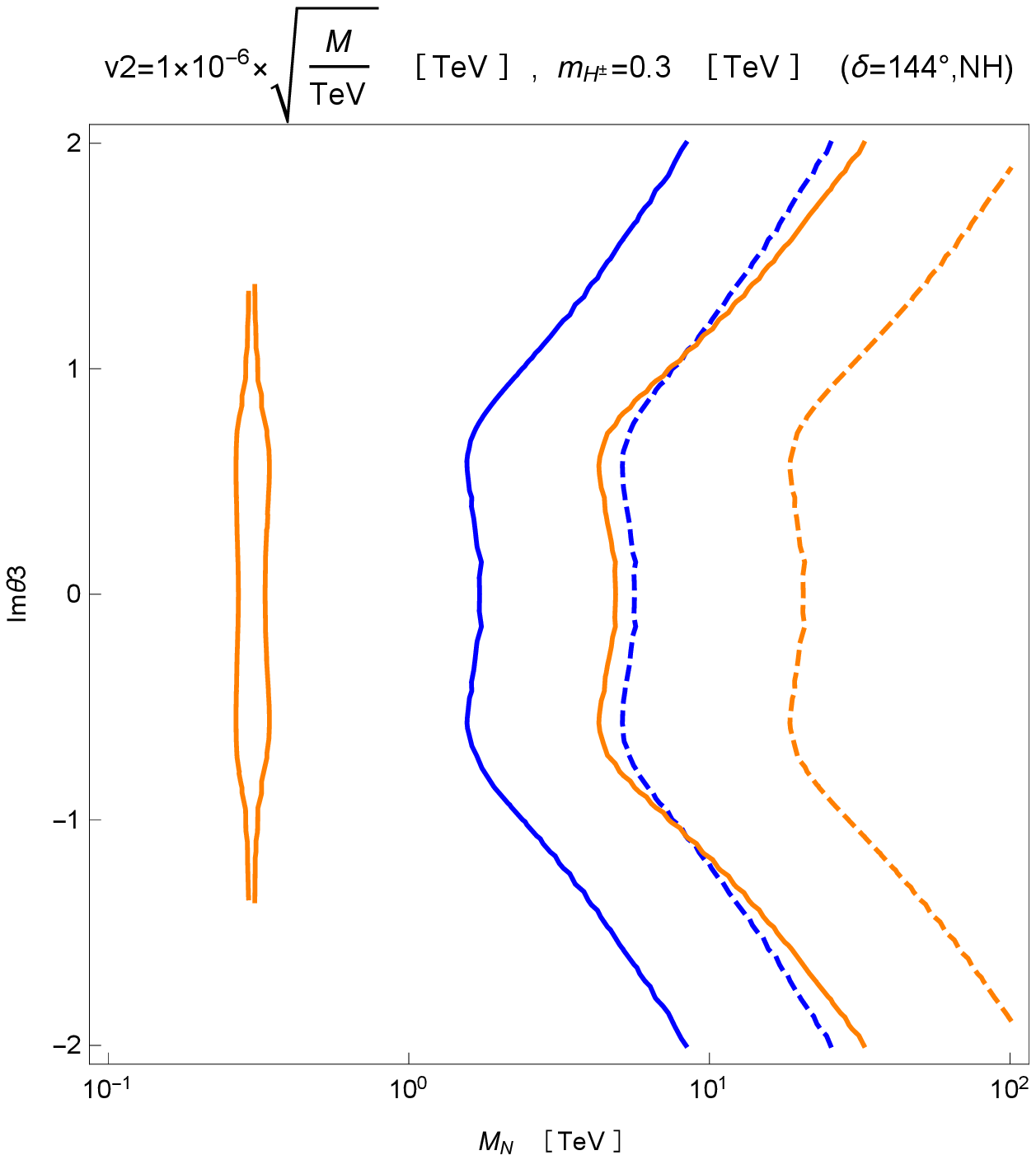}
      \end{minipage} \\
      \\
 
      \begin{minipage}{0.33\hsize}
        \centering
          \includegraphics[keepaspectratio, scale=0.44, angle=0]
                          {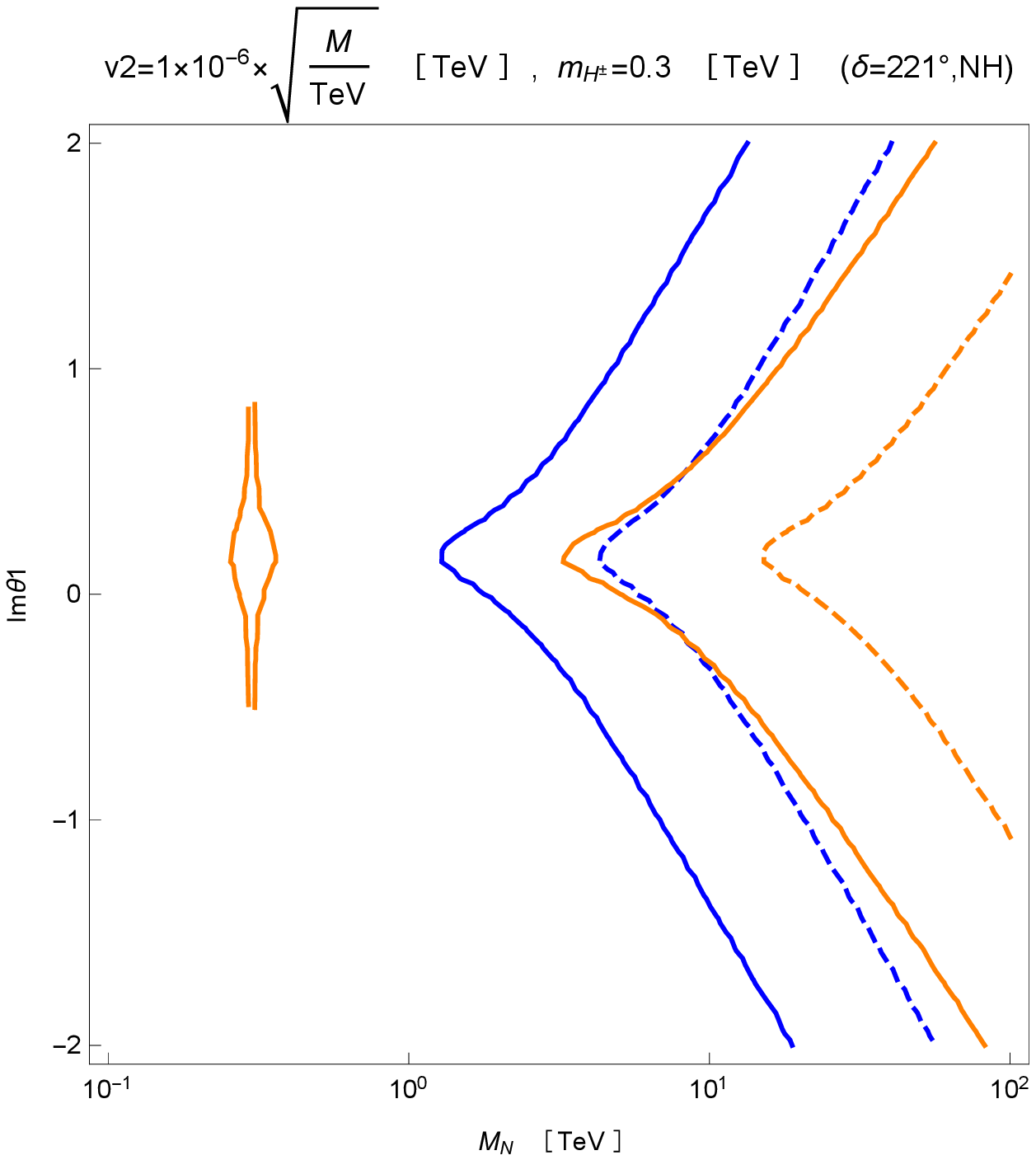}
      \end{minipage}

 
      \begin{minipage}{0.33\hsize}
        \centering
          \includegraphics[keepaspectratio, scale=0.44, angle=0]
                          {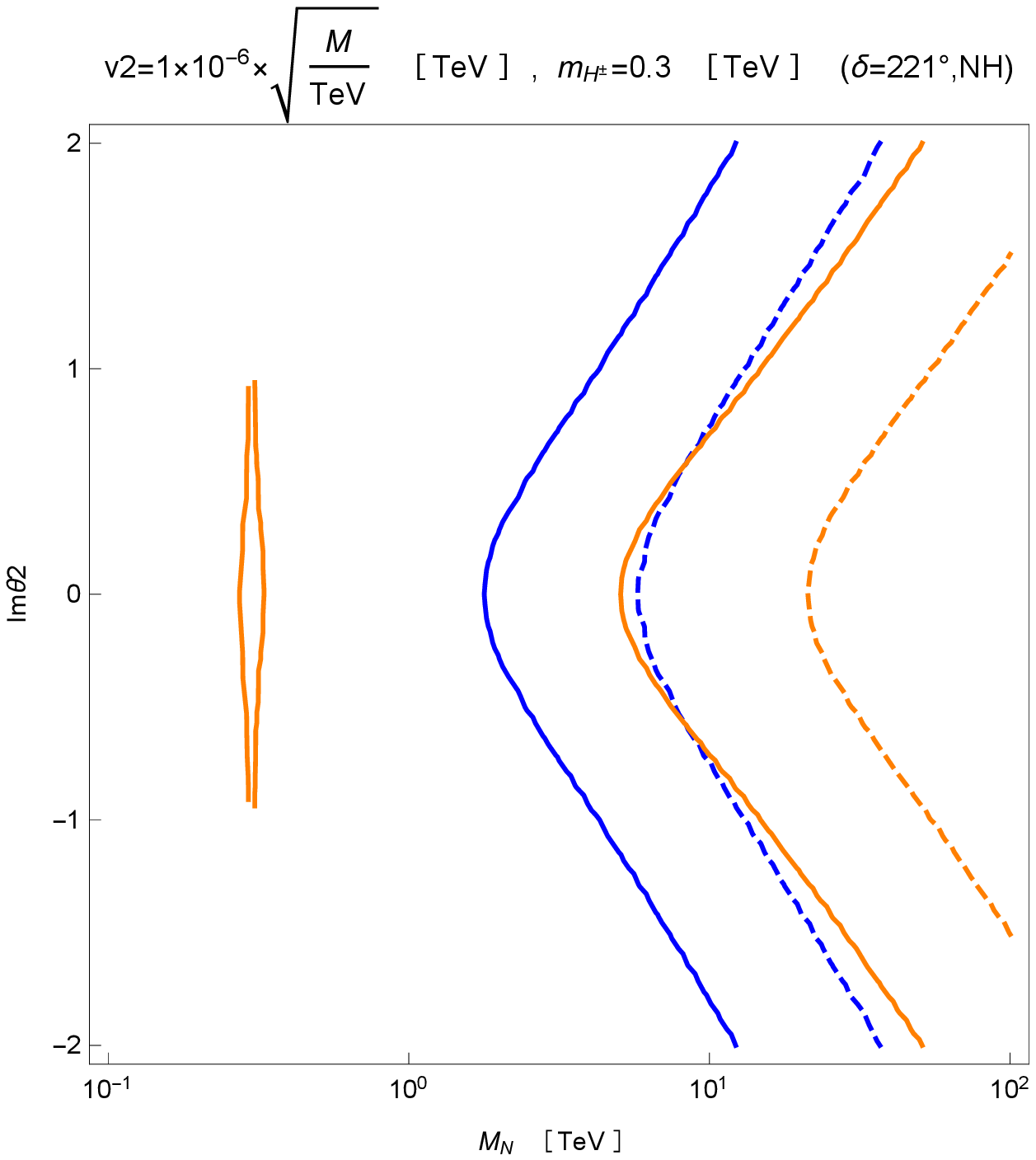}
      \end{minipage}
 
 
      \begin{minipage}{0.33\hsize}
        \centering
          \includegraphics[keepaspectratio, scale=0.44, angle=0]
                          {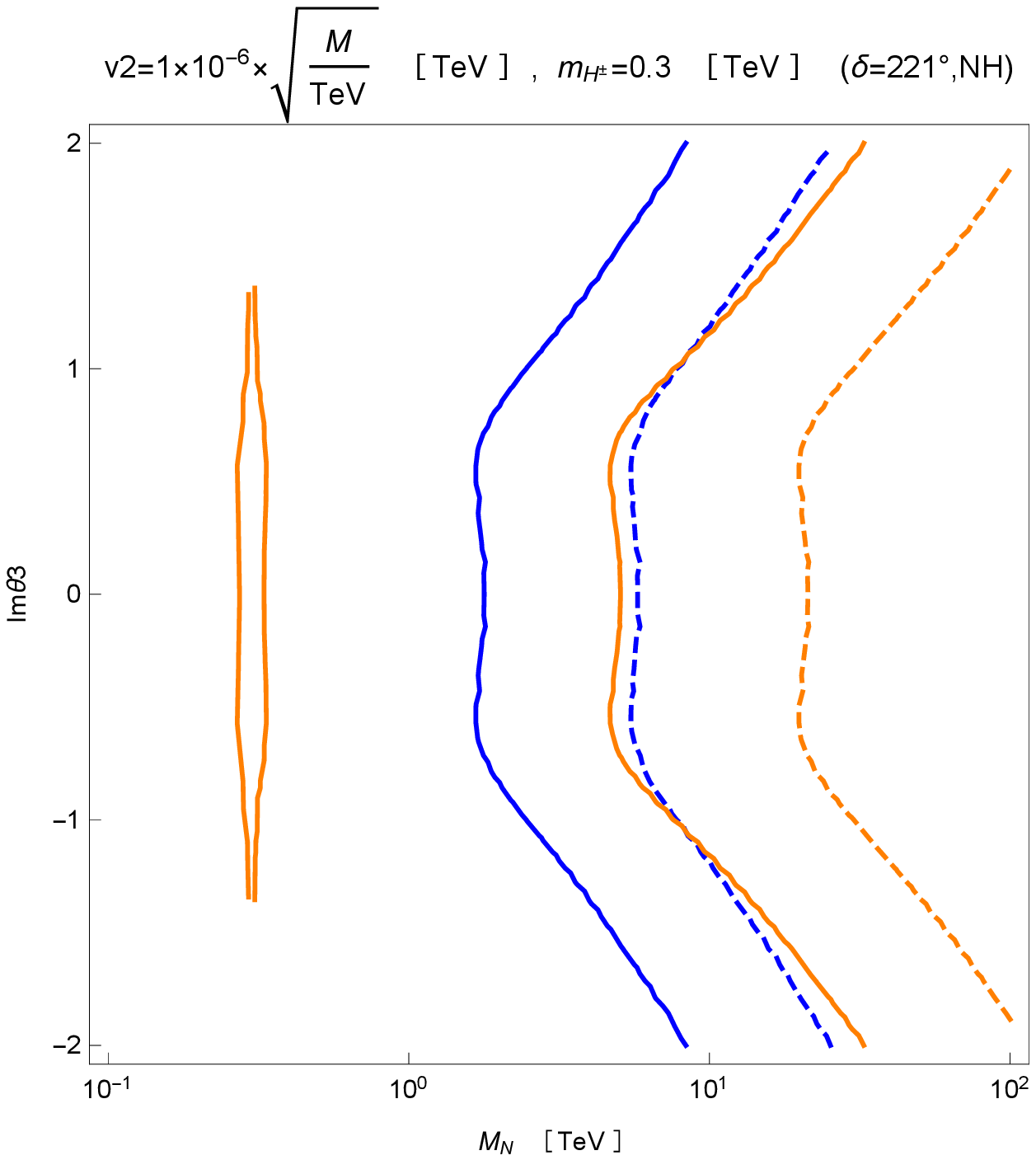}
      \end{minipage} \\ 
      \\
 
      \begin{minipage}{0.33\hsize}
        \centering
          \includegraphics[keepaspectratio, scale=0.44, angle=0]
                          {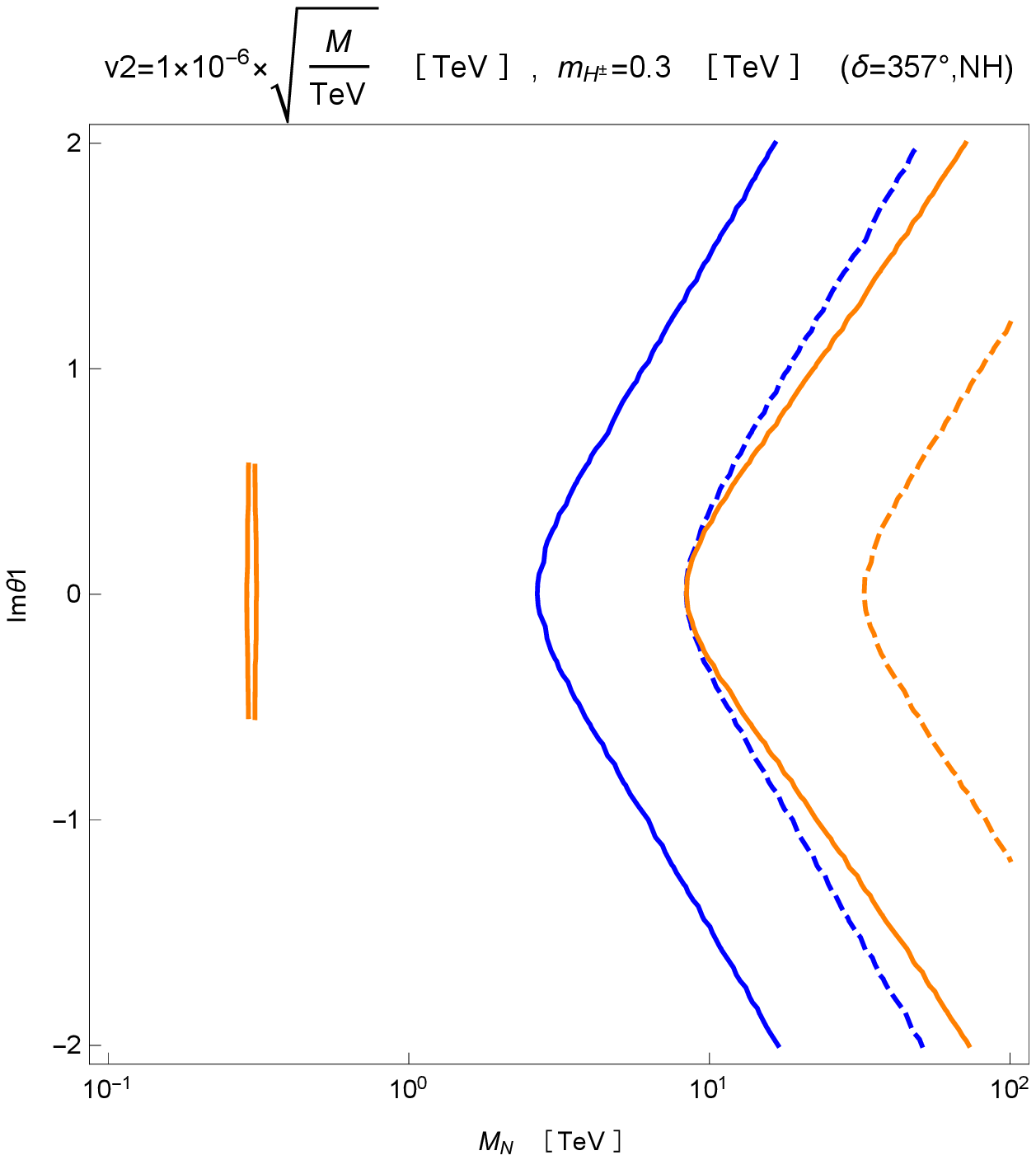}
      \end{minipage}

 
      \begin{minipage}{0.33\hsize}
        \centering
          \includegraphics[keepaspectratio, scale=0.44, angle=0]
                          {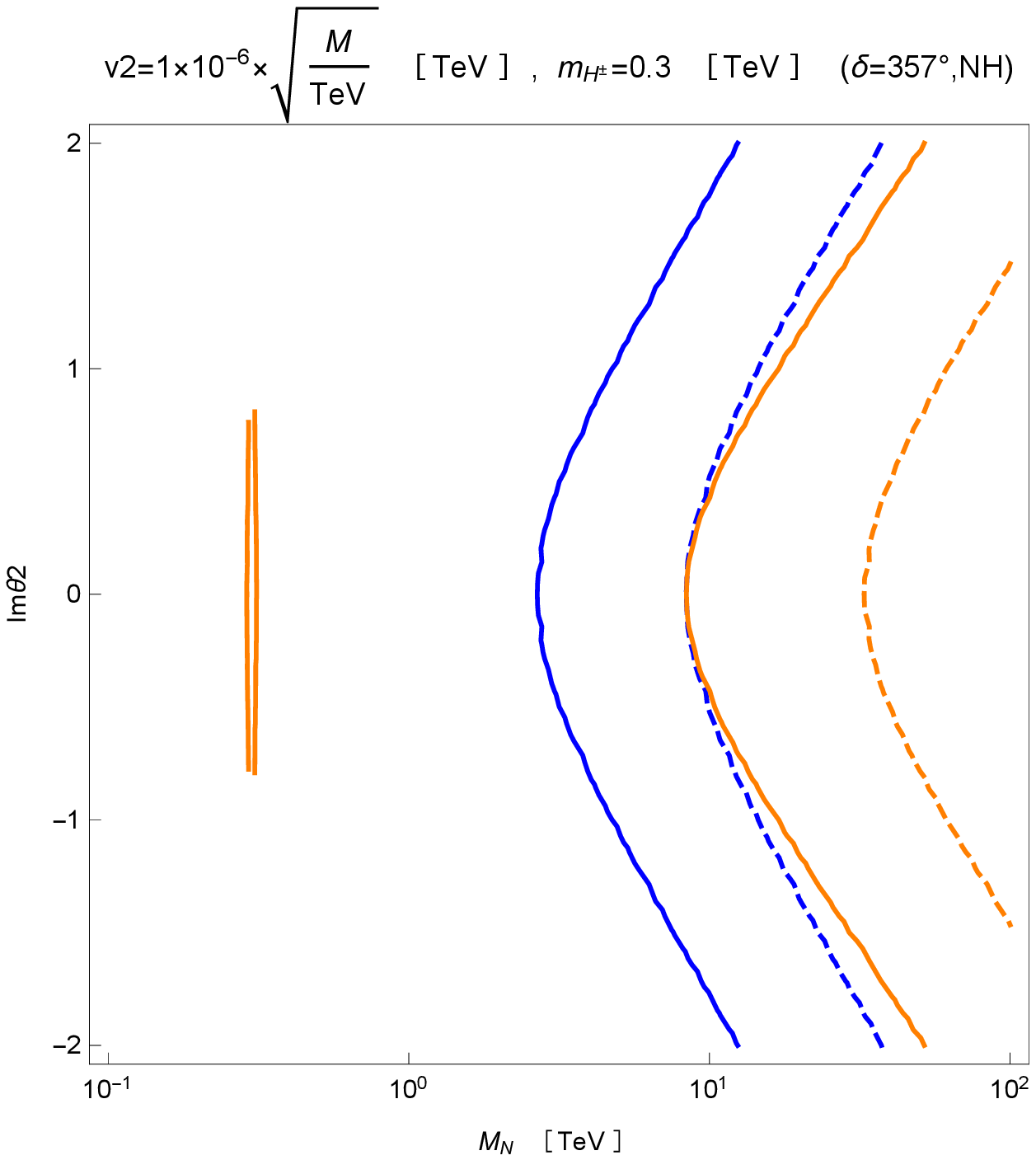}
      \end{minipage}
 
 
      \begin{minipage}{0.33\hsize}
        \centering
          \includegraphics[keepaspectratio, scale=0.44, angle=0]
                          {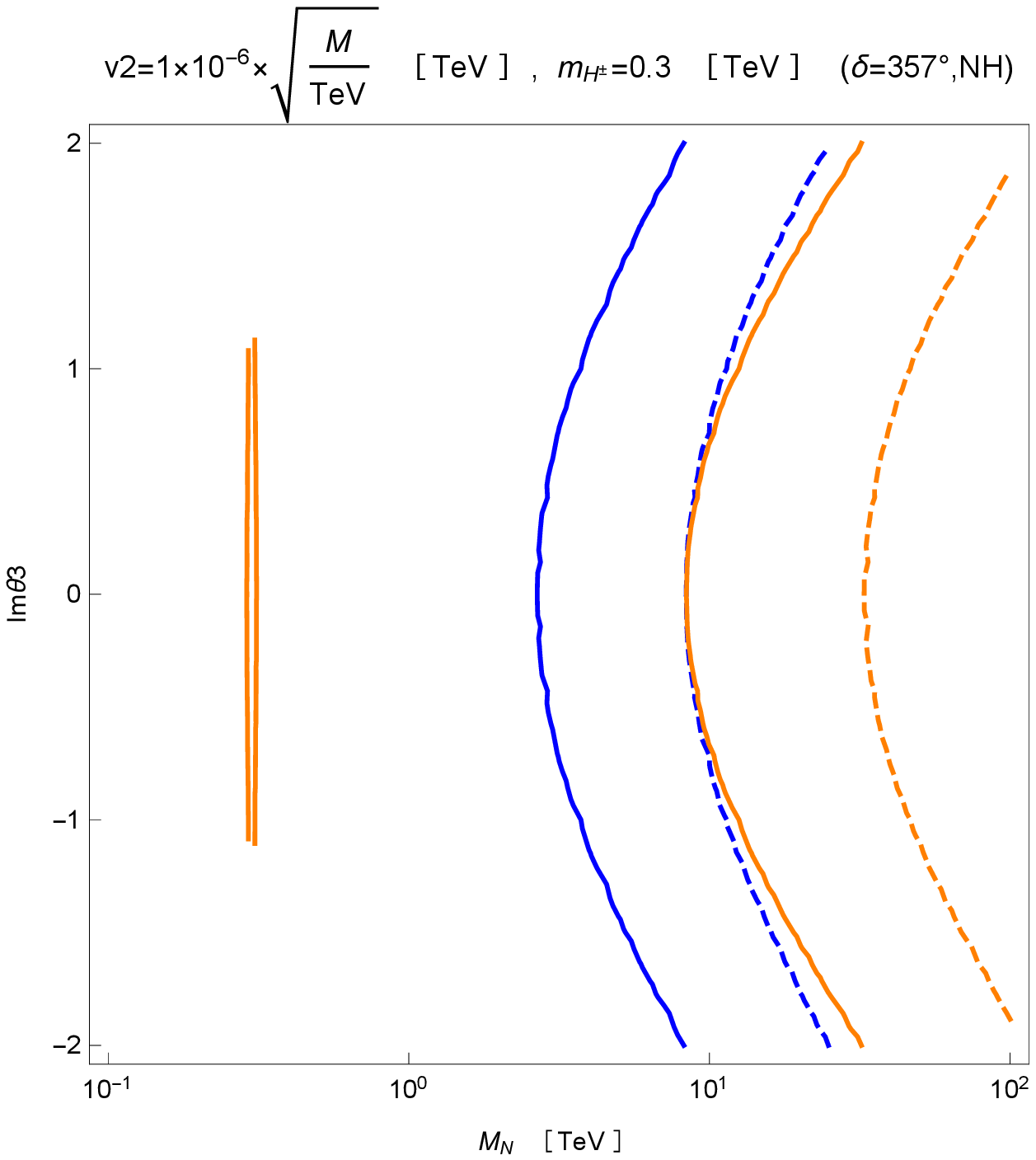}
      \end{minipage} 

    \end{tabular}
     \caption{\footnotesize Prediction for $CR(\mu +{\rm Al}\to e+{\rm Al})$, along with the values of $Br(\mu\to e\gamma)$. 
     The neutrino mass hierarchy is Normal Hierarchy, and we fix $m_{H^\pm}=0.3$ TeV.
     We take $\delta=144^\circ,~221^\circ~{\rm and}~357^\circ$ in the first, second and third rows. In the first column, we vary Im$\theta_1\neq0$ while fixing Im$\theta_2$=Im$\theta_3=0$. 
In the second column, we vary Im$\theta_2\neq0$ while fixing Im$\theta_1$=Im$\theta_3=0$. In the third column, we vary Im$\theta_3\neq0$ while fixing Im$\theta_1$=Im$\theta_2=0$.
The solid blue line corresponds to $Br(\mu\to e\gamma)=4.2\times10^{-13}$ for $v_2$ in Eq.~(\ref{v2-nh}), 
 and the region on the left of the solid blue line is excluded by the search for $Br(\mu\to e\gamma)$.
The solid orange line corresponds to $CR(\mu +{\rm Al}\to e+{\rm Al})=2\times10^{-17}$, the future sensitivity, for $v_2$ in Eq.~(\ref{v2-nh}).
The dashed blue line corresponds to $Br(\mu\to e\gamma)=4.2\times10^{-13}$ and the dashed orange line corresponds to　$CR(\mu +{\rm Al}\to e+{\rm Al})=2\times10^{-17}$
 when $v_2$ is multiplied by $1/3$.}
 \label{figprecalN}
\end{figure}

\begin{figure}[H]
  \centering
      \begin{tabular}{c}
 
 
      \begin{minipage}{0.33\hsize}
        \centering
          \includegraphics[keepaspectratio, scale=0.35, angle=0]
                          {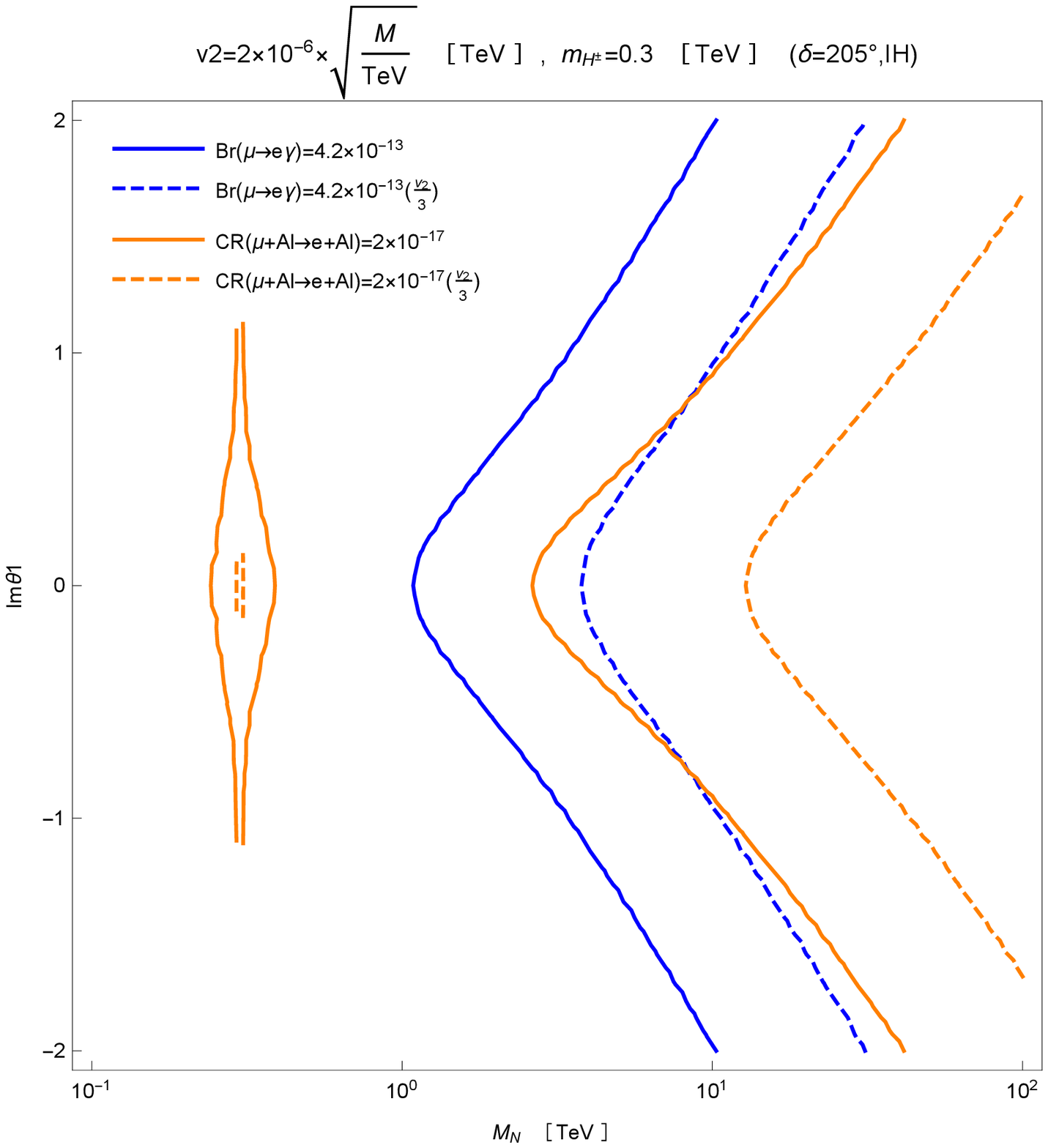}
      \end{minipage}

 
      \begin{minipage}{0.33\hsize}
        \centering
          \includegraphics[keepaspectratio, scale=0.44, angle=0]
                          {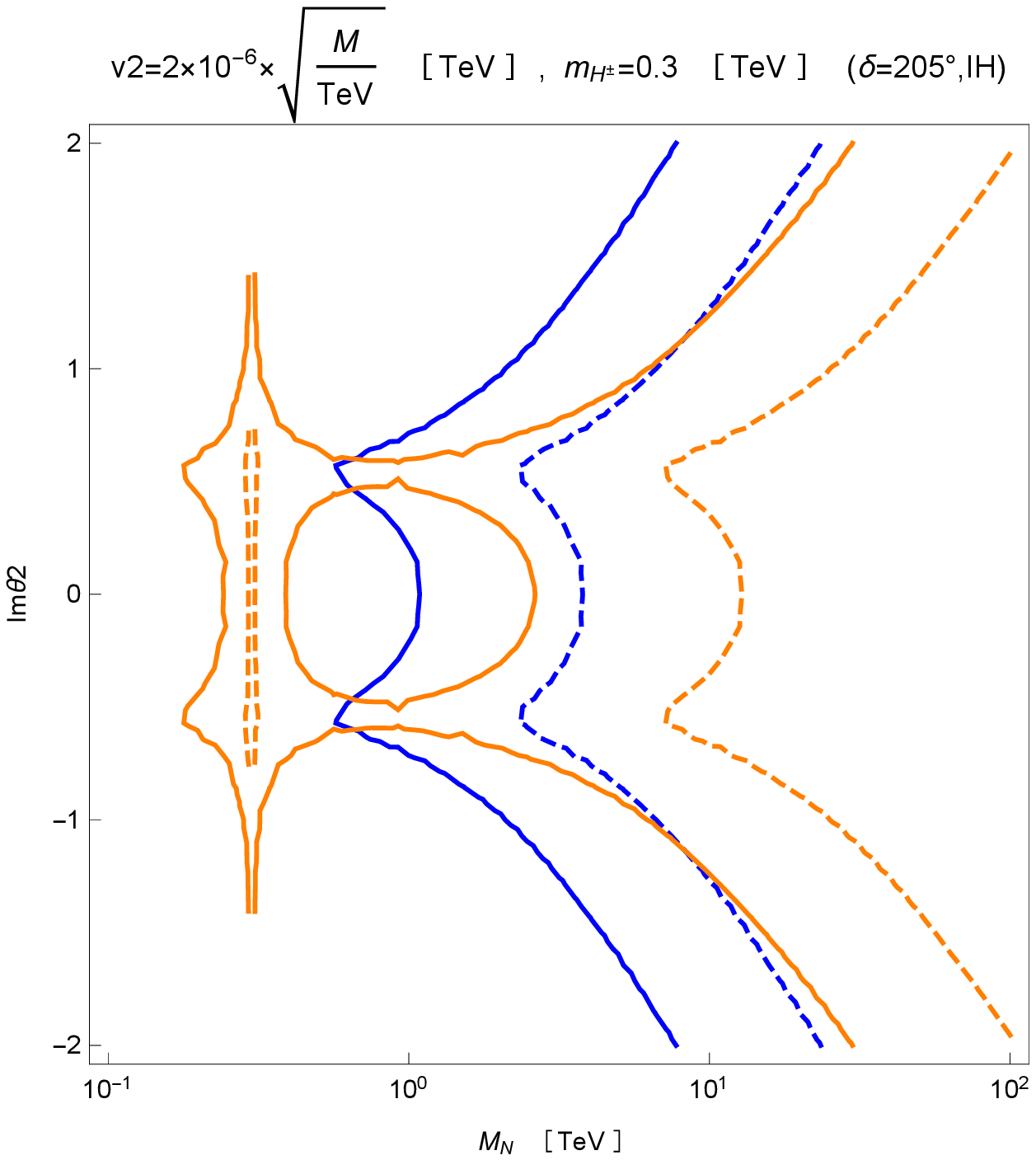}

      \end{minipage}
 
 
      \begin{minipage}{0.33\hsize}
        \centering
          \includegraphics[keepaspectratio, scale=0.44, angle=0]
                          {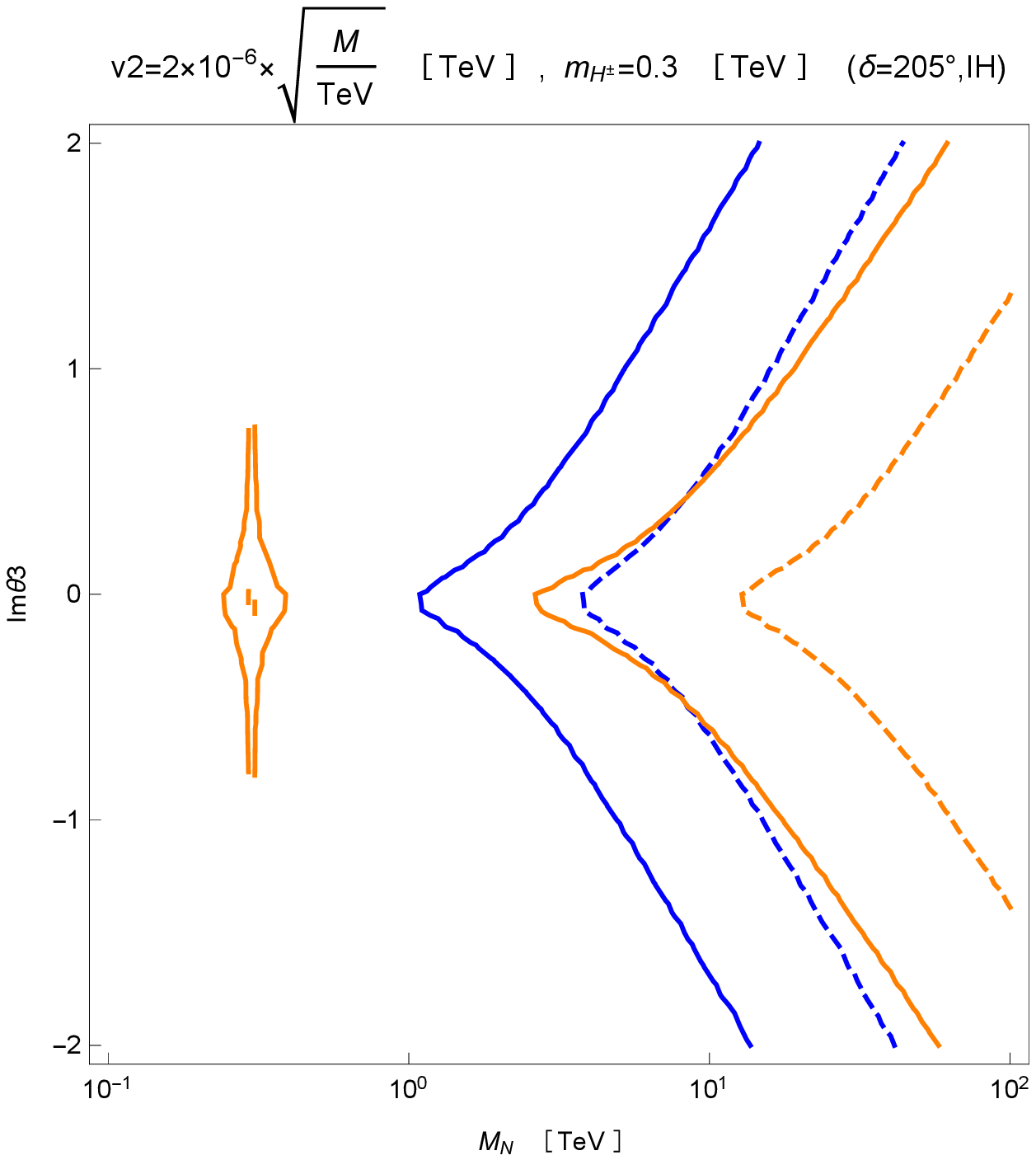}
      \end{minipage} \\
      \\
 
      \begin{minipage}{0.33\hsize}
        \centering
          \includegraphics[keepaspectratio, scale=0.44, angle=0]
                          {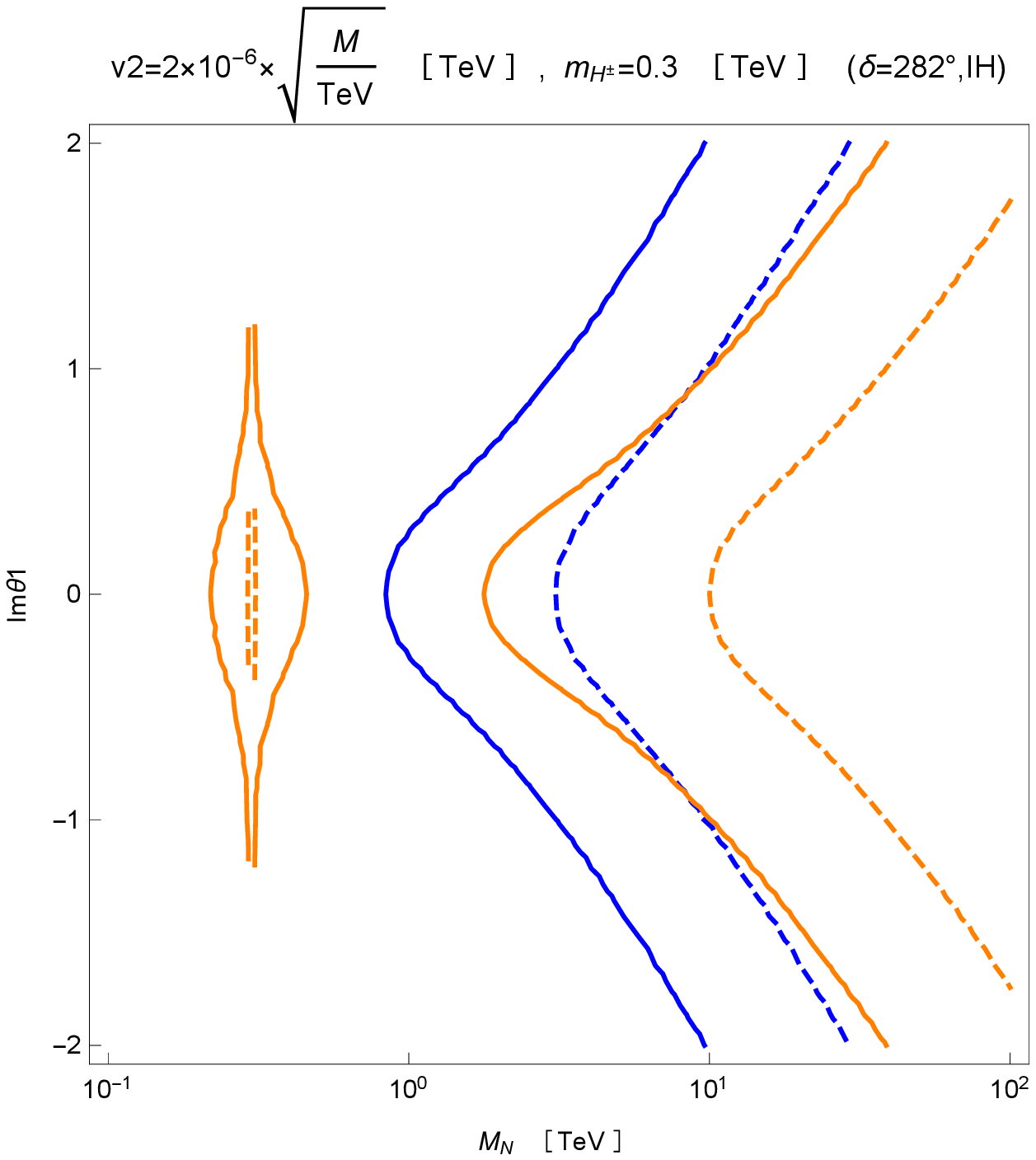}
      \end{minipage}

 
      \begin{minipage}{0.33\hsize}
        \centering
          \includegraphics[keepaspectratio, scale=0.44, angle=0]
                          {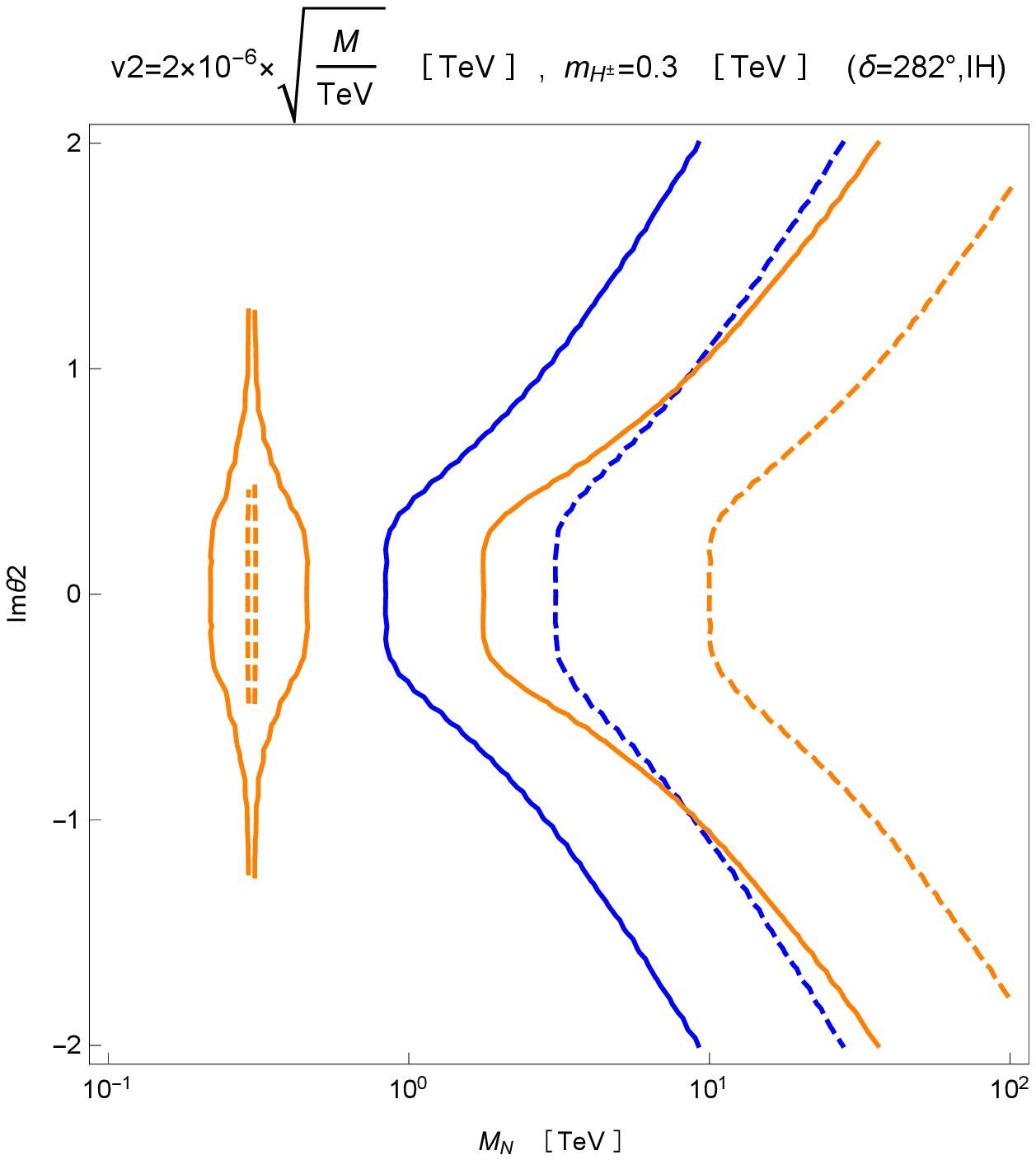}
      \end{minipage}
 
 
      \begin{minipage}{0.33\hsize}
        \centering
          \includegraphics[keepaspectratio, scale=0.44, angle=0]
                          {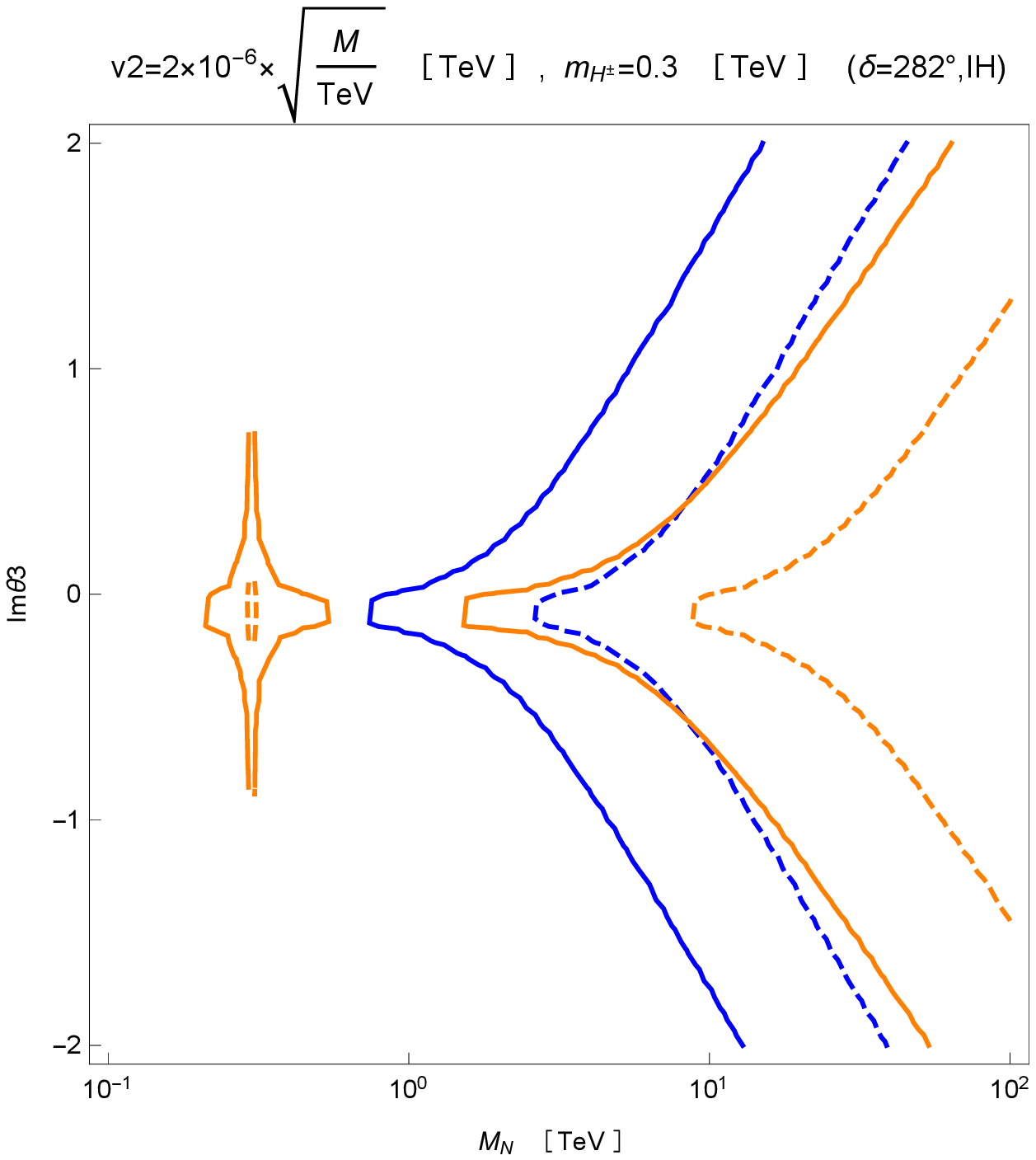}
      \end{minipage} \\ 
      \\
 
      \begin{minipage}{0.33\hsize}
        \centering
          \includegraphics[keepaspectratio, scale=0.44, angle=0]
                          {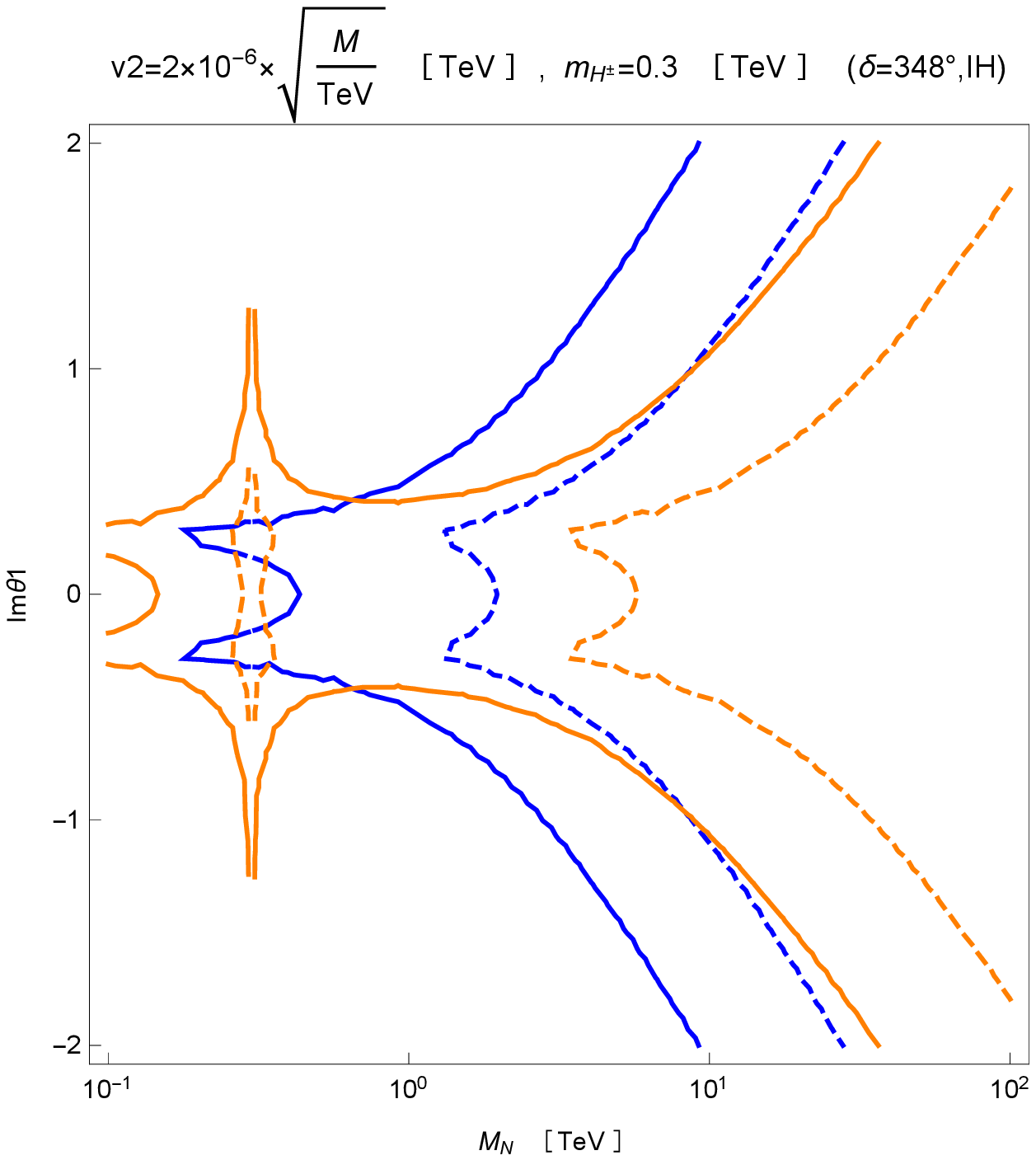}
      \end{minipage}

 
      \begin{minipage}{0.33\hsize}
        \centering
          \includegraphics[keepaspectratio, scale=0.44, angle=0]
                          {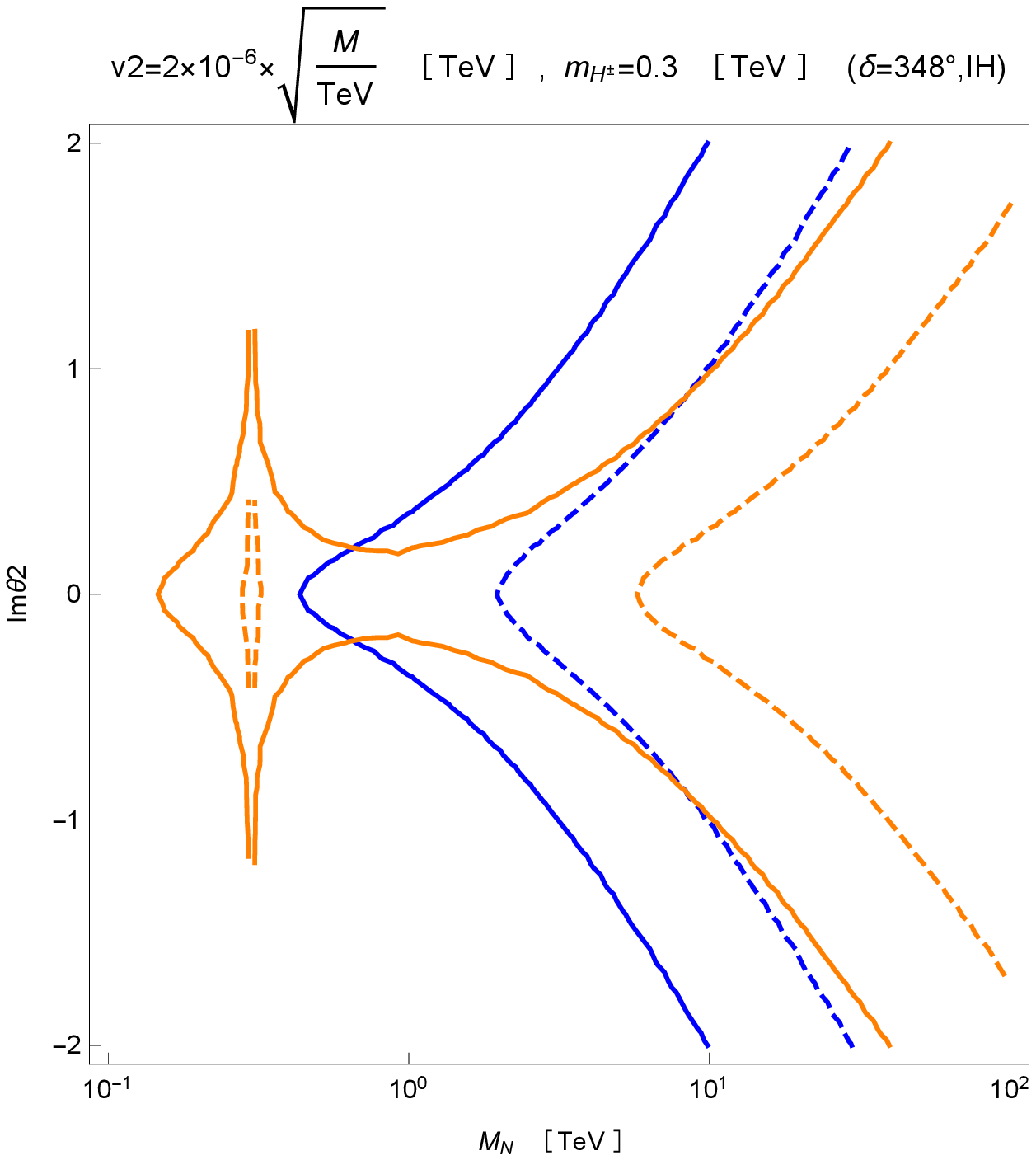}
      \end{minipage}
 
 
      \begin{minipage}{0.33\hsize}
        \centering
          \includegraphics[keepaspectratio, scale=0.44, angle=0]
                          {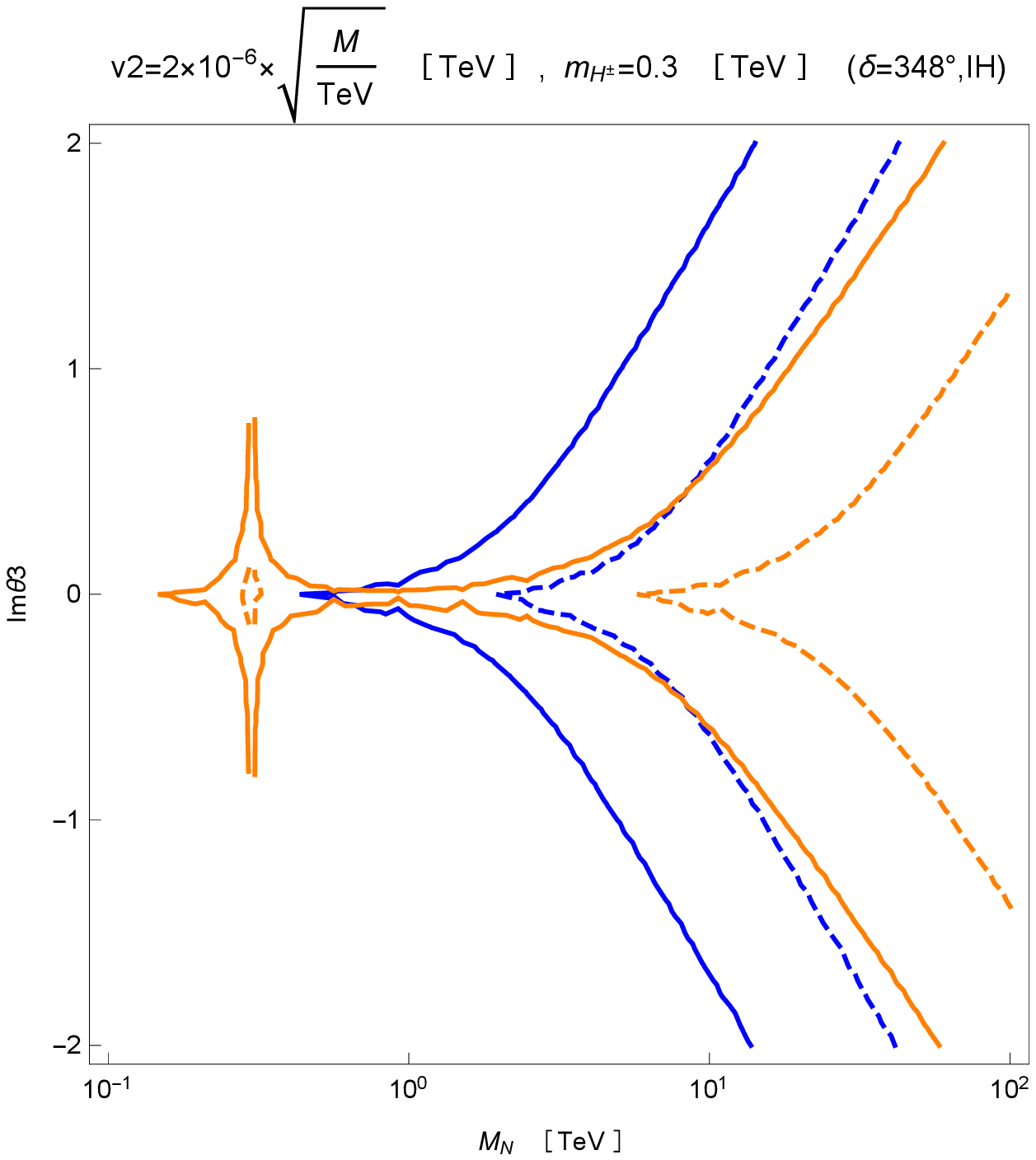}
      \end{minipage} 

    \end{tabular}
     \caption{\footnotesize
     Same as figure~\ref{figprecalN} except that the neutrino mass hierarchy is Inverted Hierarchy and $v_2$ is given in Eq.~(\ref{v2-ih}).
     }
 \label{figprecalI}
\end{figure}

In Fig.~\ref{figprecalN}, the solid orange line agrees with $CR(\mu +{\rm Al}\to e+{\rm Al})=2\times10^{-17}$, the future sensitivity, for NH and $v_2$ in Eq.~(\ref{v2-nh}).
Therefore, in the region between the solid blue line and the solid orange line (we neglect the orange line near $M_N=0.3$ TeV), 
 the $\mu +{\rm Al}\to e+{\rm Al}$ process can be detected in the future.
Figure~\ref{figprecalI} is the corresponding plot for IH and $v_2$ in Eq.~(\ref{v2-ih}).

In the same figures, the dashed orange line agrees with $CR(\mu +{\rm Al}\to e+{\rm Al})=2\times10^{-17}$ when $v_2$ is multiplied by $1/3$,
 and the $\mu +{\rm Al}\to e+{\rm Al}$ process can be detected in the region between the dashed blue line and the dashed orange line for this $v_2$.
Since the dipole and non-dipole operators $A_D,A_{ND}$ are both proportional to $Y_D^2$, $Br(\mu\to e\gamma)$ and $CR(\mu +{\rm Al}\to e+{\rm Al})$
 both scale with $1/v_2^4$. 
Hence, the relative location of the contours of $Br(\mu\to e\gamma)$ and $CR(\mu +{\rm Al}\to e+{\rm Al})$ does not depend on $v_2$.

\begin{figure}[H]
  \centering
    \begin{tabular}{c}
 
 
      \begin{minipage}{0.33\hsize}
        \centering
          \includegraphics[keepaspectratio, scale=0.45, angle=0]
                          {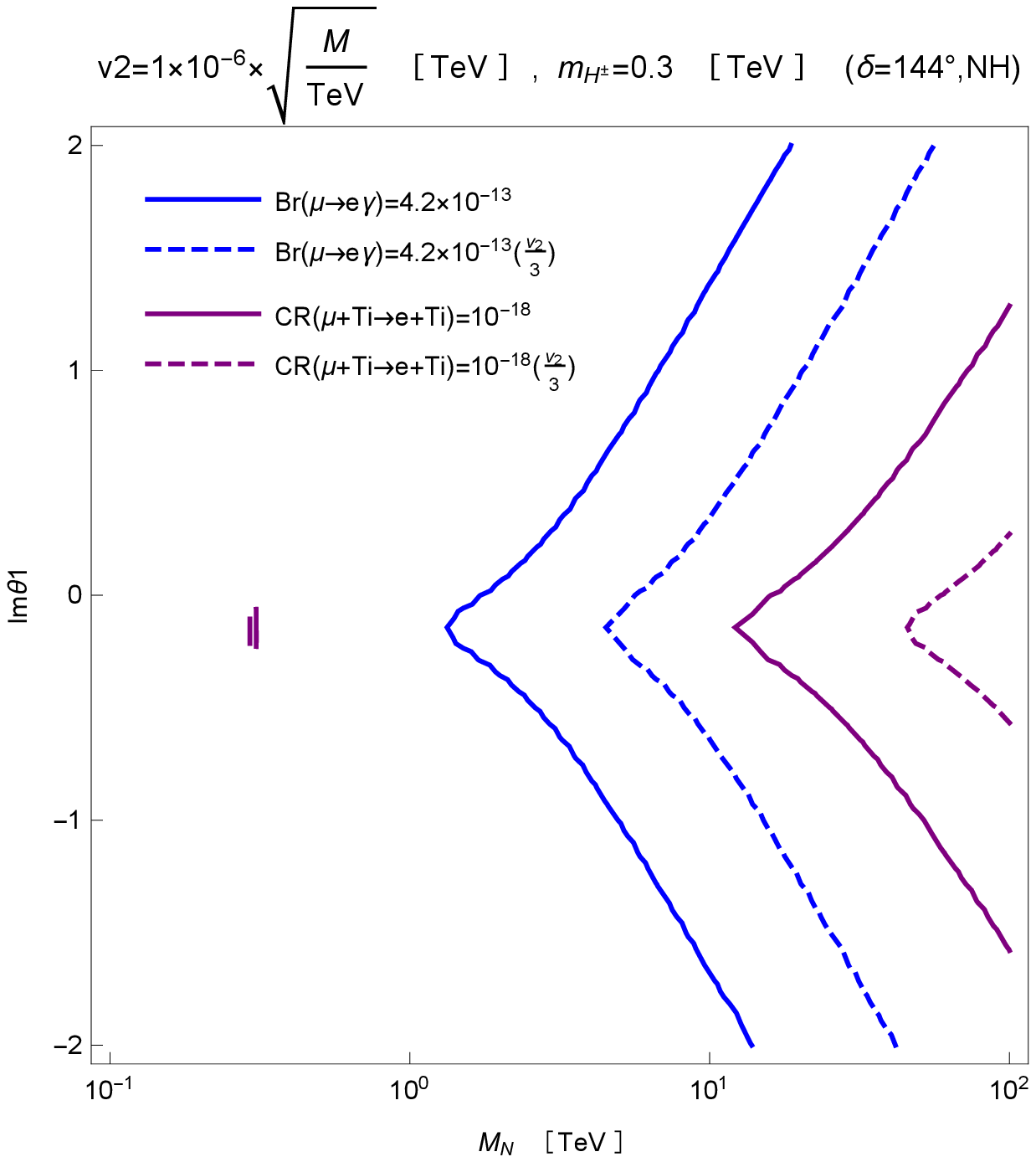}
      \end{minipage}

 
      \begin{minipage}{0.33\hsize}
        \centering
          \includegraphics[keepaspectratio, scale=0.44, angle=0]
                          {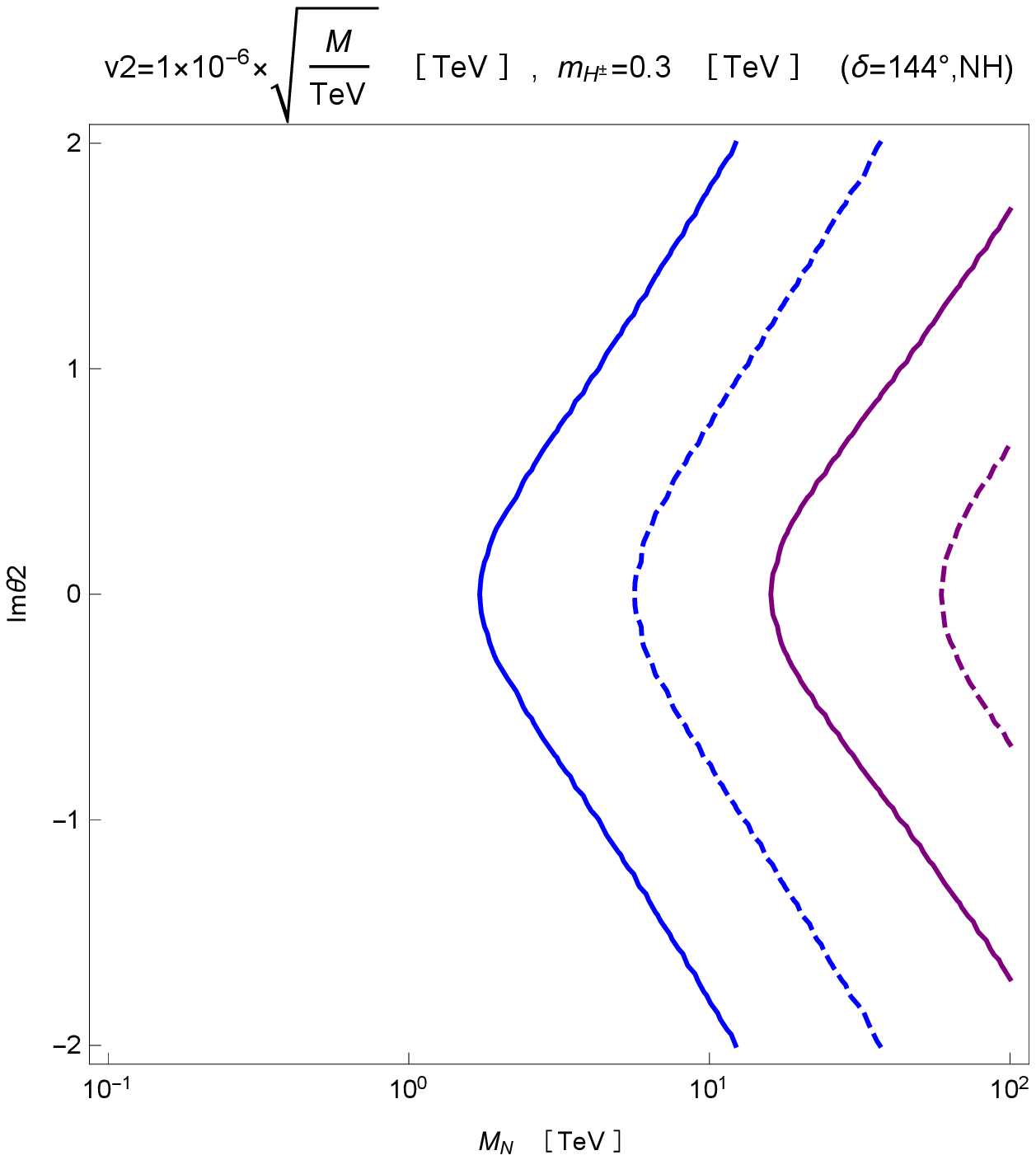}

      \end{minipage}
 
 
      \begin{minipage}{0.33\hsize}
        \centering
          \includegraphics[keepaspectratio, scale=0.44, angle=0]
                          {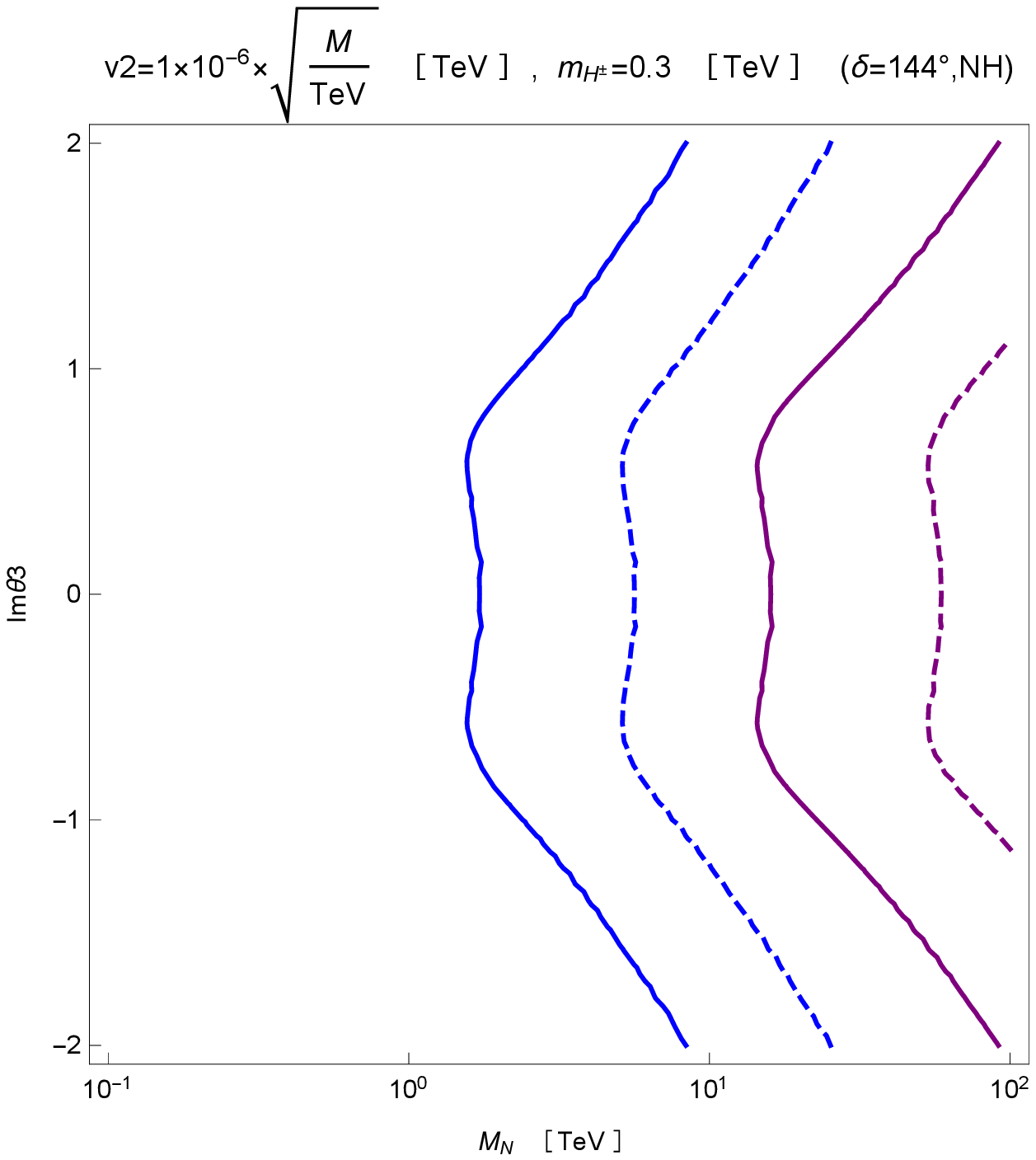}
      \end{minipage} \\
      \\
 
      \begin{minipage}{0.33\hsize}
        \centering
          \includegraphics[keepaspectratio, scale=0.44, angle=0]
                          {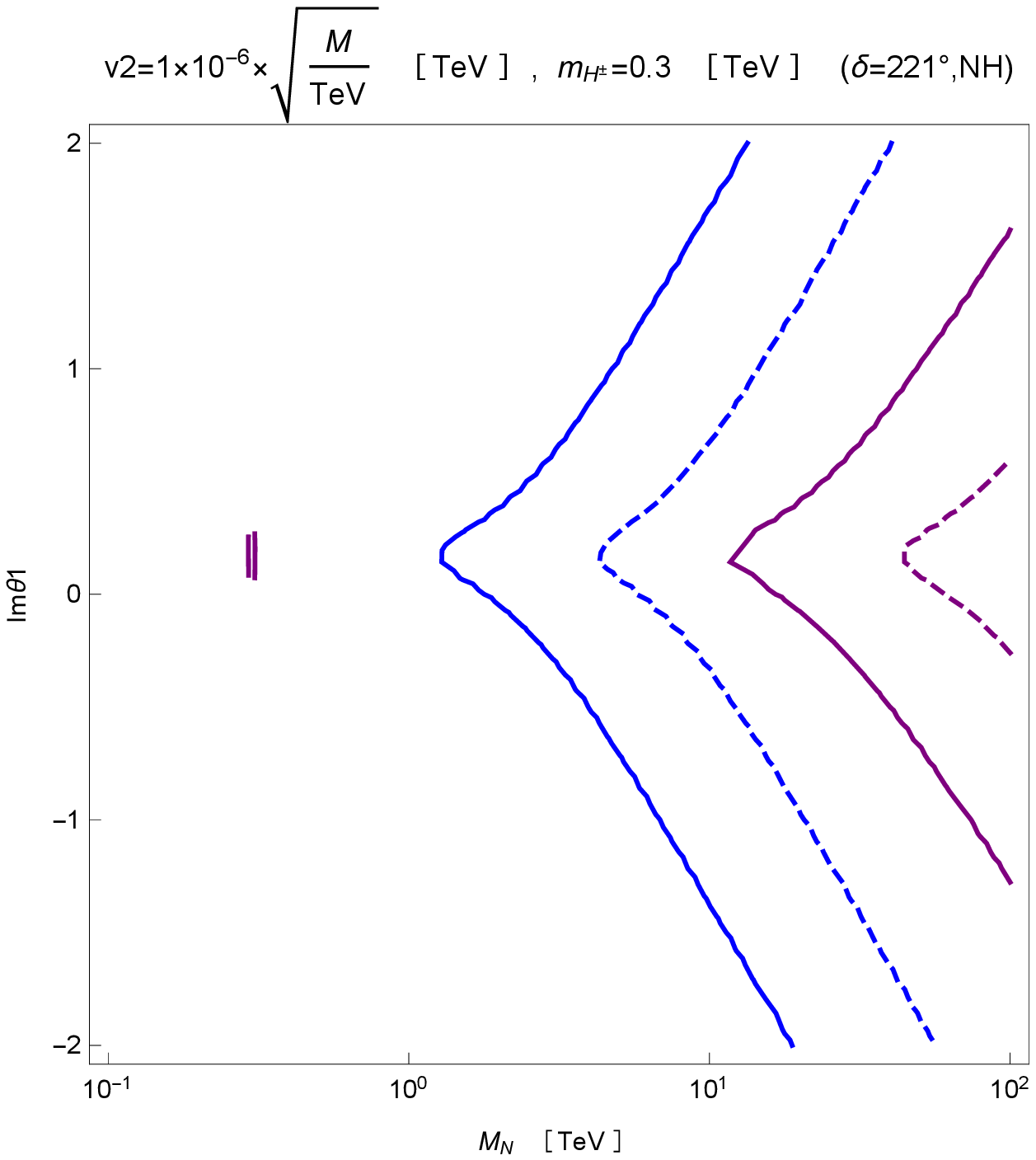}
      \end{minipage}

 
      \begin{minipage}{0.33\hsize}
        \centering
          \includegraphics[keepaspectratio, scale=0.44, angle=0]
                          {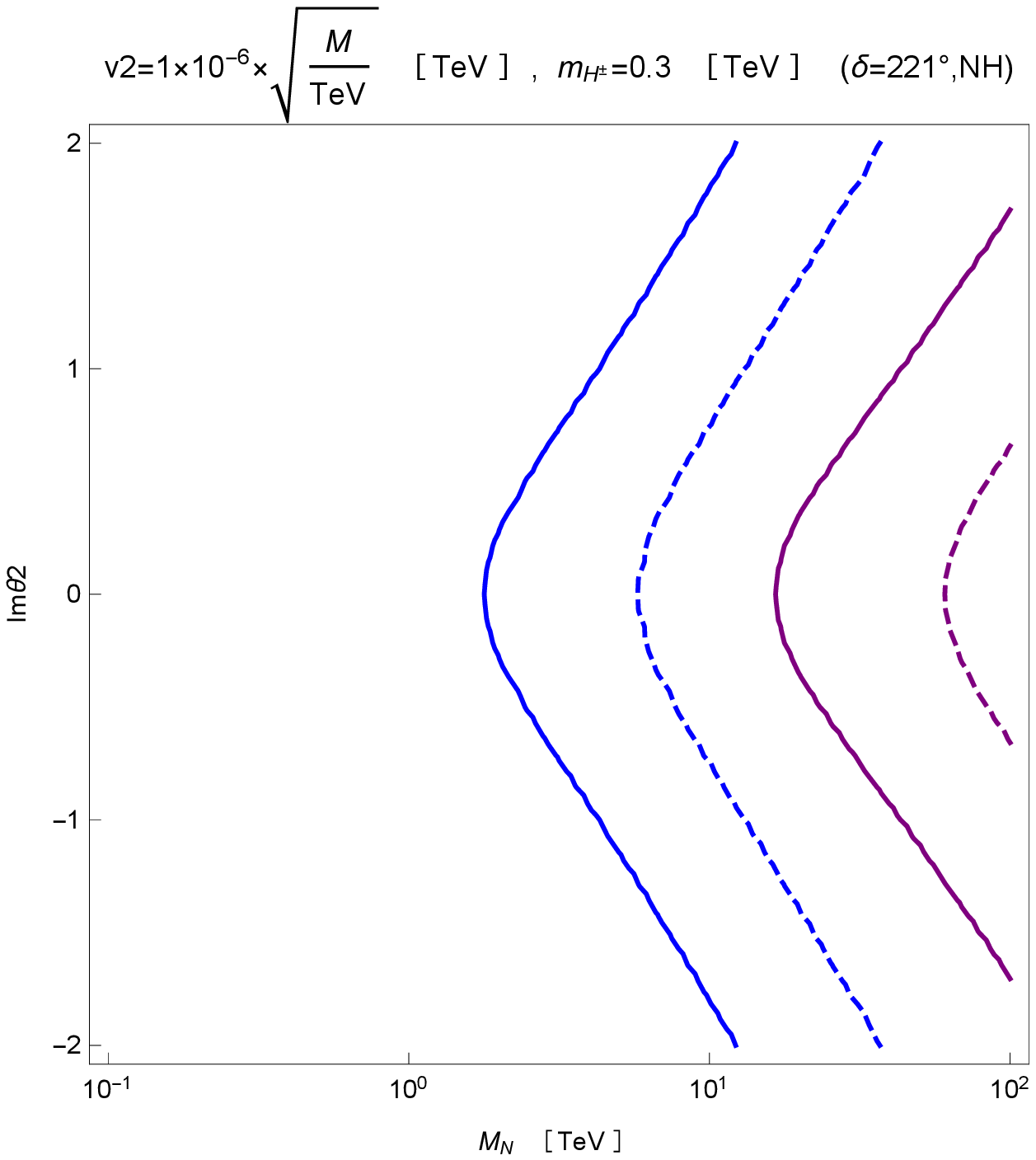}
      \end{minipage}
 
 
      \begin{minipage}{0.33\hsize}
        \centering
          \includegraphics[keepaspectratio, scale=0.44, angle=0]
                          {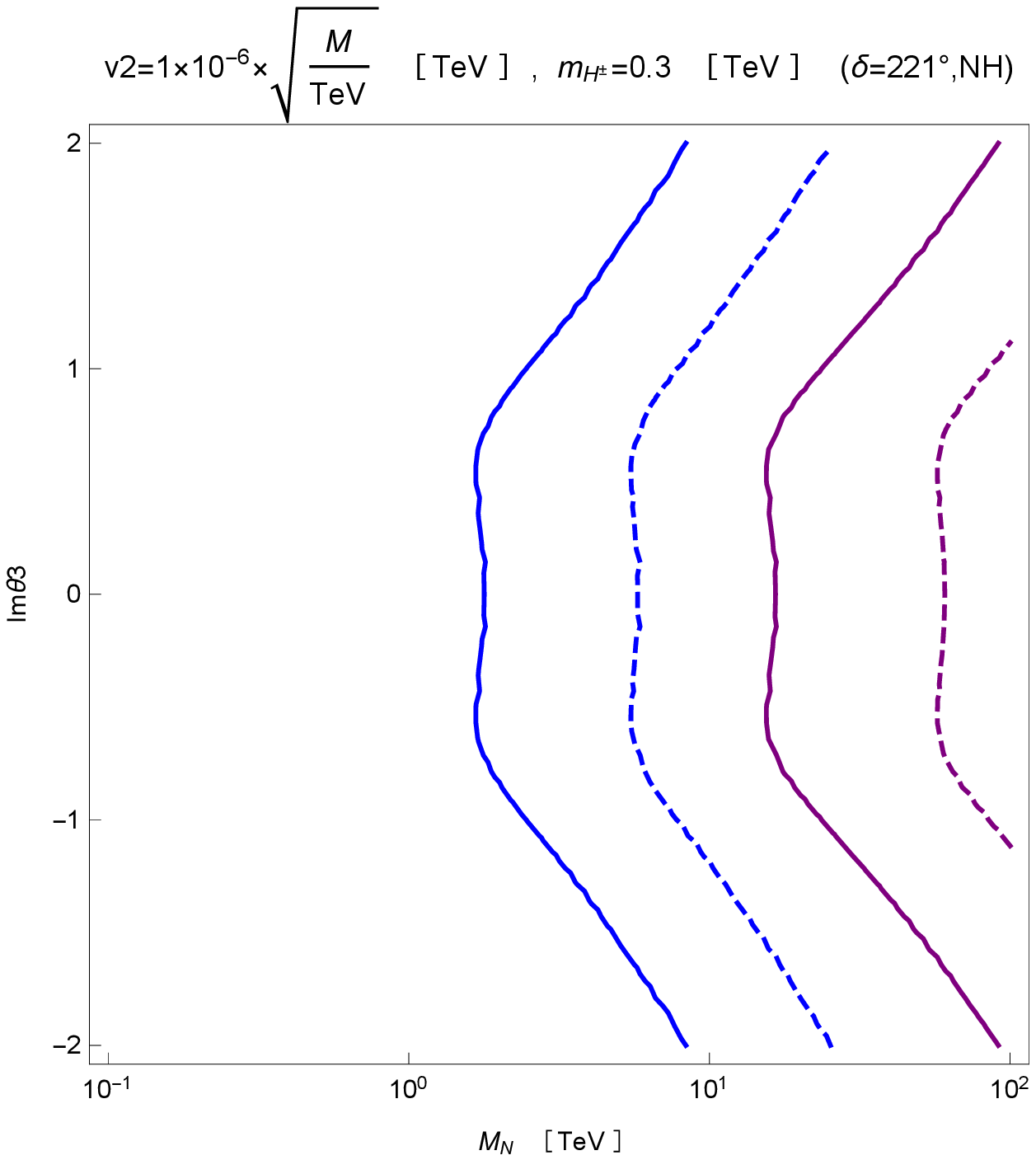}
      \end{minipage} \\ 
      \\
 
      \begin{minipage}{0.33\hsize}
        \centering
          \includegraphics[keepaspectratio, scale=0.44, angle=0]
                          {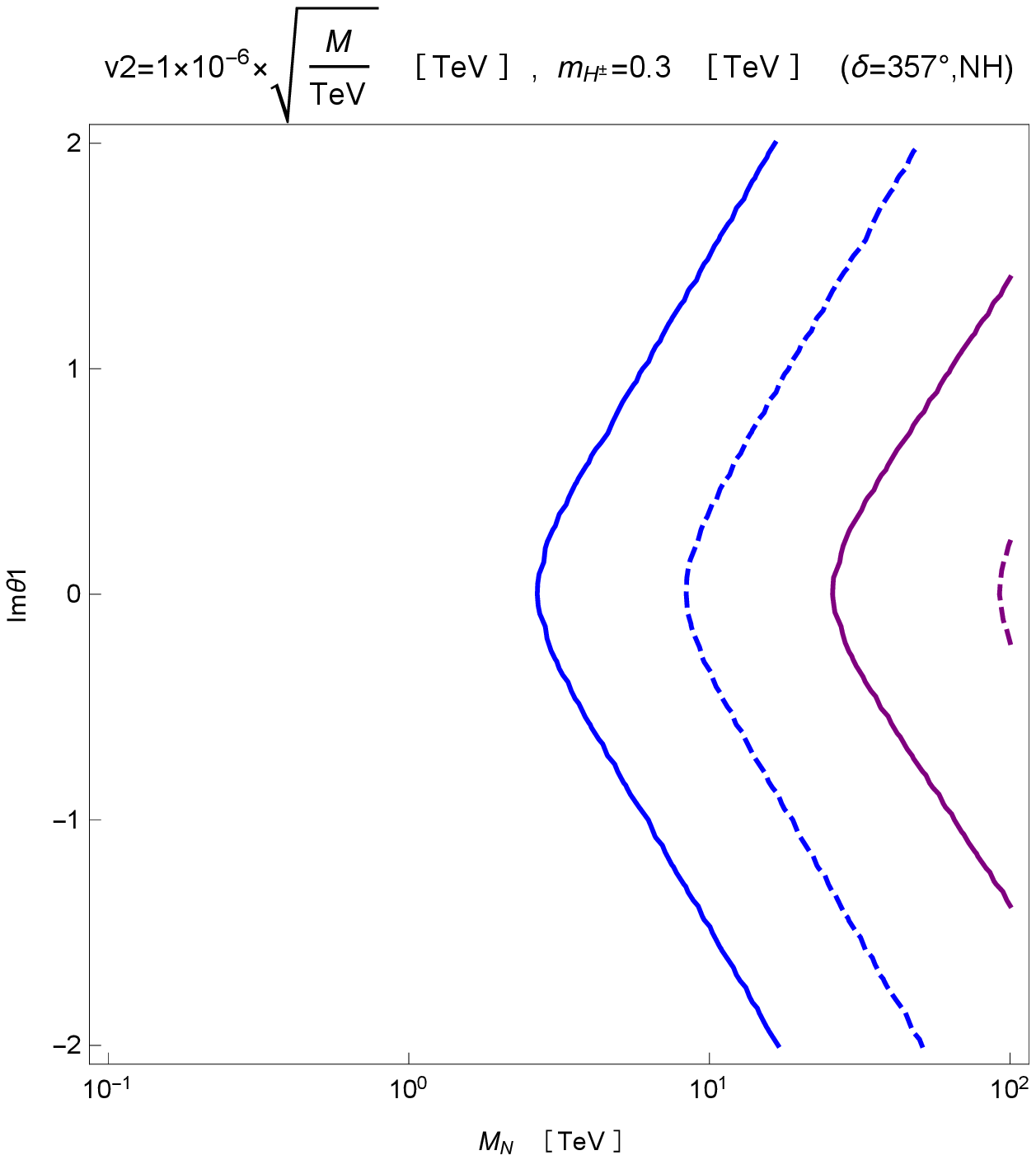}
      \end{minipage}

 
      \begin{minipage}{0.33\hsize}
        \centering
          \includegraphics[keepaspectratio, scale=0.44, angle=0]
                          {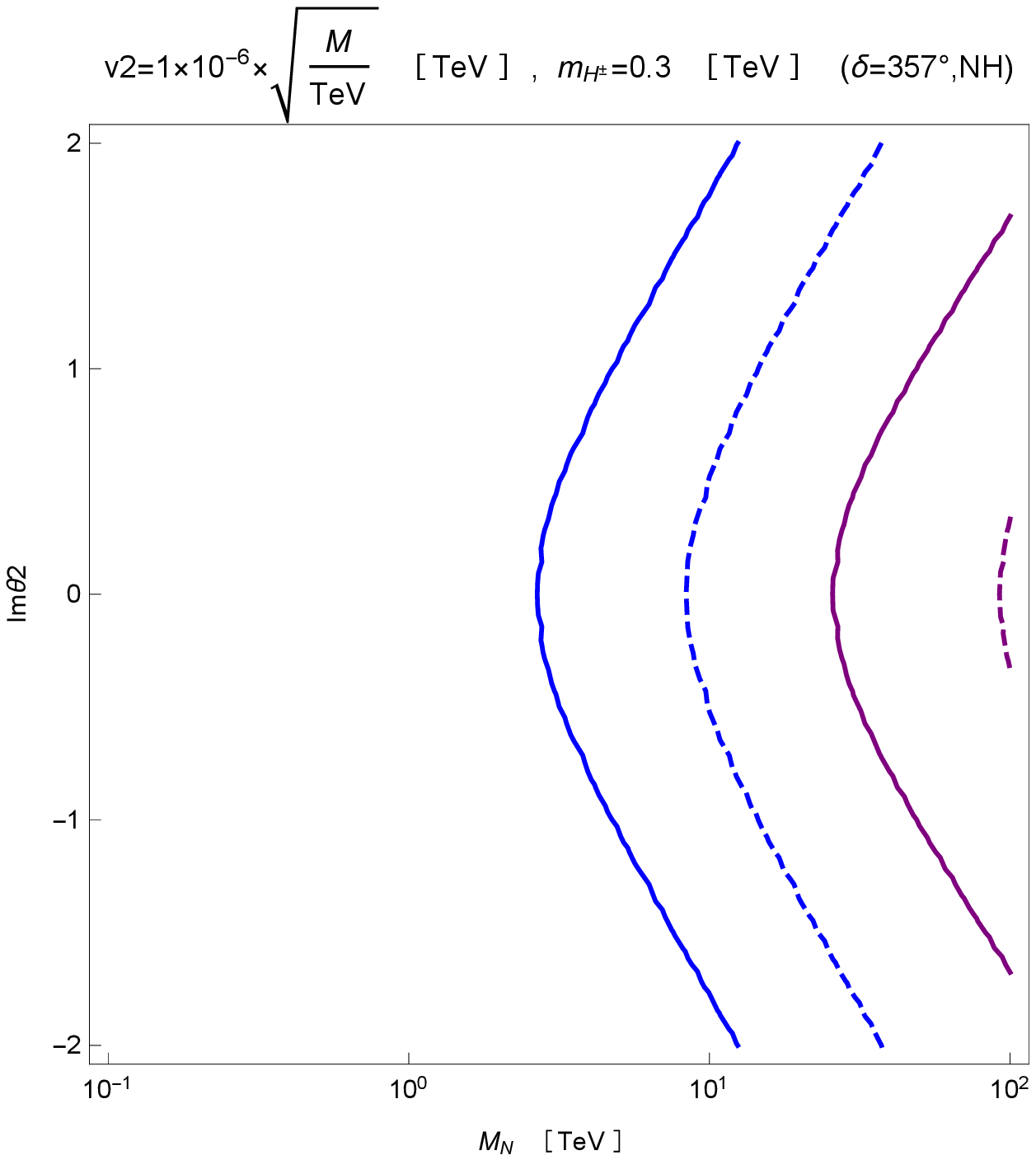}
      \end{minipage}
 
 
      \begin{minipage}{0.33\hsize}
        \centering
          \includegraphics[keepaspectratio, scale=0.44, angle=0]
                          {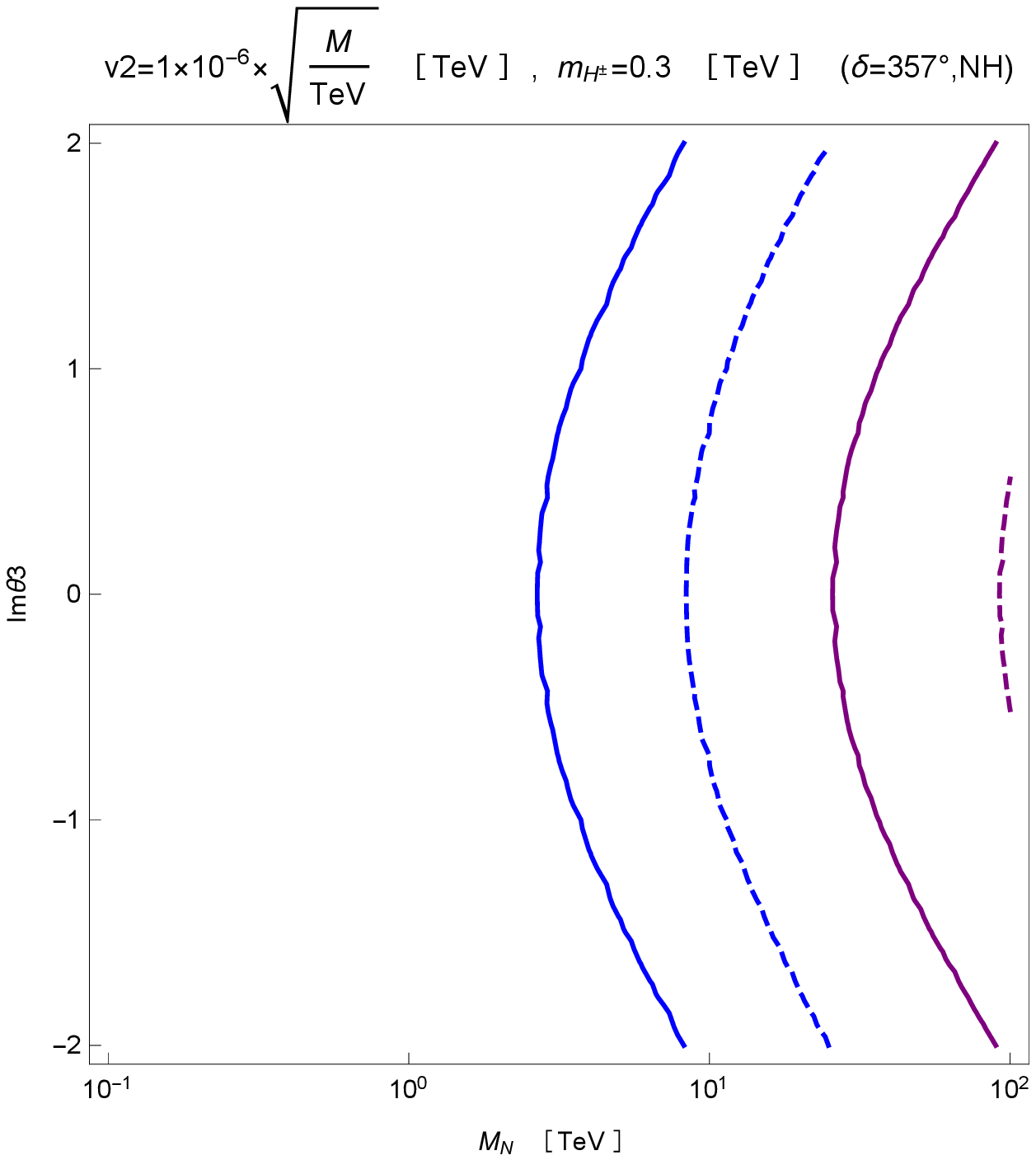}
      \end{minipage} 

    \end{tabular}
     \caption{\footnotesize  Same as figure~\ref{figprecalN} except that the prediction for $CR(\mu +{\rm Ti}\to e+{\rm Ti})$
      is presented by the purple lines, for Normal Hierarchy.
      The solid purple line corresponds to $CR(\mu +{\rm Ti}\to e+{\rm Ti})=10^{-18}$ for $v_2$ in Eq.~(\ref{v2-nh}),
       and the dashed purple line corresponds to $CR(\mu +{\rm Ti}\to e+{\rm Ti})=10^{-18}$ when $v_2$ is multiplied by $1/3$.
      }
 \label{figprectiN}
\end{figure}

\begin{figure}[H]
  \centering
      \begin{tabular}{c}
 
 
      \begin{minipage}{0.33\hsize}
        \centering
          \includegraphics[keepaspectratio, scale=0.42, angle=0]
                          {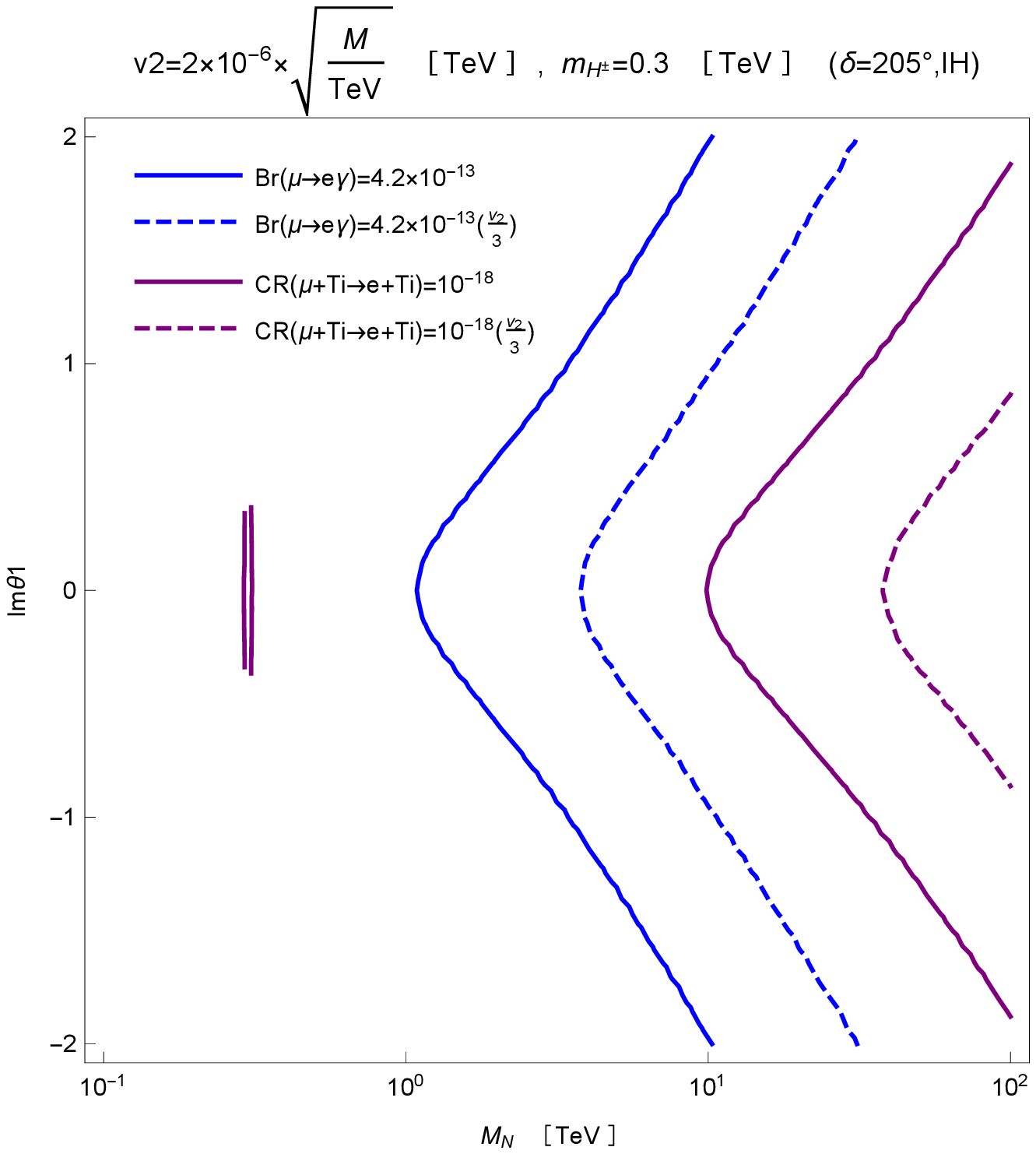}
      \end{minipage}

 
      \begin{minipage}{0.33\hsize}
        \centering
          \includegraphics[keepaspectratio, scale=0.44, angle=0]
                          {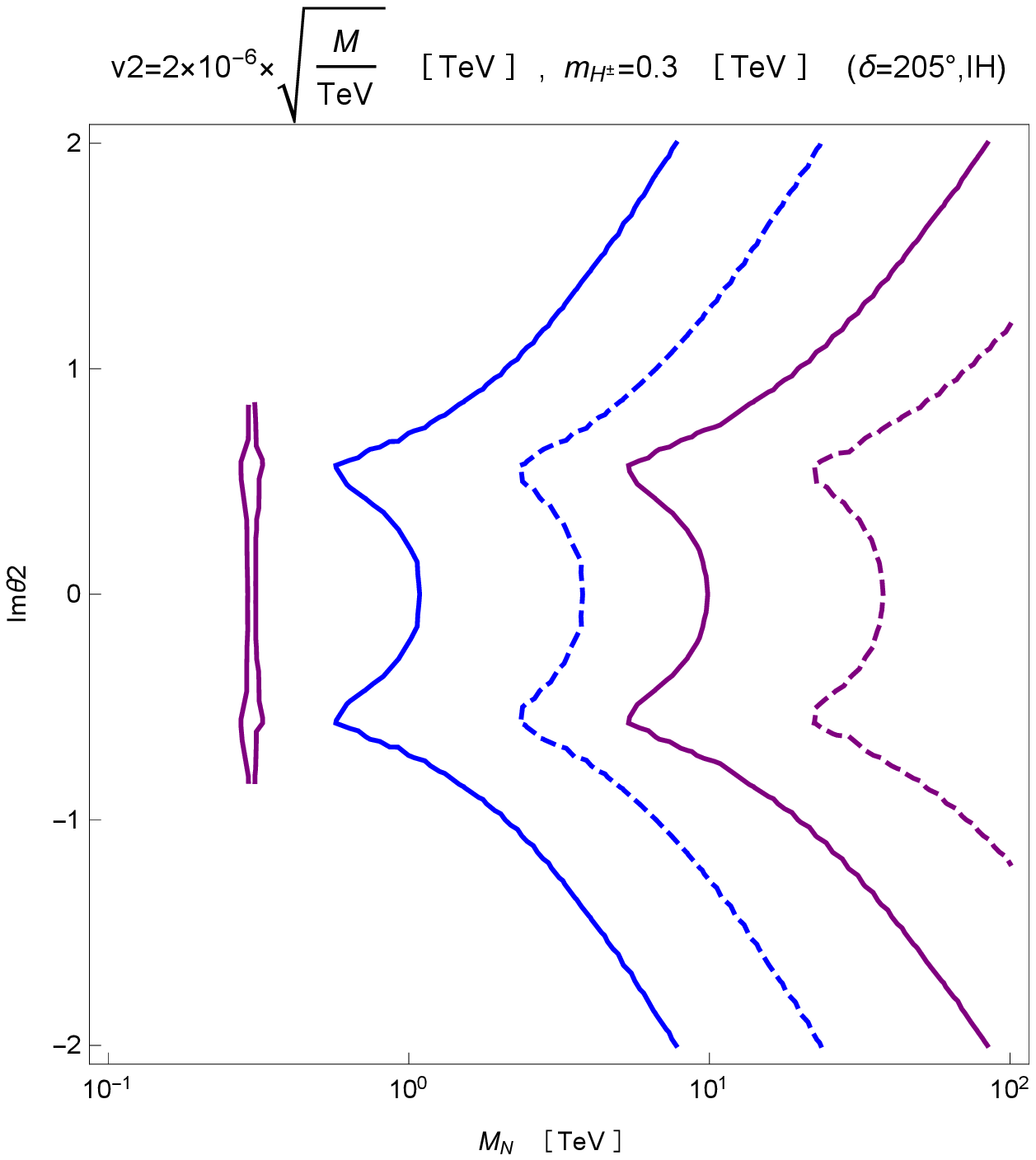}

      \end{minipage}
 
 
      \begin{minipage}{0.33\hsize}
        \centering
          \includegraphics[keepaspectratio, scale=0.44, angle=0]
                          {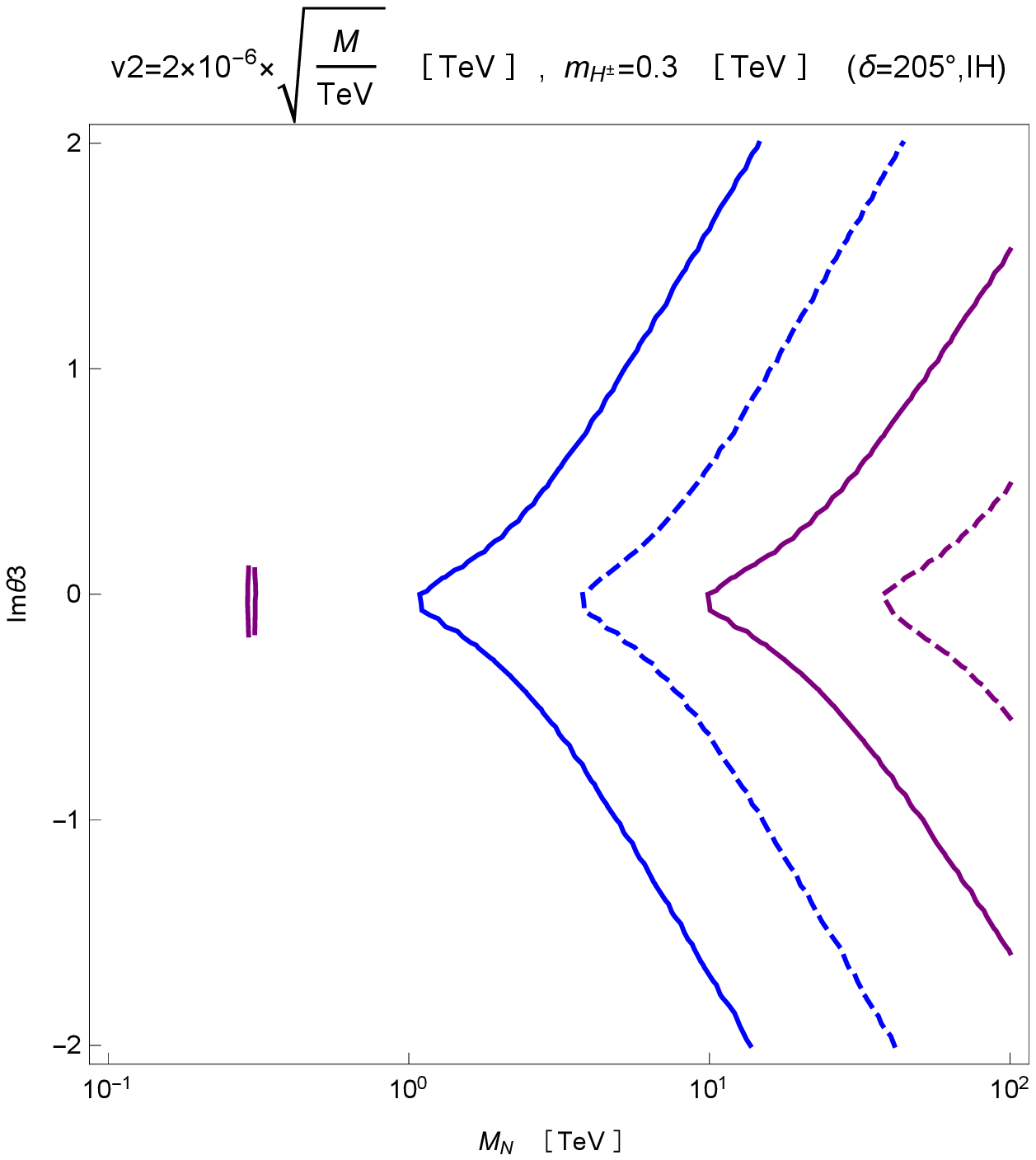}
      \end{minipage} \\
      \\
 
      \begin{minipage}{0.33\hsize}
        \centering
          \includegraphics[keepaspectratio, scale=0.44, angle=0]
                          {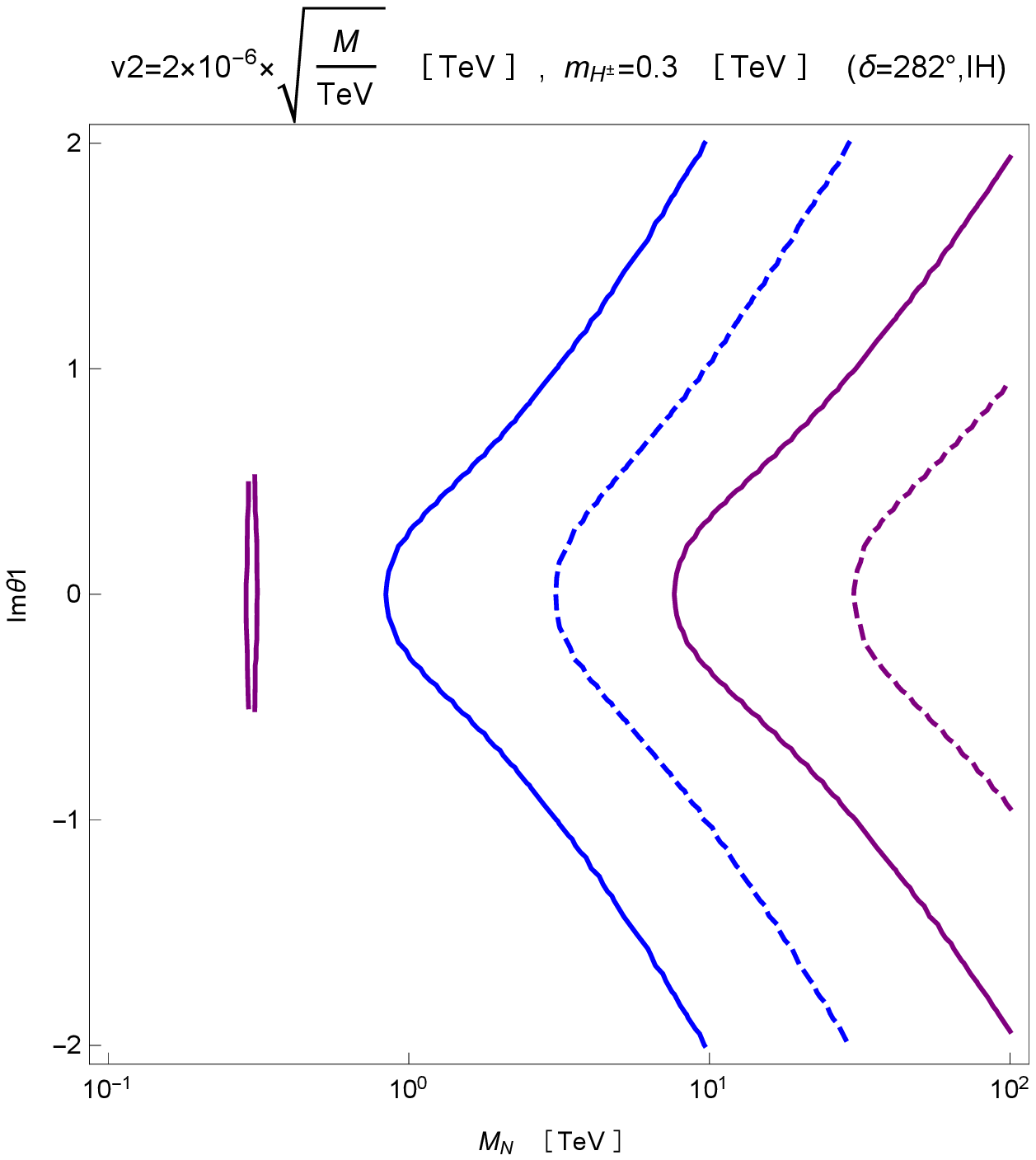}
      \end{minipage}

 
      \begin{minipage}{0.33\hsize}
        \centering
          \includegraphics[keepaspectratio, scale=0.44, angle=0]
                          {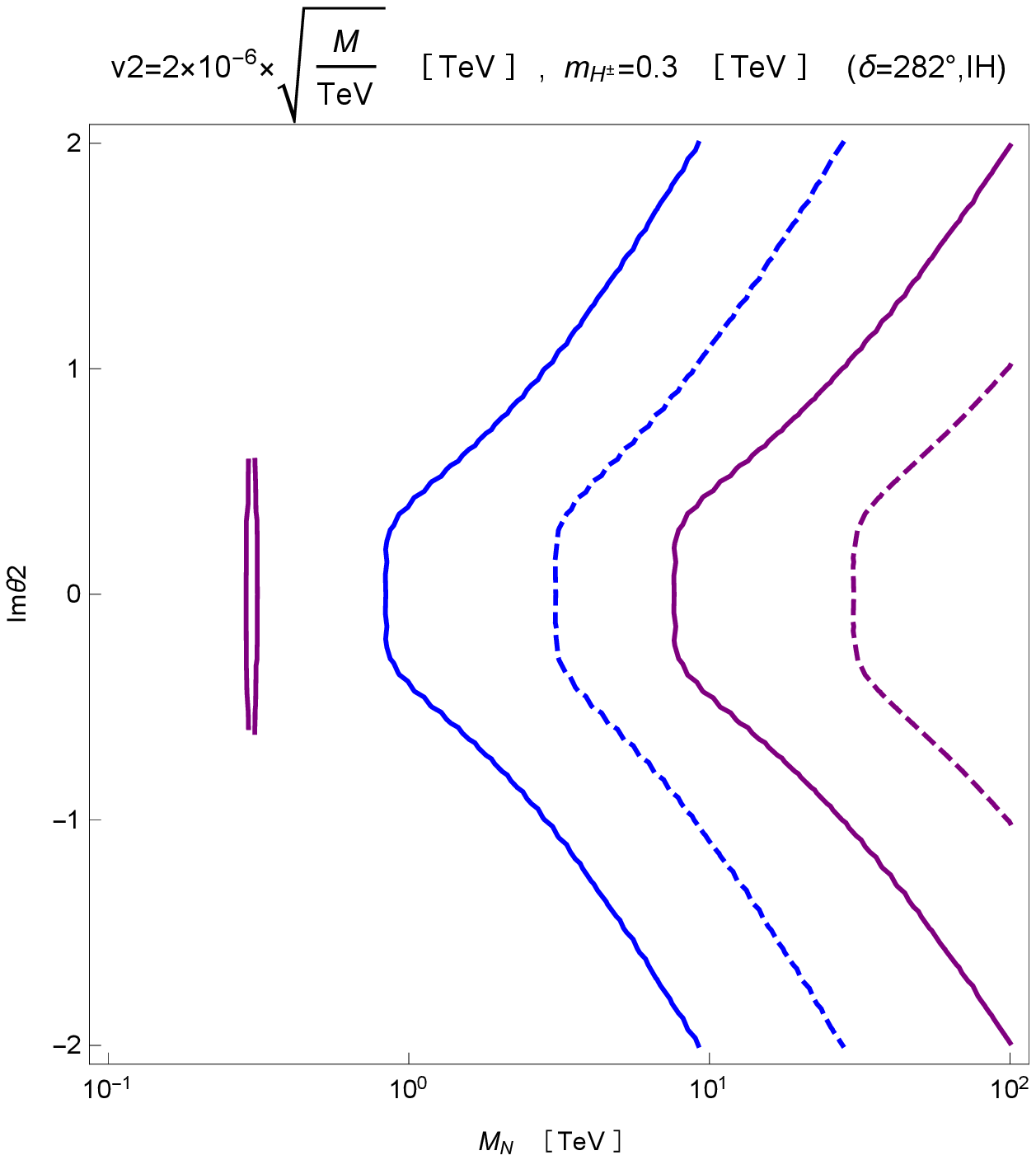}
      \end{minipage}
 
 
      \begin{minipage}{0.33\hsize}
        \centering
          \includegraphics[keepaspectratio, scale=0.44, angle=0]
                          {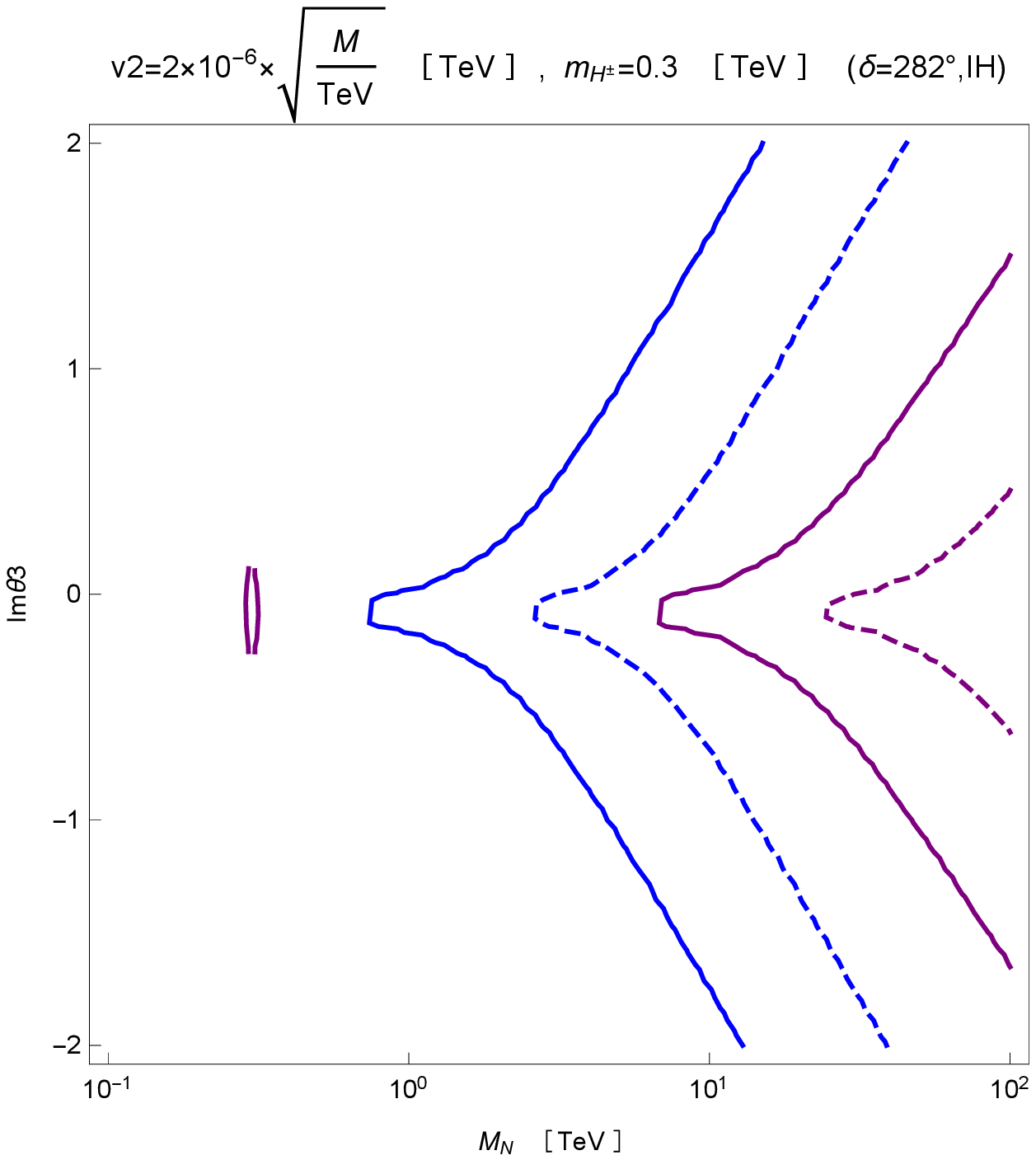}
      \end{minipage} \\ 
      \\
 
      \begin{minipage}{0.33\hsize}
        \centering
          \includegraphics[keepaspectratio, scale=0.44, angle=0]
                          {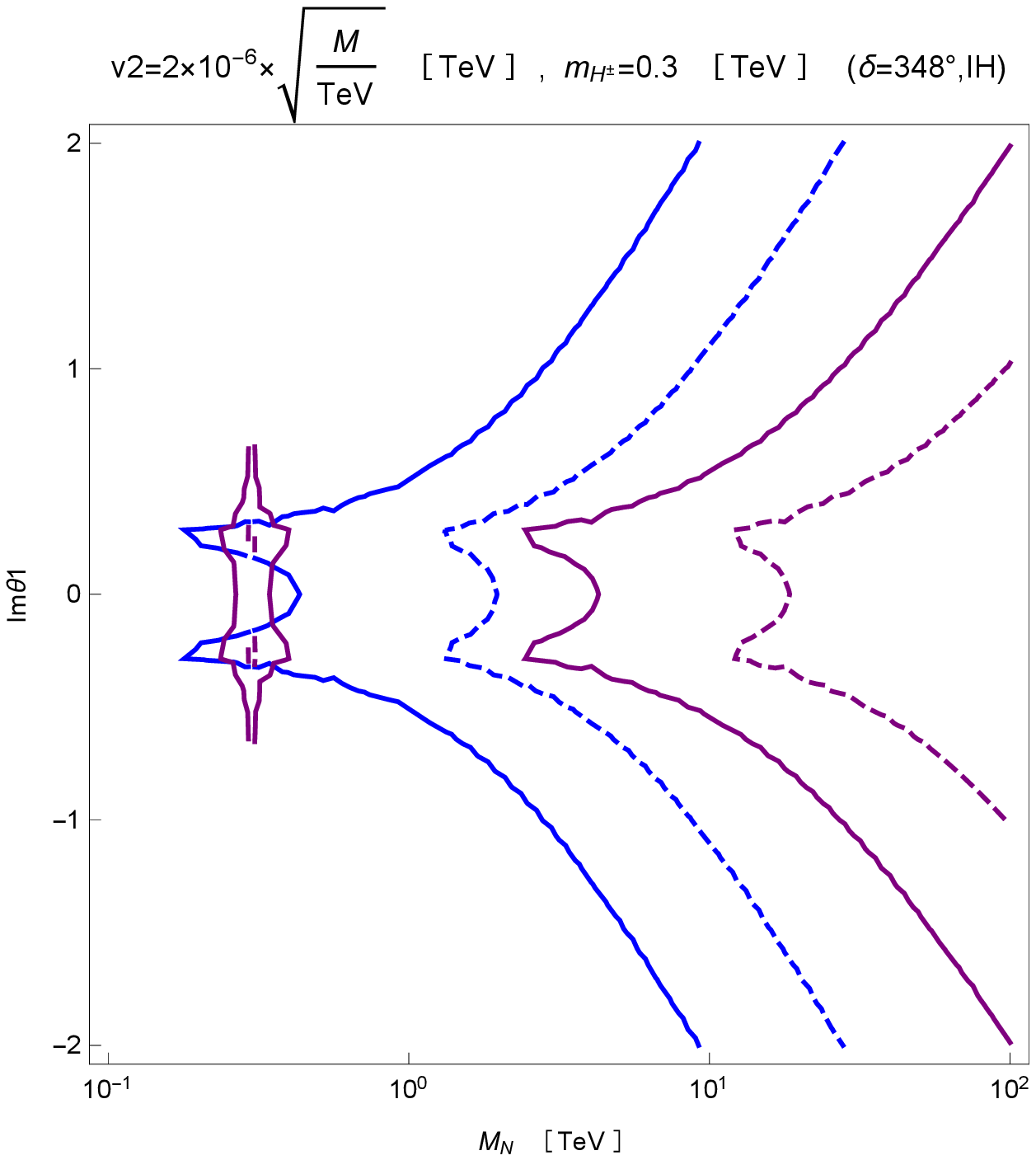}
      \end{minipage}

 
      \begin{minipage}{0.33\hsize}
        \centering
          \includegraphics[keepaspectratio, scale=0.44, angle=0]
                          {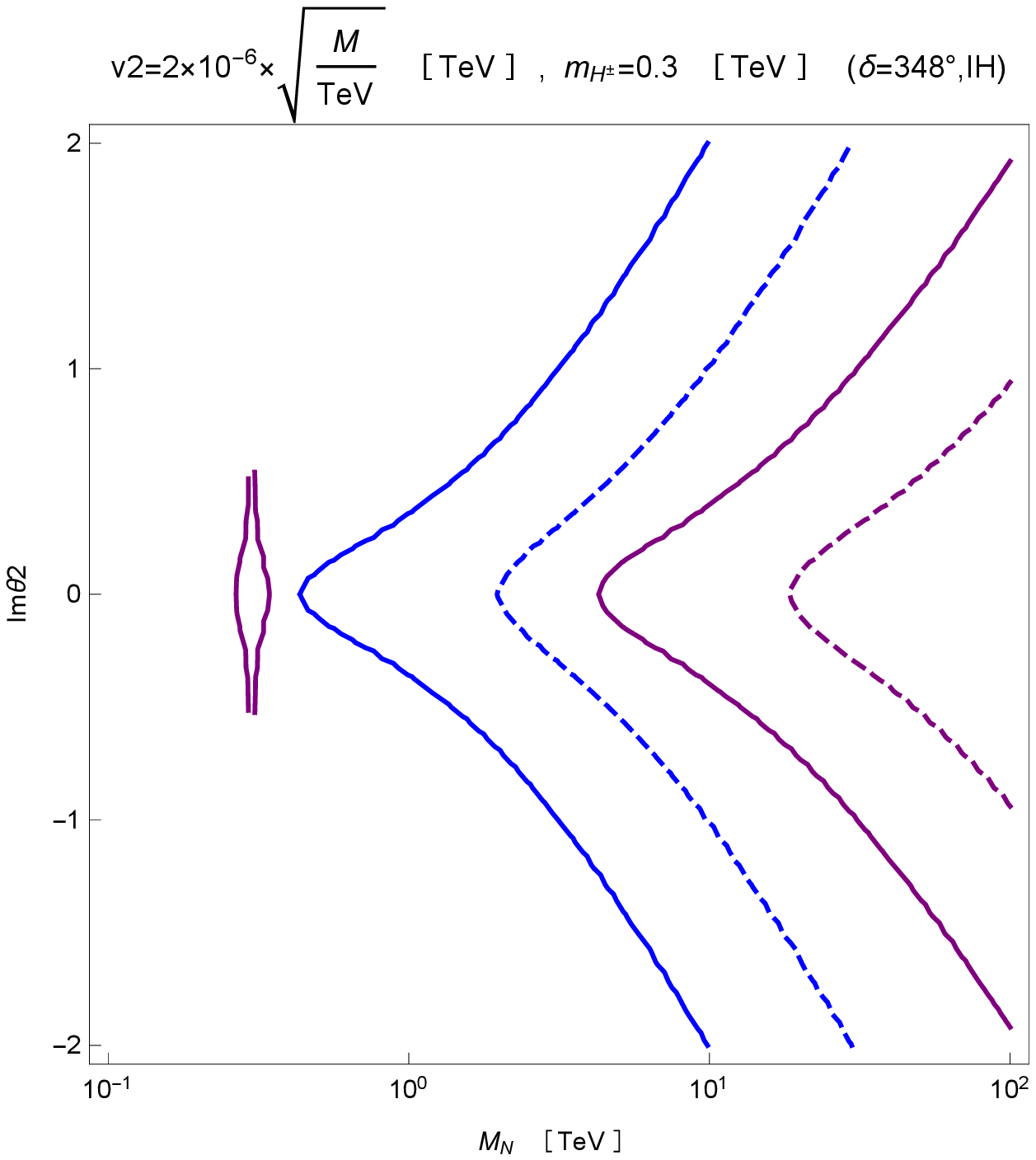}
      \end{minipage}
 
 
      \begin{minipage}{0.33\hsize}
        \centering
          \includegraphics[keepaspectratio, scale=0.44, angle=0]
                          {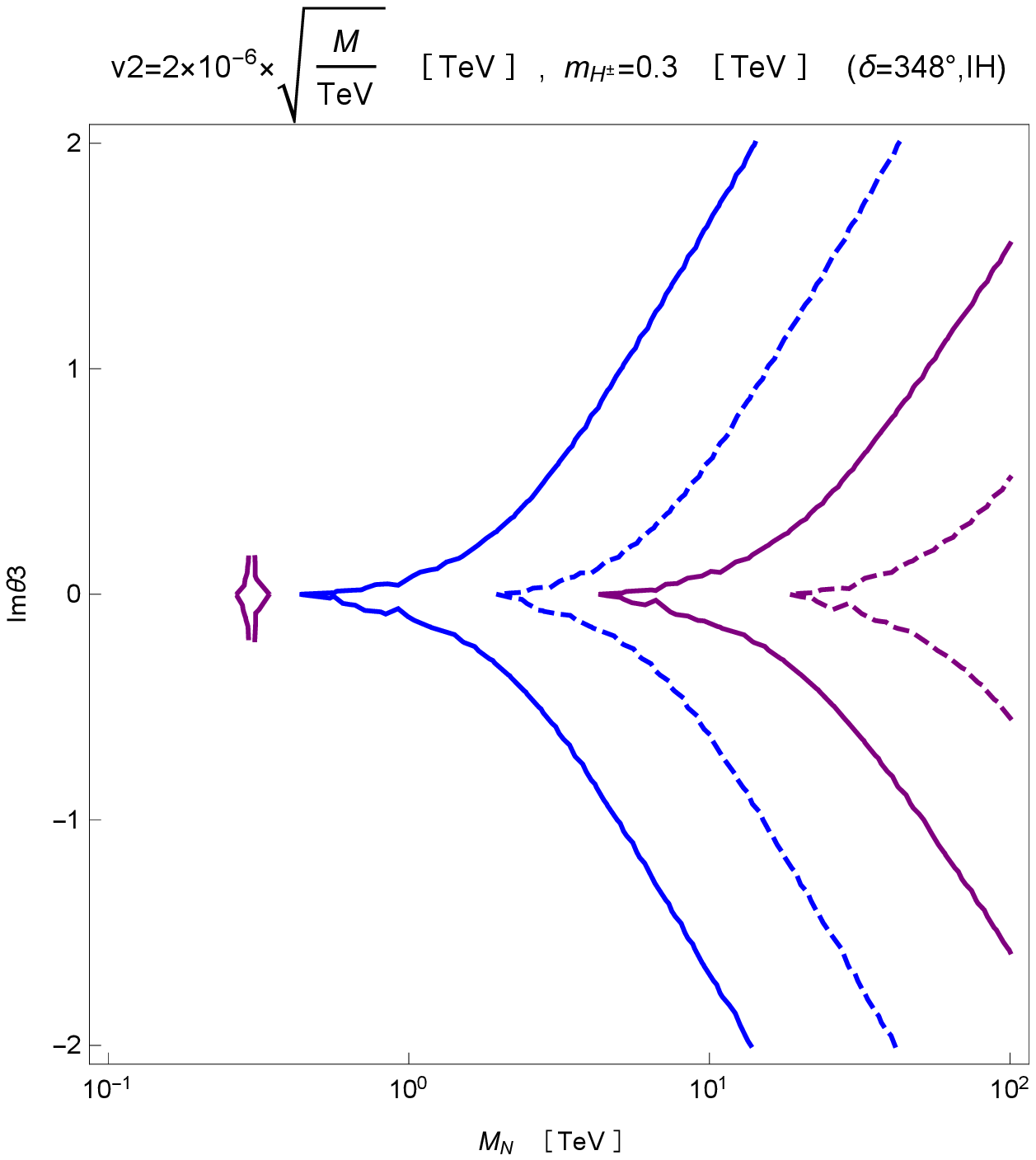}
      \end{minipage} 

    \end{tabular}
     \caption{\footnotesize Same as figure~\ref{figprectiN} except that mass hierarchy is Inverted Hierarchy and $v_2$ is given in Eq.~(\ref{v2-ih}).}
 \label{figprectiI}
\end{figure}

In Fig.~\ref{figprectiN}, the solid purple line agrees with $CR(\mu +{\rm Ti}\to e+{\rm Ti})=10^{-18}$, the future sensitivity, for NH and $v_2$ in Eq.~(\ref{v2-nh}).
Threfore, in the region between the solid blue line and the solid purple line (we neglect the purple line near $M_N=0.3$ TeV), 
 the $\mu +{\rm Ti}\to e+{\rm Ti}$ process can be detected in the future.
Figure~\ref{figprectiI} is the corresponding plot for IH and $v_2$ in Eq.~(\ref{v2-ih}).

In the same figures, the dashed purple line agrees with $CR(\mu +{\rm Ti}\to e+{\rm Ti})=10^{-18}$ when $v_2$ is multiplied by $1/3$,
 and the $\mu +{\rm Ti}\to e+{\rm Ti}$ process can be detected in the region between the dashed blue line and the dashed purple line for this $v_2$.
Just as with Al, the relative location of the contours of $Br(\mu\to e\gamma)$ and $CR(\mu +{\rm Ti}\to e+{\rm Ti})$ does not depend on $v_2$.

 \newpage
\newpage

 \newpage
\newpage

 \newpage
\newpage

 \newpage
\newpage
 
\subsubsection{$Z \to \bar{e}_\alpha e_\beta$}

\begin{figure}[H]
  \centering
  \thispagestyle{empty}
    \begin{tabular}{c}
 
 
      \begin{minipage}{0.33\hsize}
        \centering
          \includegraphics[keepaspectratio, scale=0.38, angle=0]
                          {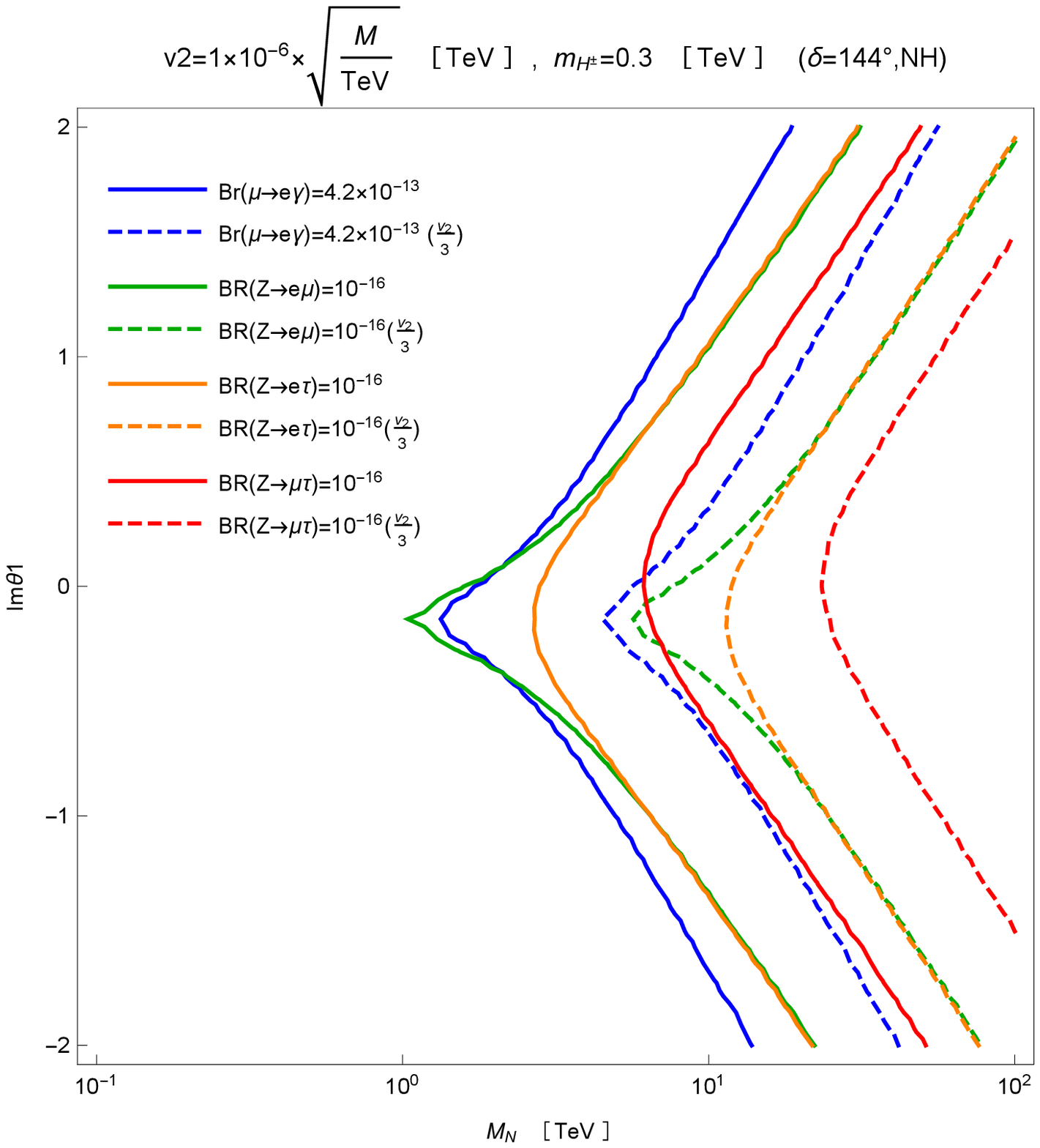}
      \end{minipage}

 
      \begin{minipage}{0.33\hsize}
        \centering
          \includegraphics[keepaspectratio, scale=0.44, angle=0]
                          {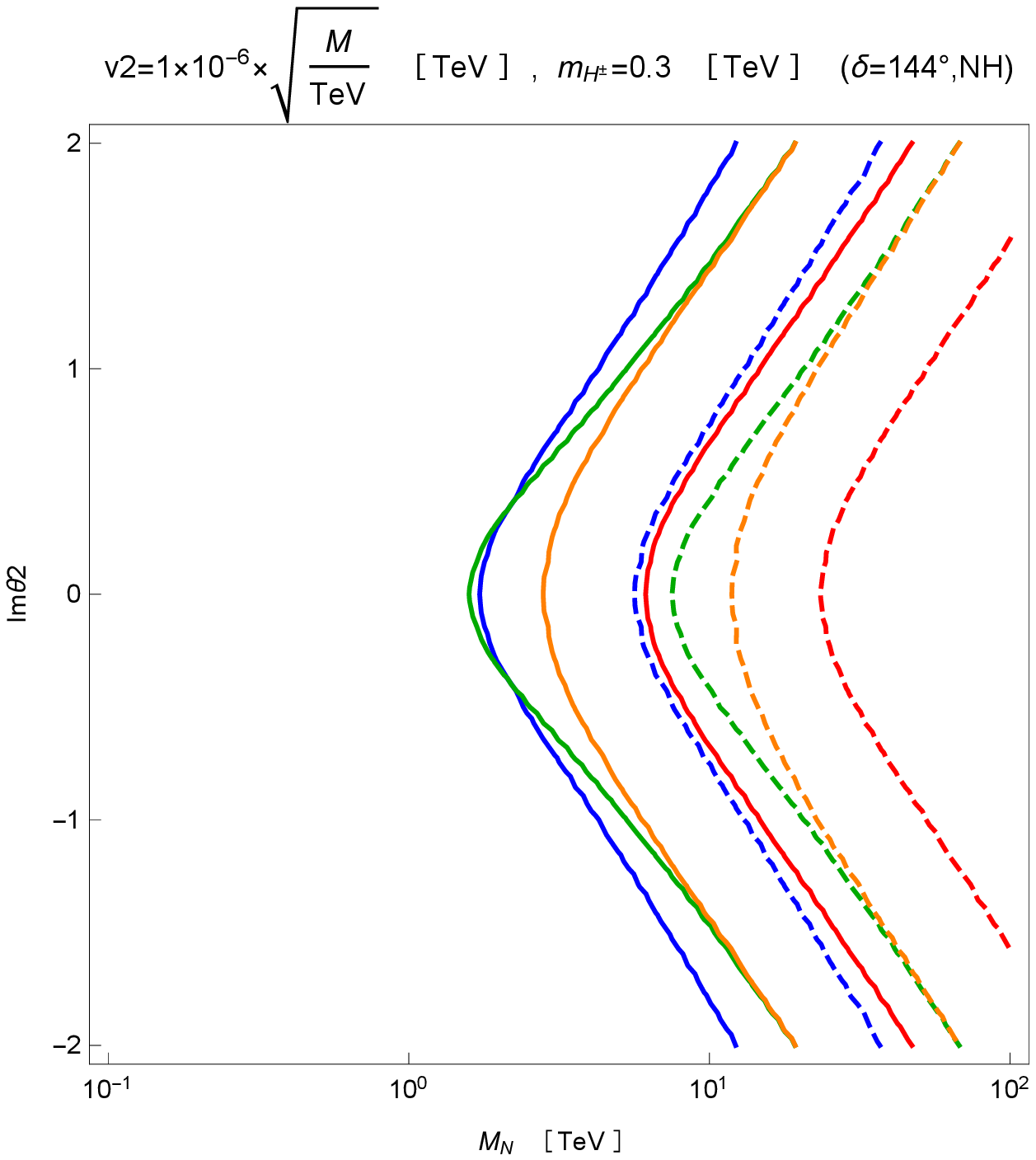}

      \end{minipage}
 
 
      \begin{minipage}{0.33\hsize}
        \centering
          \includegraphics[keepaspectratio, scale=0.44, angle=0]
                          {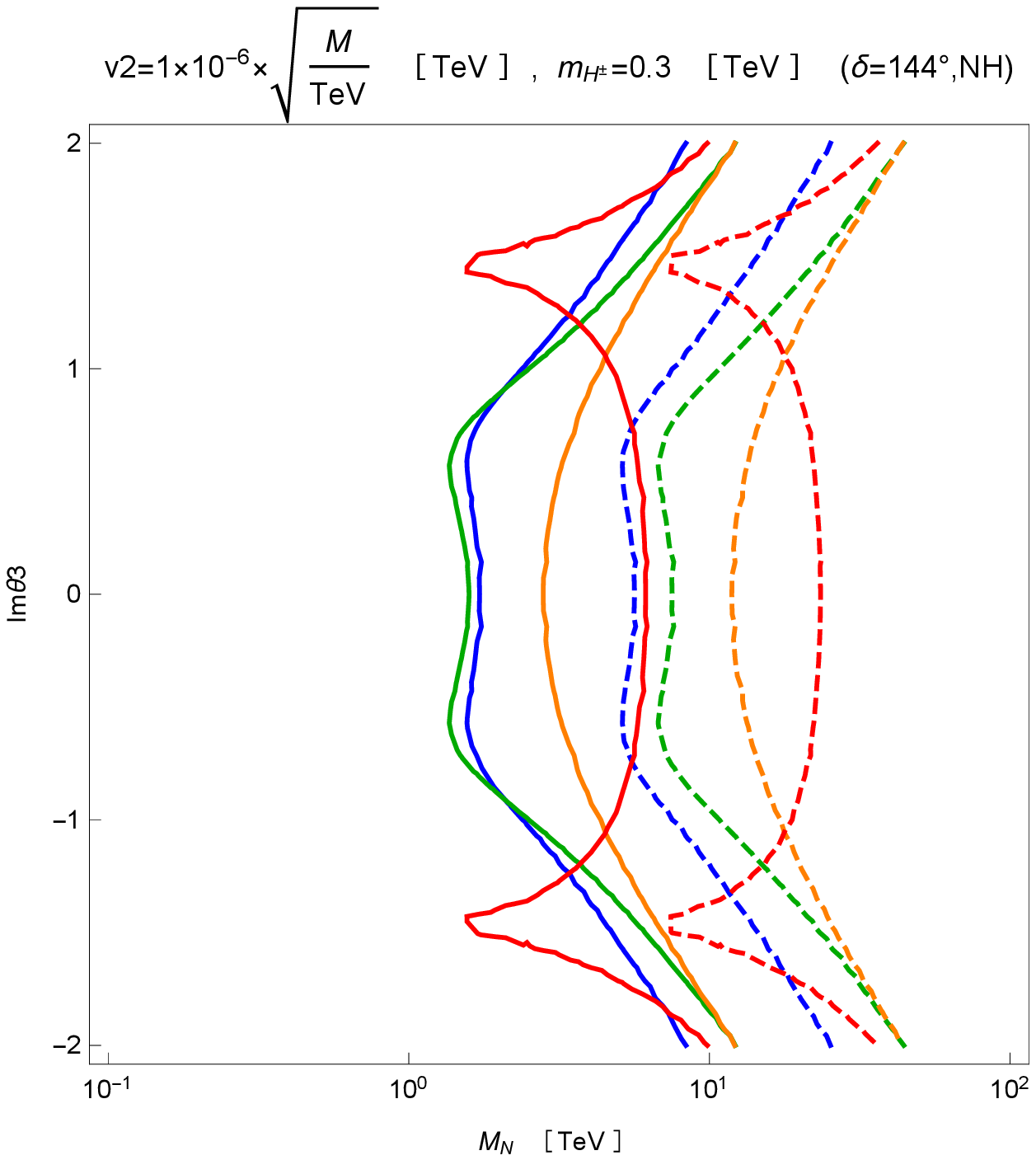}
      \end{minipage} \\
      \\
 
      \begin{minipage}{0.33\hsize}
        \centering
          \includegraphics[keepaspectratio, scale=0.44, angle=0]
                          {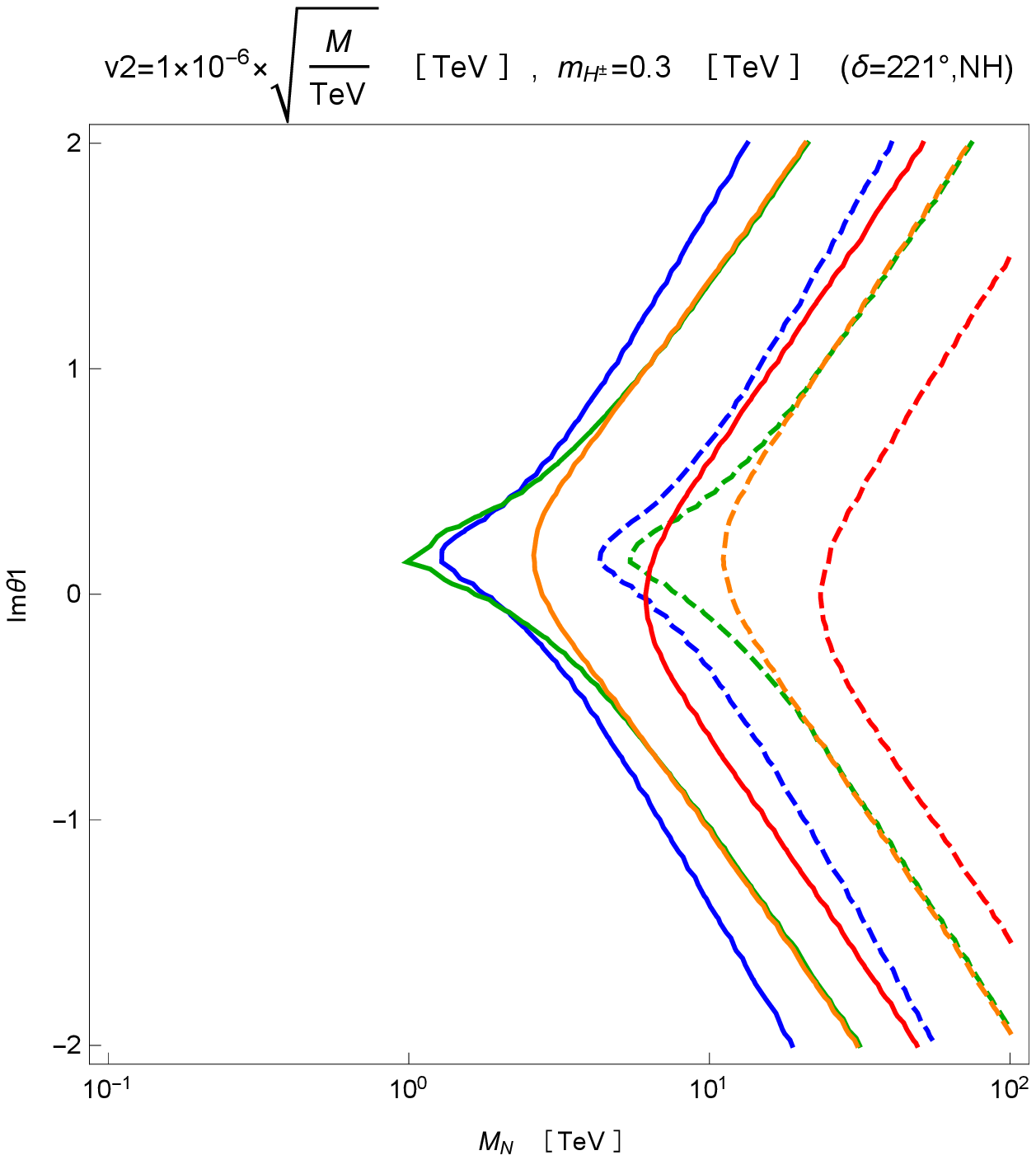}
      \end{minipage}

 
      \begin{minipage}{0.33\hsize}
        \centering
          \includegraphics[keepaspectratio, scale=0.44, angle=0]
                          {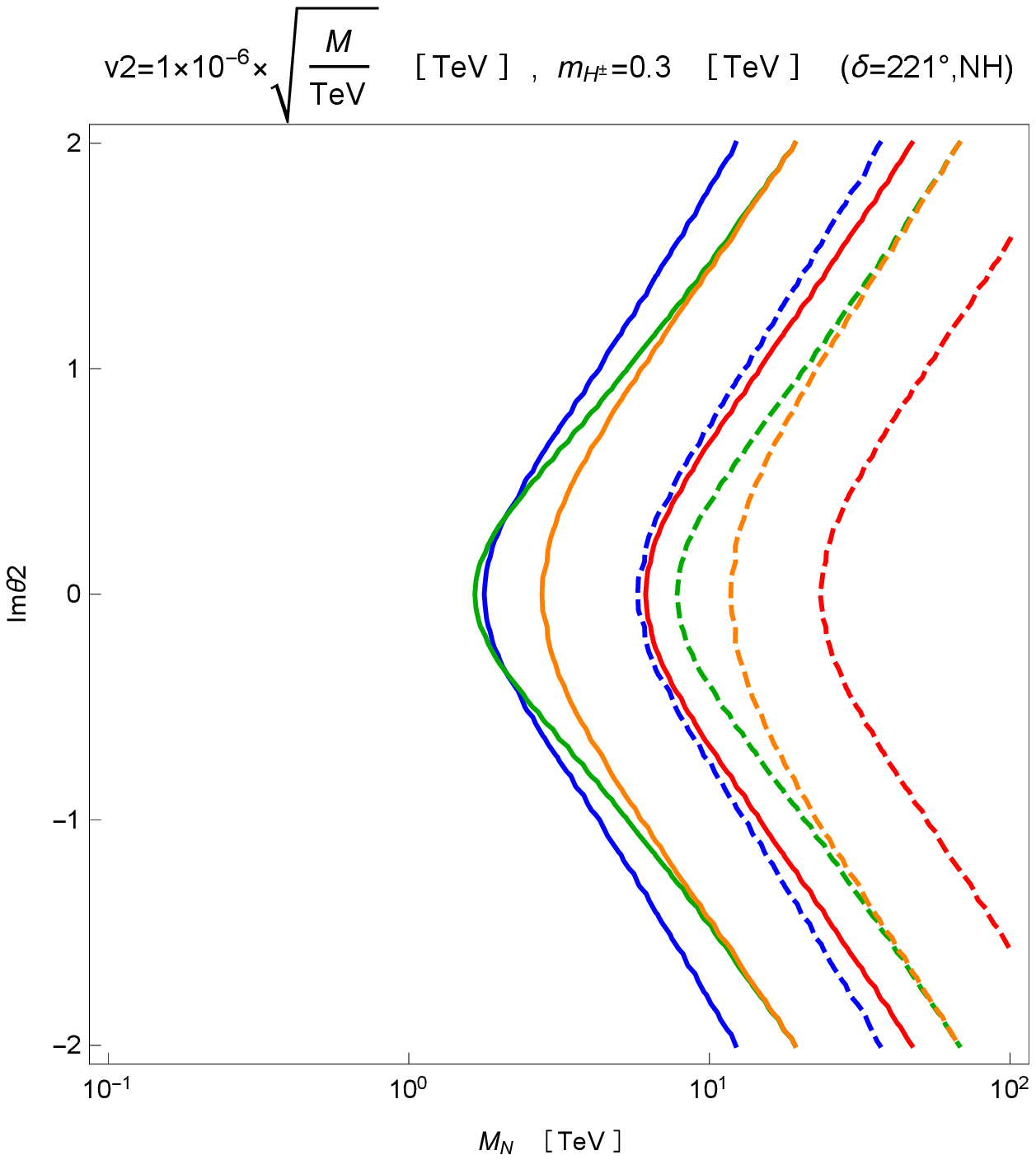}
      \end{minipage}
 
 
      \begin{minipage}{0.33\hsize}
        \centering
          \includegraphics[keepaspectratio, scale=0.44, angle=0]
                          {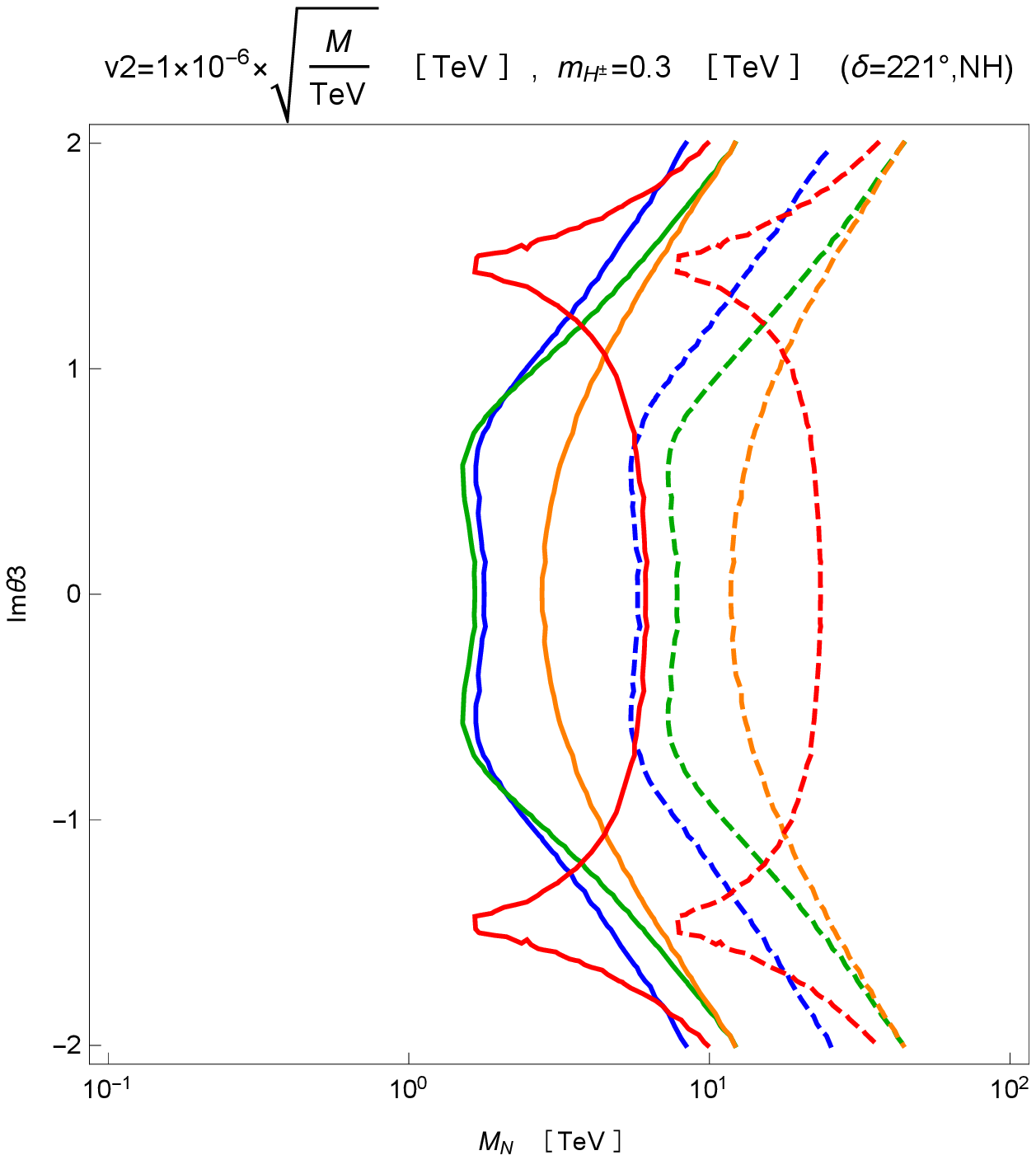}
      \end{minipage} \\ 
      \\
 
      \begin{minipage}{0.33\hsize}
        \centering
          \includegraphics[keepaspectratio, scale=0.44, angle=0]
                          {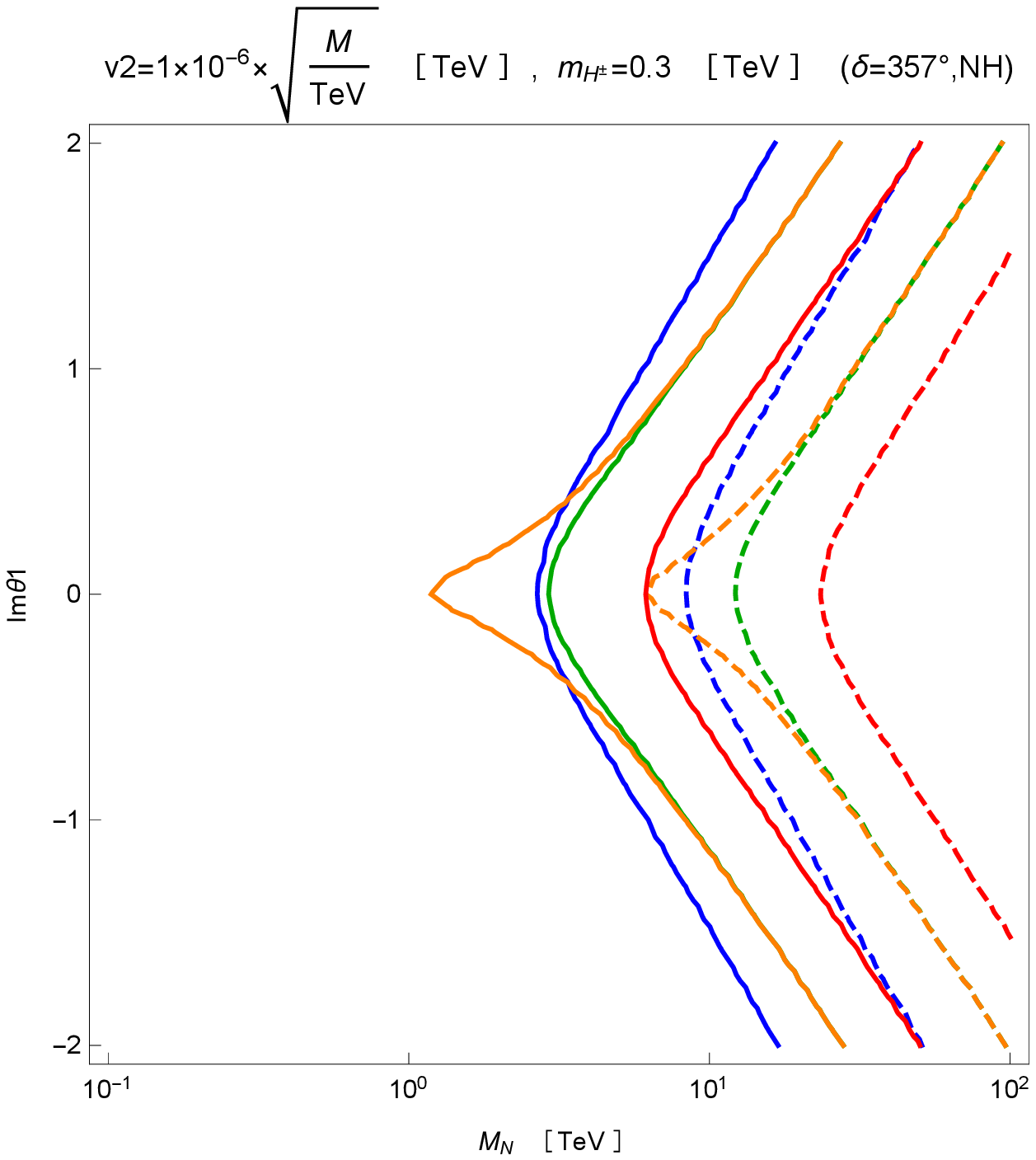}
      \end{minipage}

 
      \begin{minipage}{0.33\hsize}
        \centering
          \includegraphics[keepaspectratio, scale=0.44, angle=0]
                          {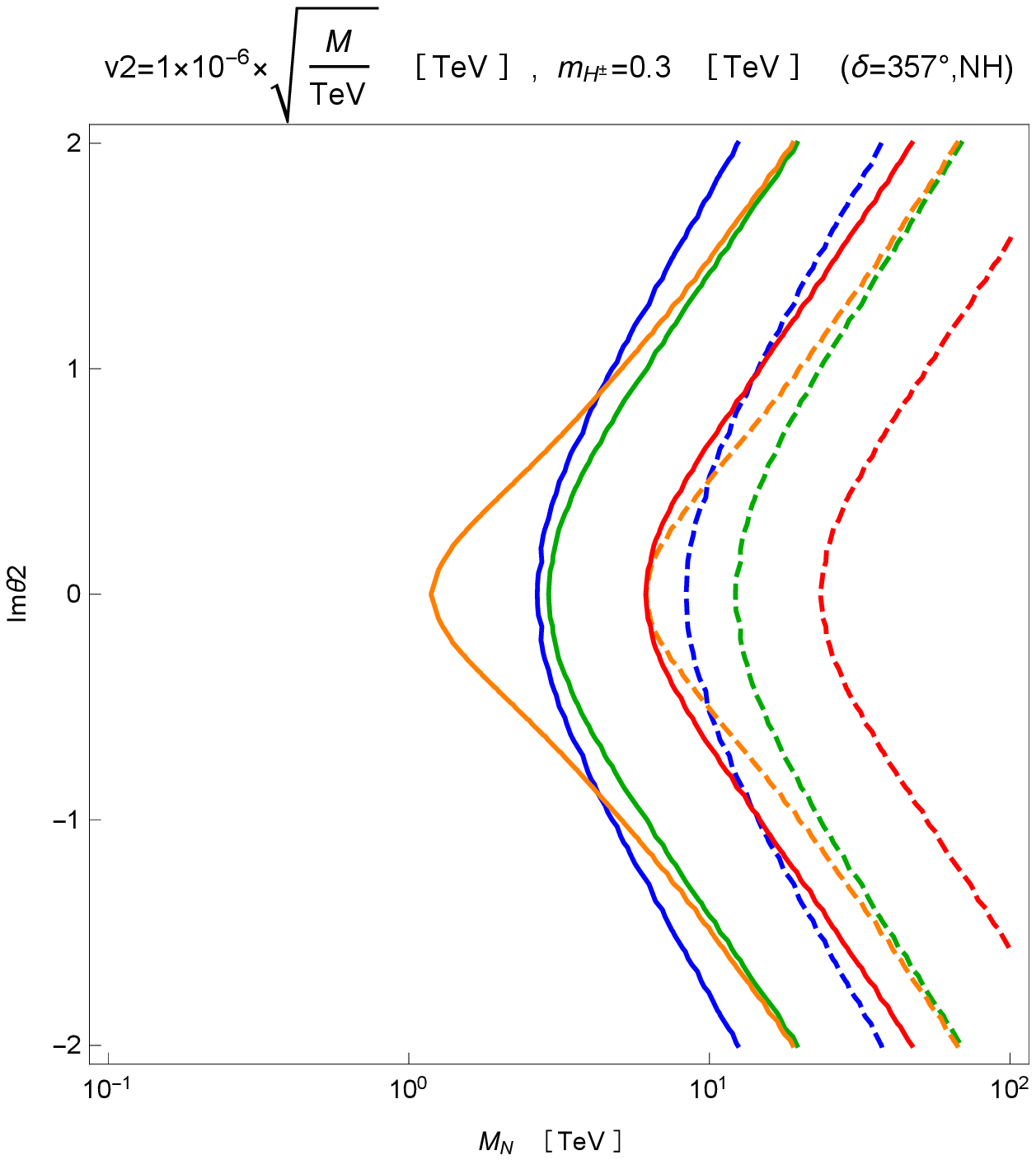}
      \end{minipage}
 
 
      \begin{minipage}{0.33\hsize}
        \centering
          \includegraphics[keepaspectratio, scale=0.44, angle=0]
                          {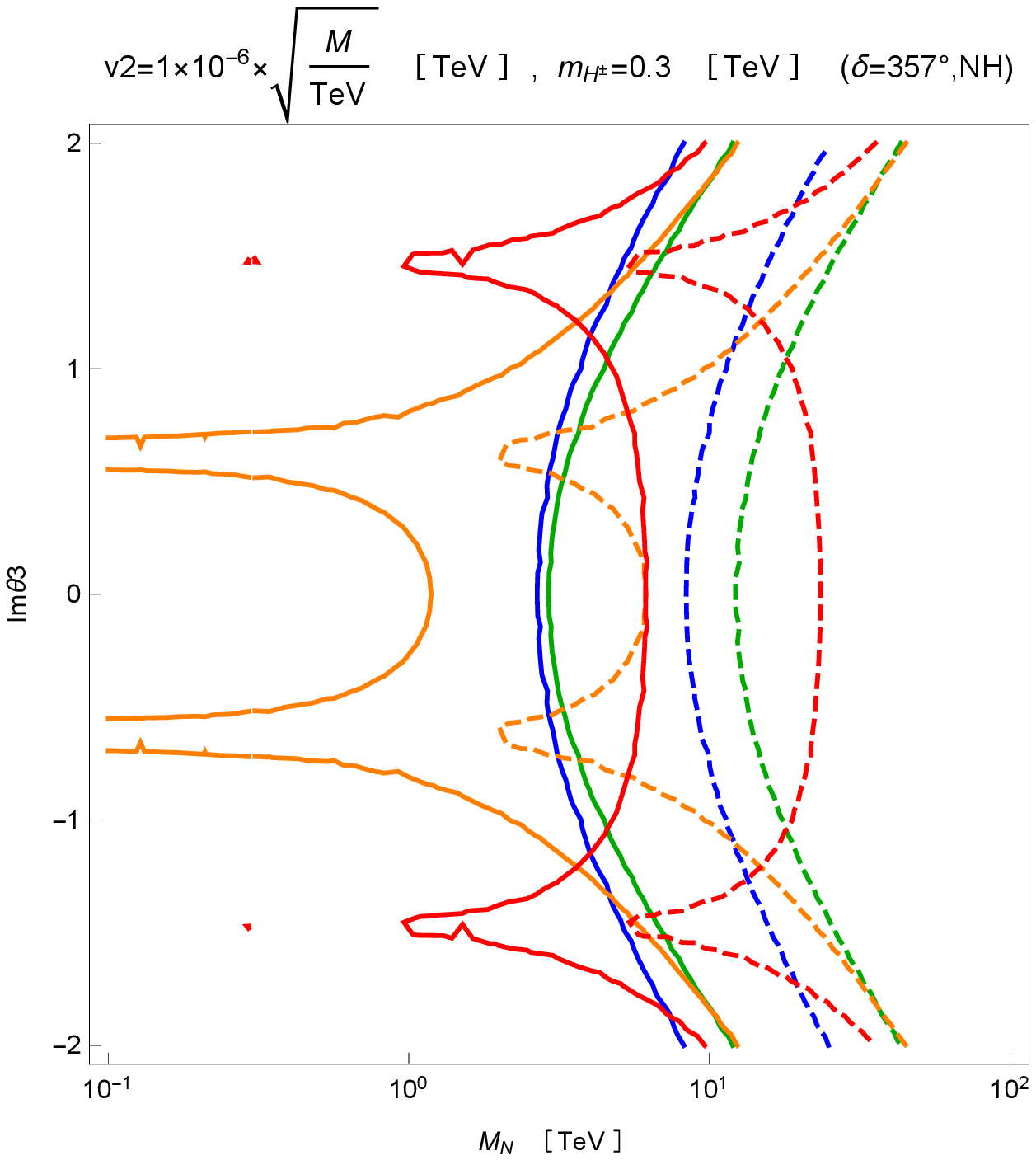}
      \end{minipage} 

    \end{tabular}
     \caption{\footnotesize Prediction for $Br(Z\to e\mu)$, $Br(Z\to e\tau)$ and $Br(Z\to \mu\tau)$, along with the values of $Br(\mu\to e\gamma)$. 
     The neutrino mass hierarchy is Normal Hierarchy, and we fix $m_{H^\pm}=0.3$ TeV. We take $\delta=144^\circ,~221^\circ~{\rm and}~357^\circ$ in the first, second and third rows. In the first column, we vary Im$\theta_1\neq0$ while fixing Im$\theta_2$=Im$\theta_3=0$. 
In the second column, we vary Im$\theta_2\neq0$ while fixing Im$\theta_1$=Im$\theta_3=0$. In the third column, we vary Im$\theta_3\neq0$ while fixing Im$\theta_1$=Im$\theta_2=0$.
The solid blue line corresponds to $Br(\mu\to e\gamma)=4.2\times10^{-13}$ for $v_2$ in Eq.~(\ref{v2-nh}), 
 and the region on the left of the solid blue line is excluded by the search for $Br(\mu\to e\gamma)$.
The solid green, orange and red lines correspond to the contours of $Br(Z\to e\mu)=10^{-16}$, $Br(Z\to e\tau)=10^{-16}$ and $Br(Z\to \mu\tau)=10^{-16}$, respectively, for $v_2$ in Eq.~(\ref{v2-nh}).
The dashed blue line corresponds to $Br(\mu\to e\gamma)=4.2\times10^{-13}$ and the dashed green, orange and red lines correspond to the contours of $Br(Z\to e\mu)=10^{-16}$, $Br(Z\to e\tau)=10^{-16}$ and $Br(Z\to \mu\tau)=10^{-16}$, respectively,
 when $v_2$ is multiplied by $1/3$.
     }
 \label{figprezN}
\end{figure}

\begin{figure}[H]
  \centering
  \thispagestyle{empty}
    \begin{tabular}{c}
 
 
      \begin{minipage}{0.33\hsize}
        \centering
          \includegraphics[keepaspectratio, scale=0.29, angle=0]
                          {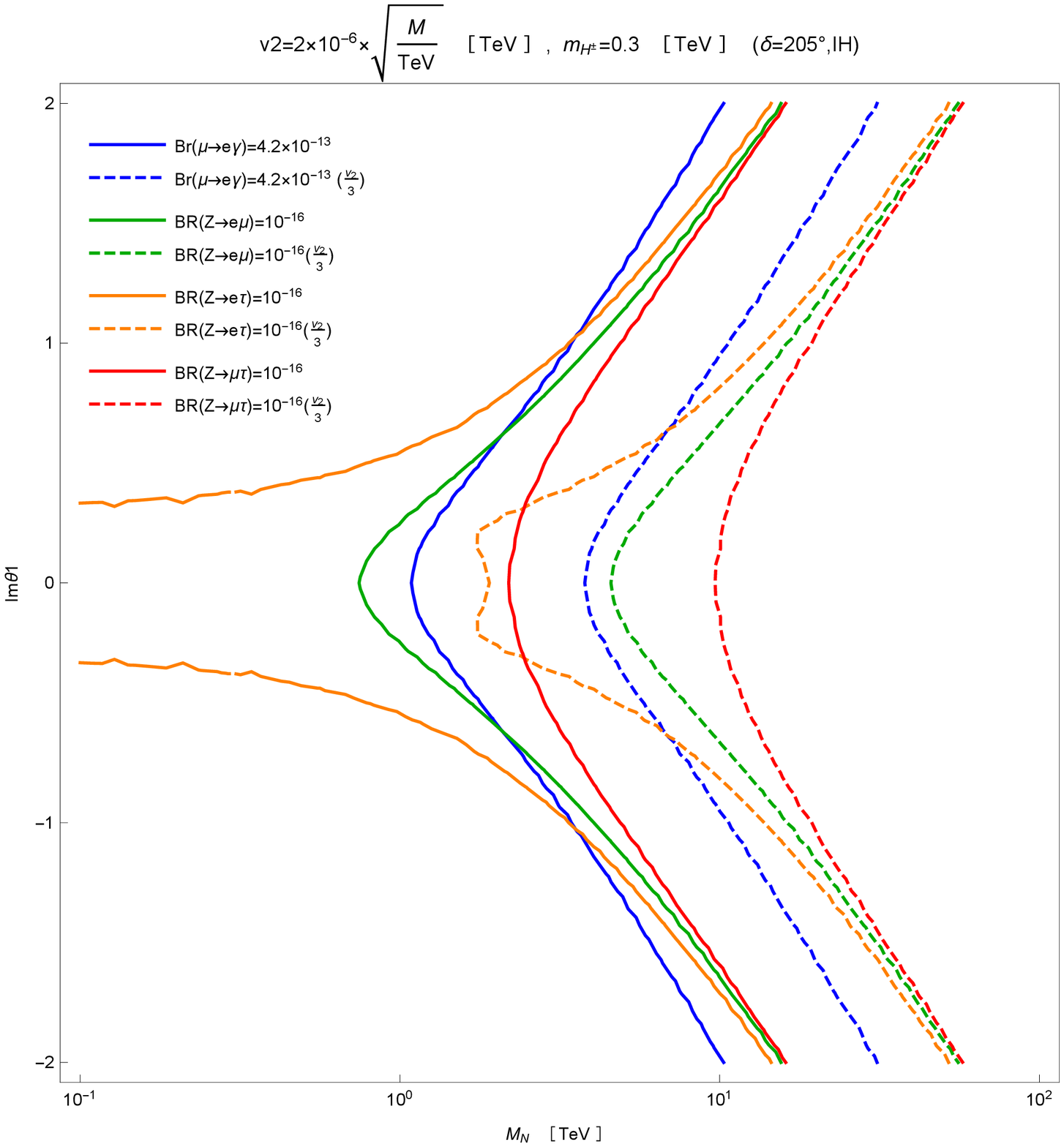}
      \end{minipage}

 
      \begin{minipage}{0.33\hsize}
        \centering
          \includegraphics[keepaspectratio, scale=0.44, angle=0]
                          {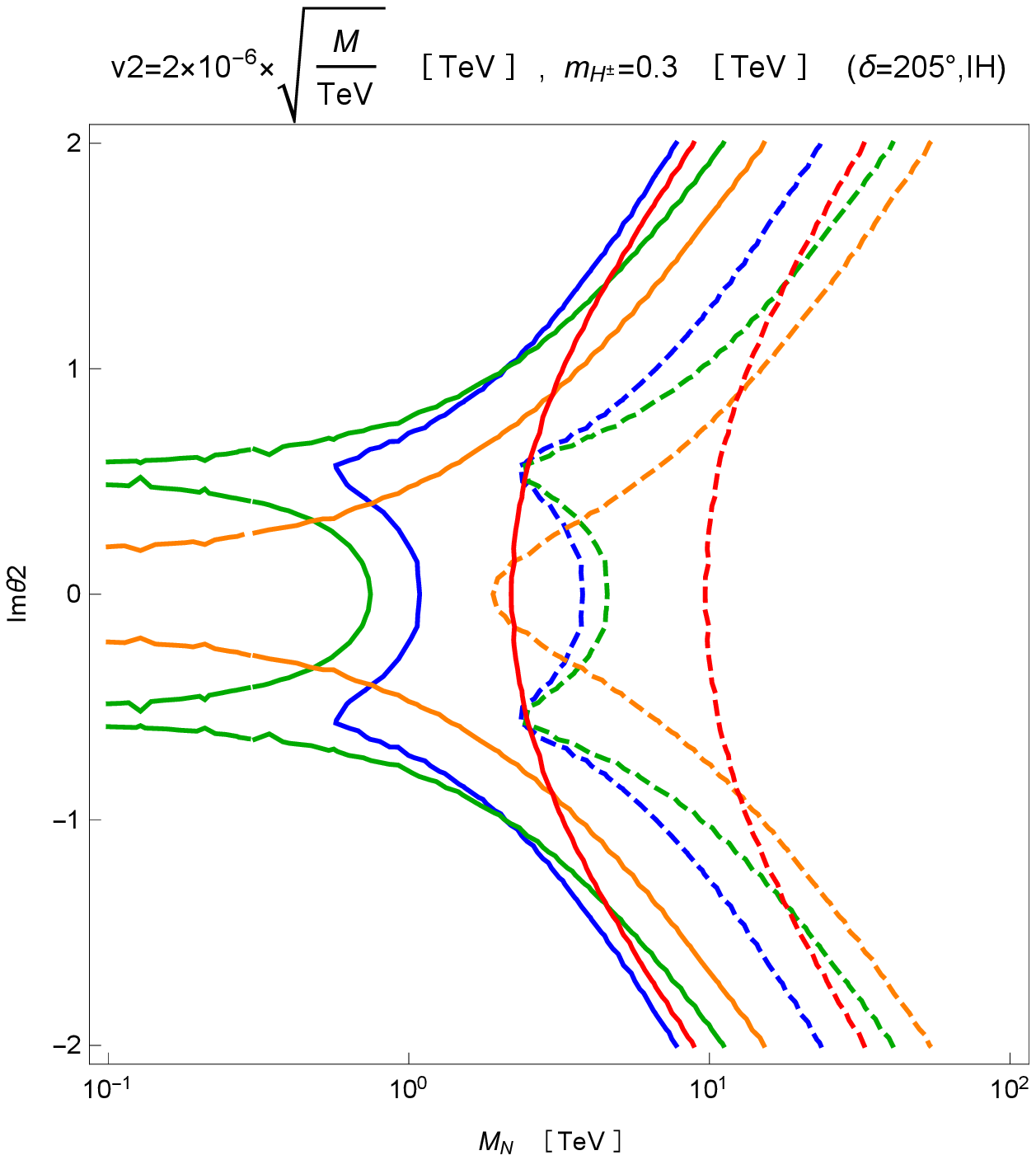}

      \end{minipage}
 
 
      \begin{minipage}{0.33\hsize}
        \centering
          \includegraphics[keepaspectratio, scale=0.44, angle=0]
                          {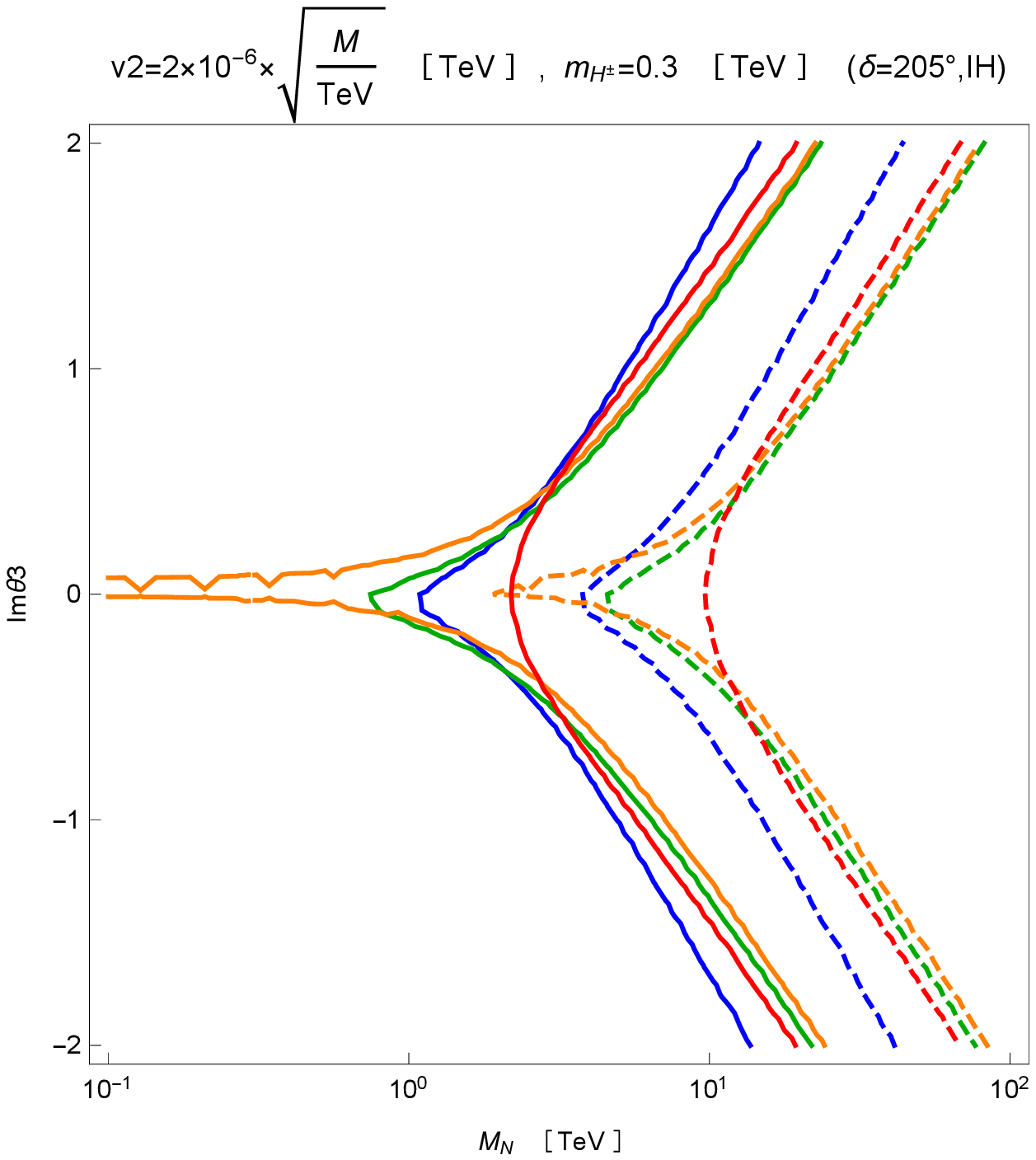}
      \end{minipage} \\
      \\
 
      \begin{minipage}{0.33\hsize}
        \centering
          \includegraphics[keepaspectratio, scale=0.44, angle=0]
                          {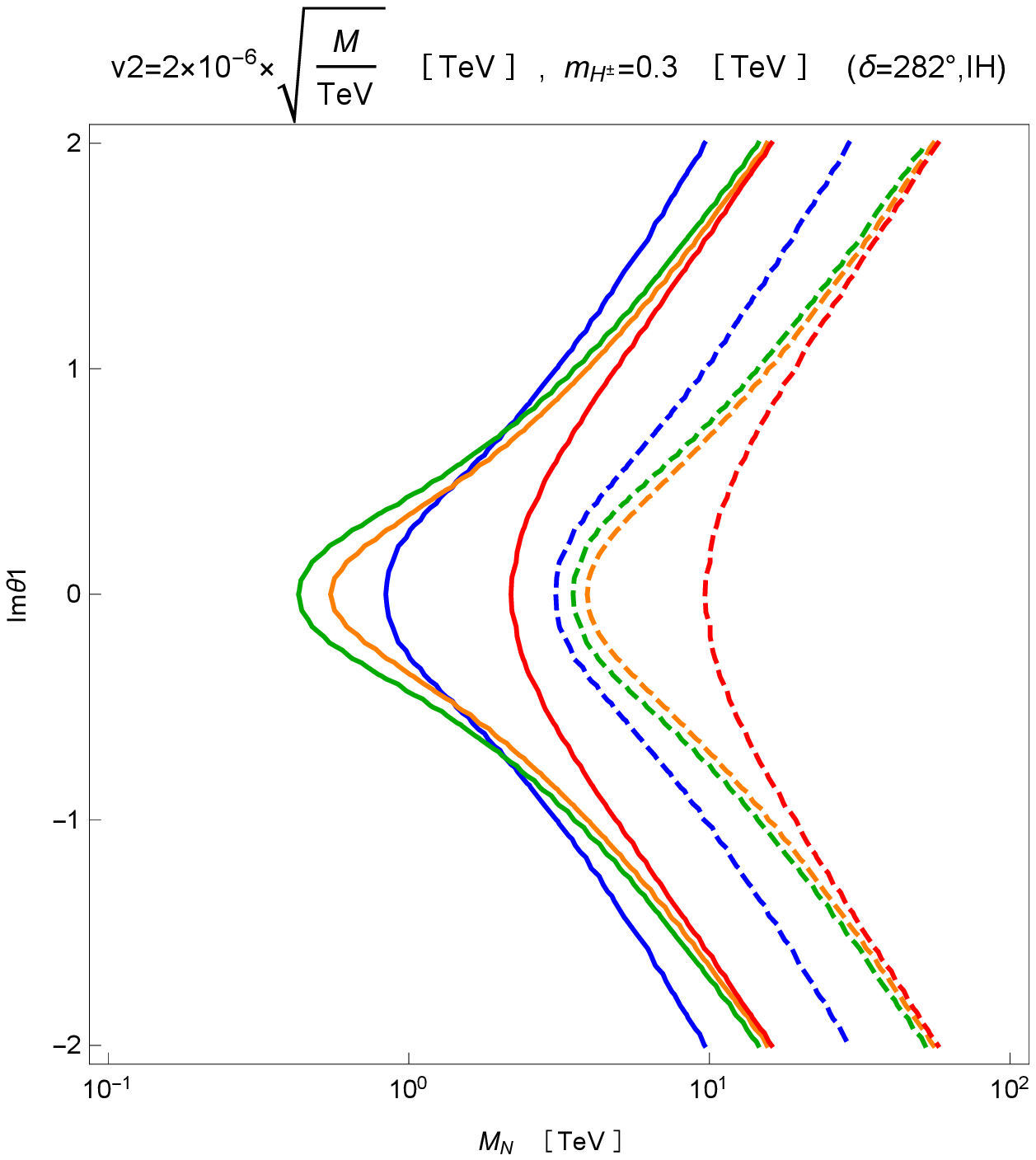}
      \end{minipage}

 
      \begin{minipage}{0.33\hsize}
        \centering
          \includegraphics[keepaspectratio, scale=0.44, angle=0]
                          {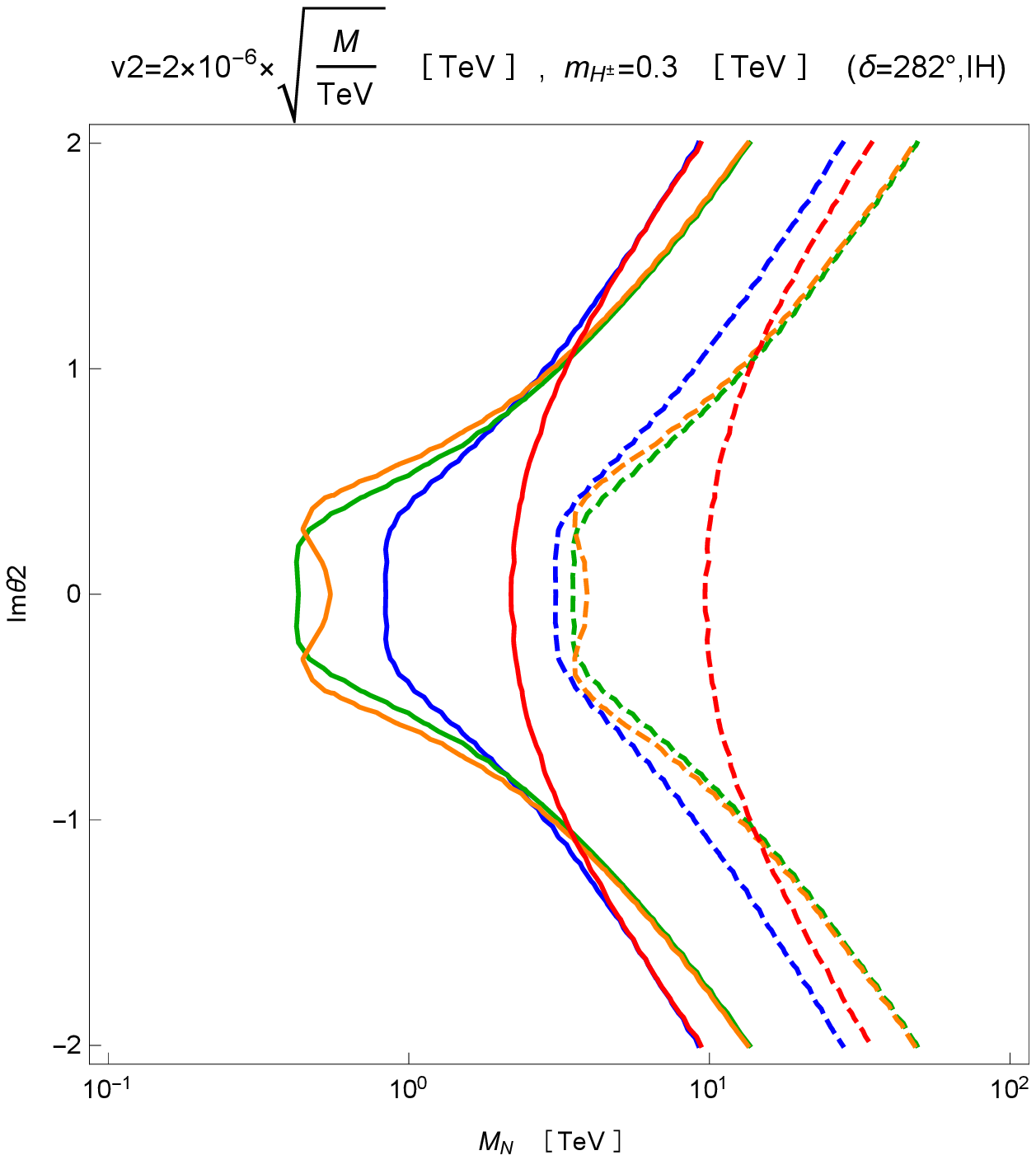}
      \end{minipage}
 
 
      \begin{minipage}{0.33\hsize}
        \centering
          \includegraphics[keepaspectratio, scale=0.44, angle=0]
                          {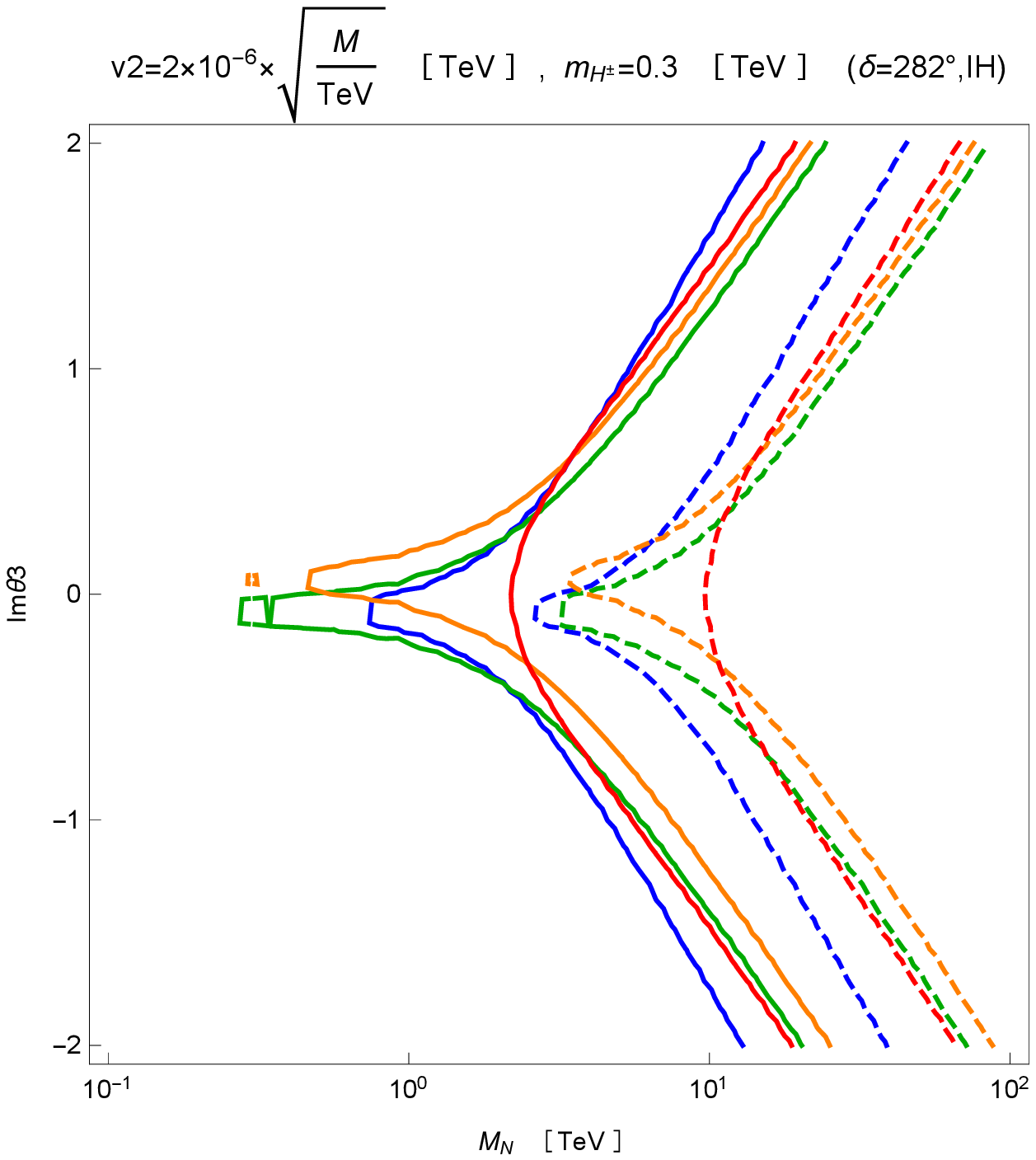}
      \end{minipage} \\ 
      \\
 
      \begin{minipage}{0.33\hsize}
        \centering
          \includegraphics[keepaspectratio, scale=0.44, angle=0]
                          {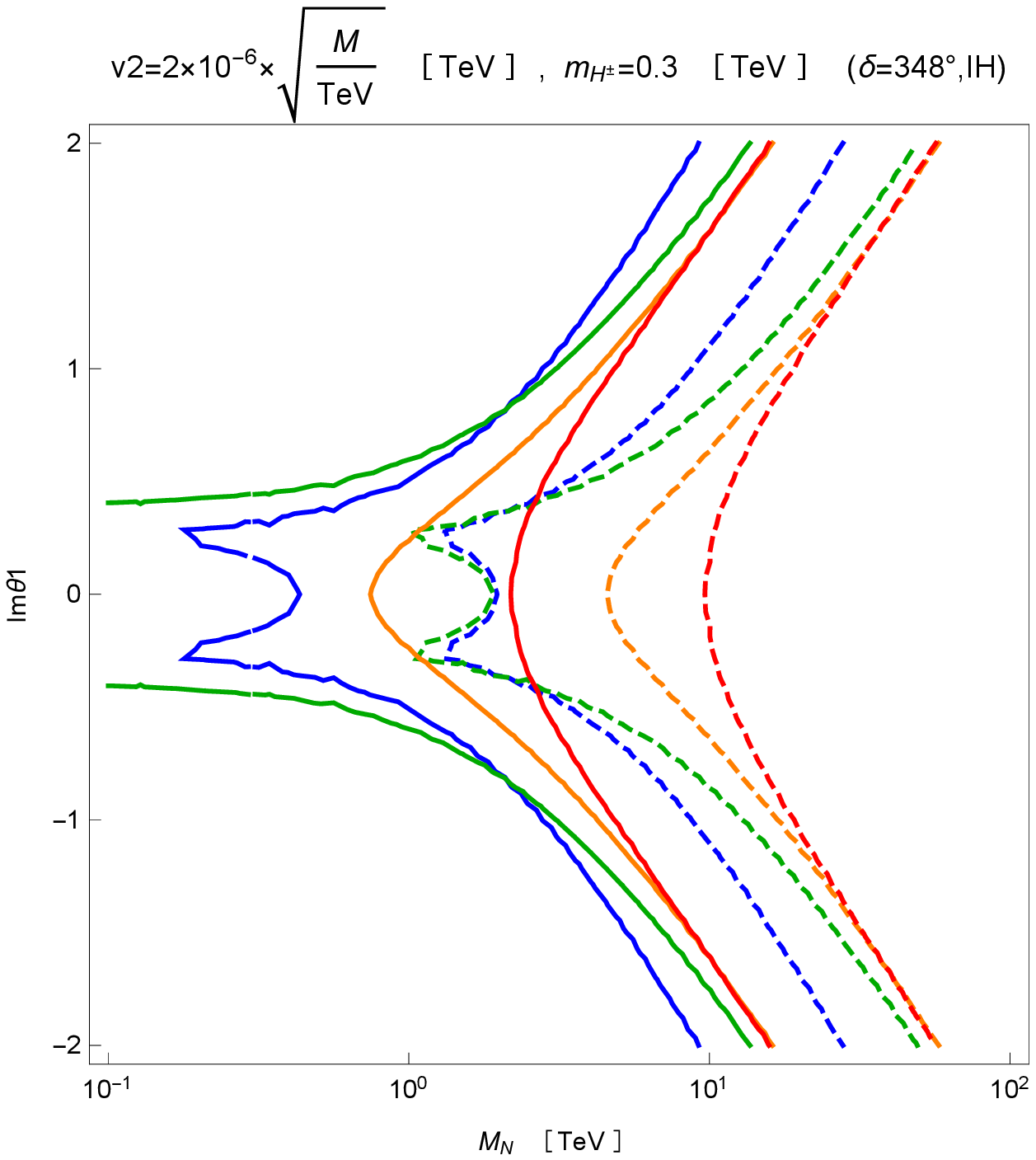}
      \end{minipage}

 
      \begin{minipage}{0.33\hsize}
        \centering
          \includegraphics[keepaspectratio, scale=0.44, angle=0]
                          {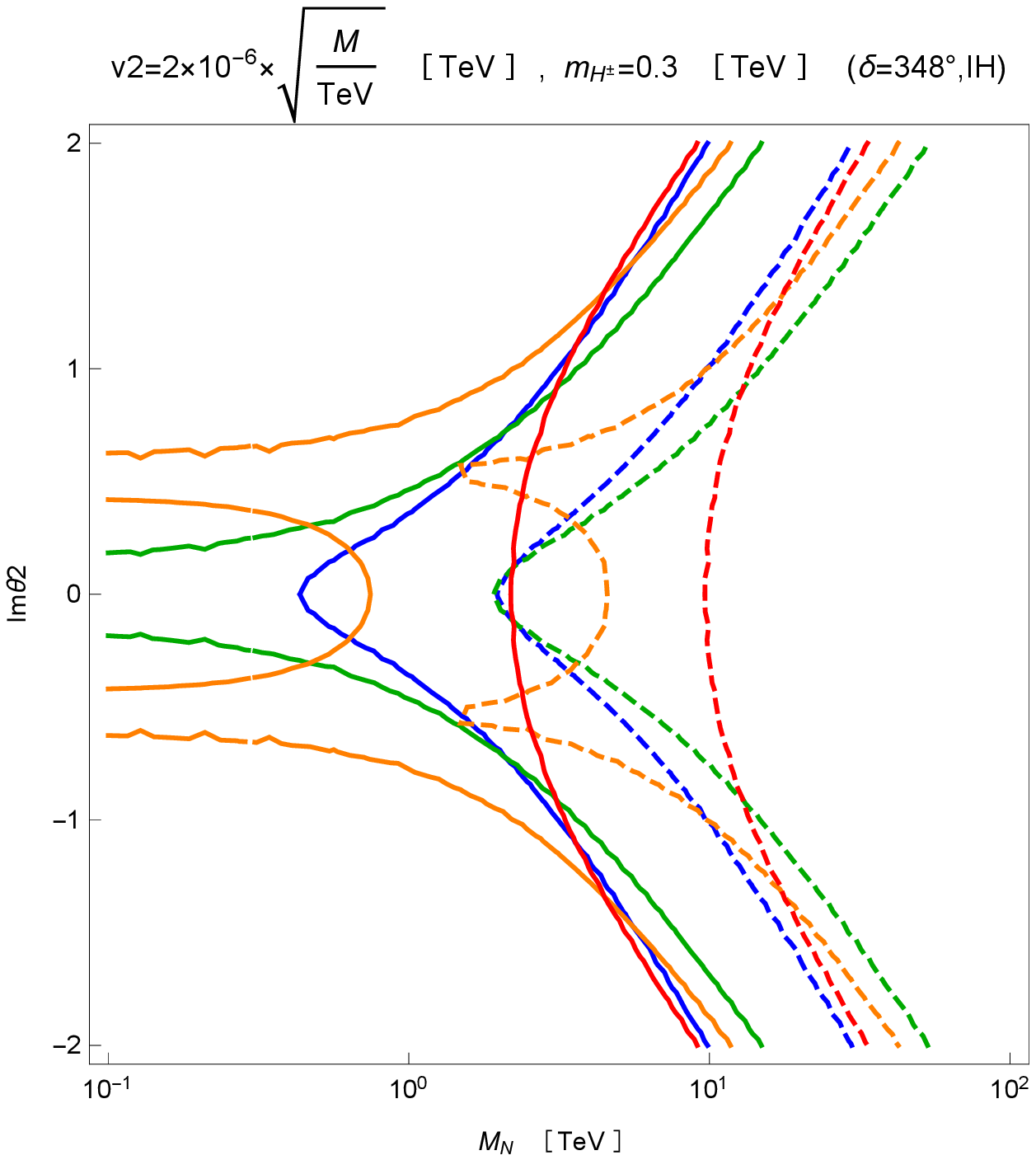}
      \end{minipage}
 
 
      \begin{minipage}{0.33\hsize}
        \centering
          \includegraphics[keepaspectratio, scale=0.44, angle=0]
                          {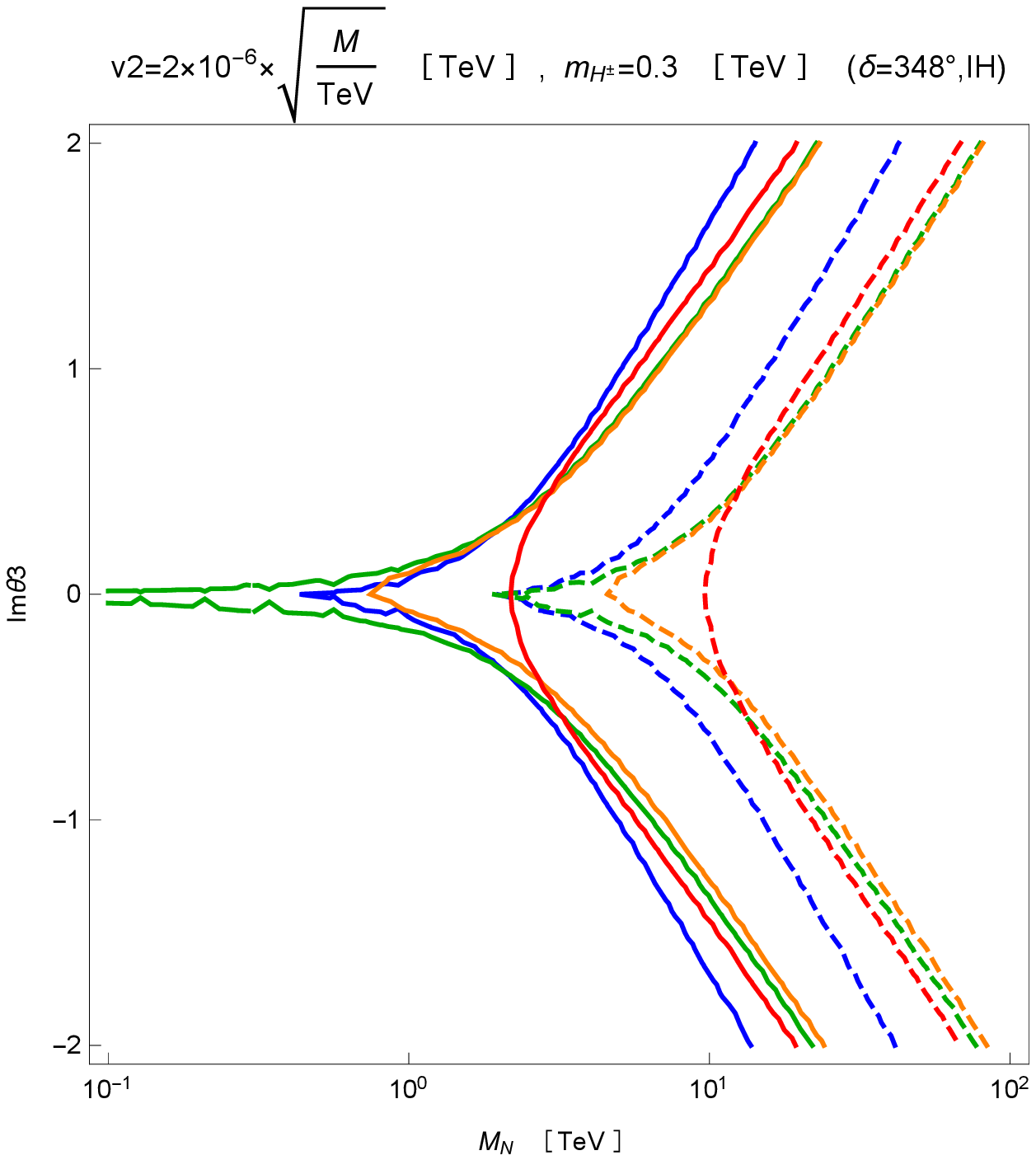}
      \end{minipage} 

    \end{tabular}
     \caption{\footnotesize Same as figure~\ref{figprezN} except that the neutrino mass hierarchy is Inverted Hierarchy and $v_2$ is given in Eq.~(\ref{v2-ih}).}
 \label{figprezI}
\end{figure}

There are three modes, $Z\to e\mu,\,Z\to e\tau$ and $Z\to \mu\tau$.
In Fig.~\ref{figprezN}, the solid green, orange and red lines correspond to the contours of $Br(Z\to e\mu)=10^{-16}$, $Br(Z\to e\tau)=10^{-16}$ and $Br(Z\to \mu\tau)=10^{-16}$, respectively, for NH and $v_2$ in Eq.~(\ref{v2-nh}).
Figure~\ref{figprezI} is the corresponding figure for IH　and $v_2$ in Eq.~(\ref{v2-ih}).

In the same figure, the dashed green, orange and red lines correspond to the contours of $Br(Z\to e\mu)=10^{-16}$, $Br(Z\to e\tau)=10^{-16}$ and $Br(Z\to \mu\tau)=10^{-16}$, respectively, when $v_2$ is multiplied by $1/3$.
Since the dipole and non-dipole operators $A_D$ and $A_{ND}$ are both proportional to $Y_D^2$, the branching ratios
 $Br(\mu\to e\gamma)$ and $Br(Z \to \bar{e}_\alpha e_\beta)$ both scale with $1/v_2^4$. 
Hence, the relative location of the contours of $Br(\mu\to e\gamma)$ and $Br(Z\to e\mu),Br(Z\to e\tau) $ and $ Br(Z\to \mu\tau)$ do not depend on $v_2$.

We observe that in all cases, all of the $Z\to e\mu,~Z\to e\tau$ and $Z\to \mu\tau$ decays can be detected at a rate of about $10^{-16}$ 
 even when the model satisfies the current experimental bound on $Br(\mu \to e\gamma)$.
Unfortunately, the rate $10^{-16}$ is much lower than the future sensitivity of a high-luminosity $Z$-factory proposed in Ref.~\cite{Abada:2014cca}.

 \newpage
\newpage

 \newpage
\newpage

\subsubsection{$h \to e \tau$ and $h \to \mu \tau$}

Among the $h \to \bar{e}_\alpha e_\beta$ ($\alpha\neq\beta$) decay modes,
  the diagrams of $h\to e\tau$ and $h\to \mu\tau$ involve the large $\tau$ Yukawa coupling and so 
  these modes have much larger branching ratios than $h\to e\mu$.
Therefore, we concentrate on the former two.
$Br(h \to e\tau)$ and $Br(h \to \mu\tau)$ involve one unknown coupling constant, that is, $\lambda_3$.
We present our prediction by assuming $\lambda_3=1$. Since the prediction scales with $\lambda_3^2$, it is straightforward
 to consider cases with other values of $\lambda_3$.

\begin{figure}[H]
  \centering
  \thispagestyle{empty}
    \begin{tabular}{c}
 
 
      \begin{minipage}{0.33\hsize}
        \centering
          \includegraphics[keepaspectratio, scale=0.39, angle=0]
                          {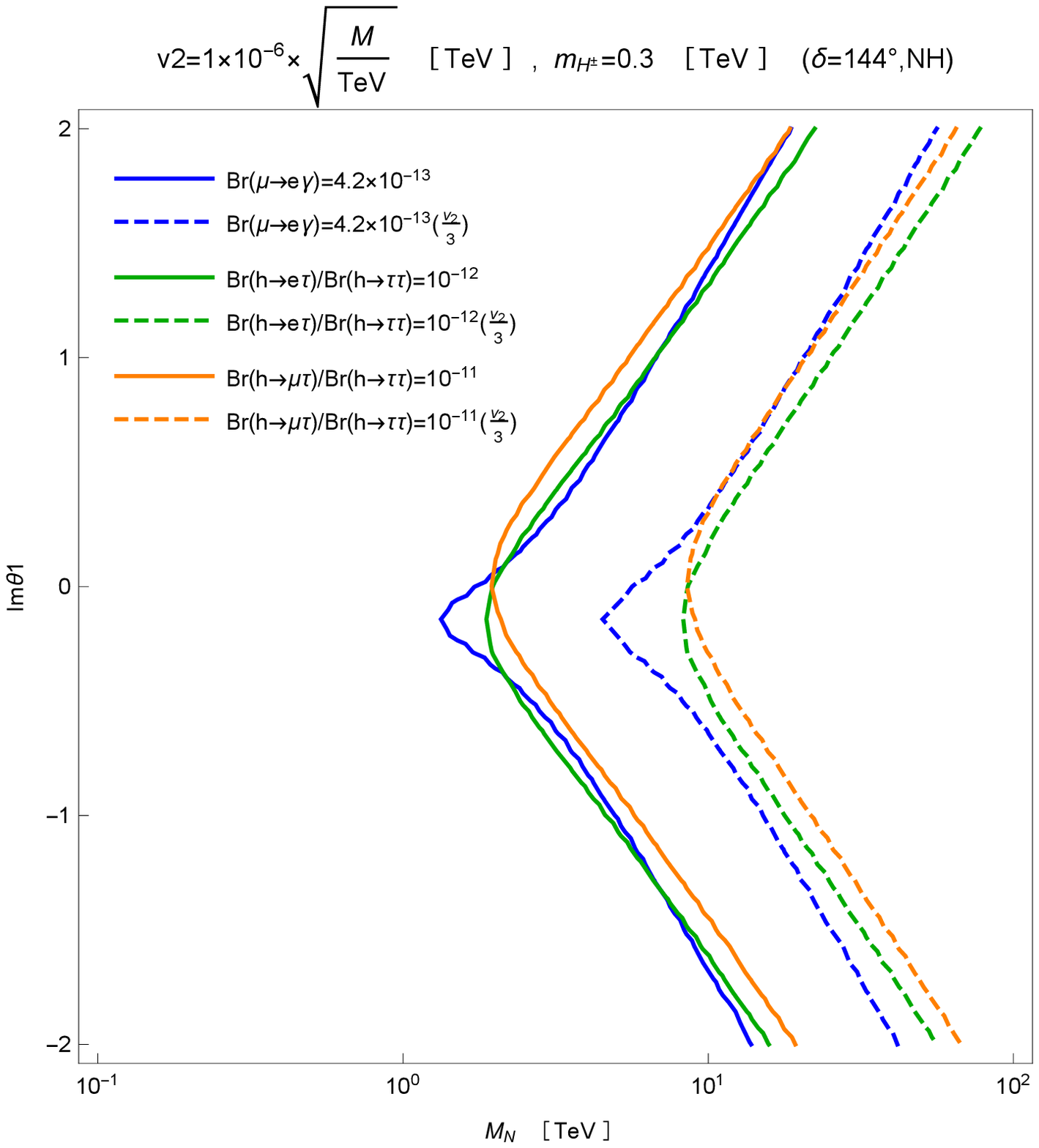}
      \end{minipage}

 
      \begin{minipage}{0.33\hsize}
        \centering
          \includegraphics[keepaspectratio, scale=0.44, angle=0]
                          {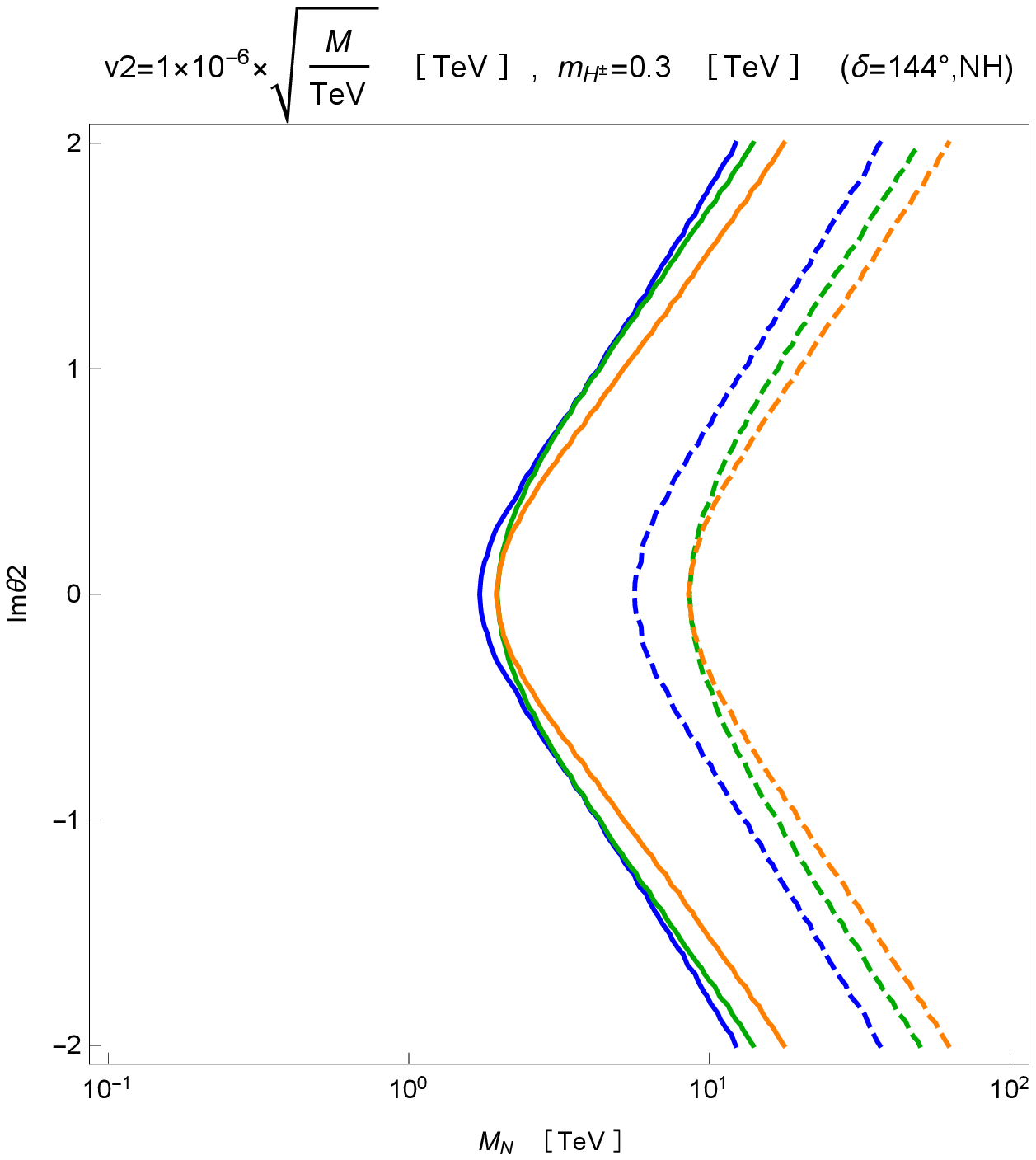}

      \end{minipage}
 
 
      \begin{minipage}{0.33\hsize}
        \centering
          \includegraphics[keepaspectratio, scale=0.44, angle=0]
                          {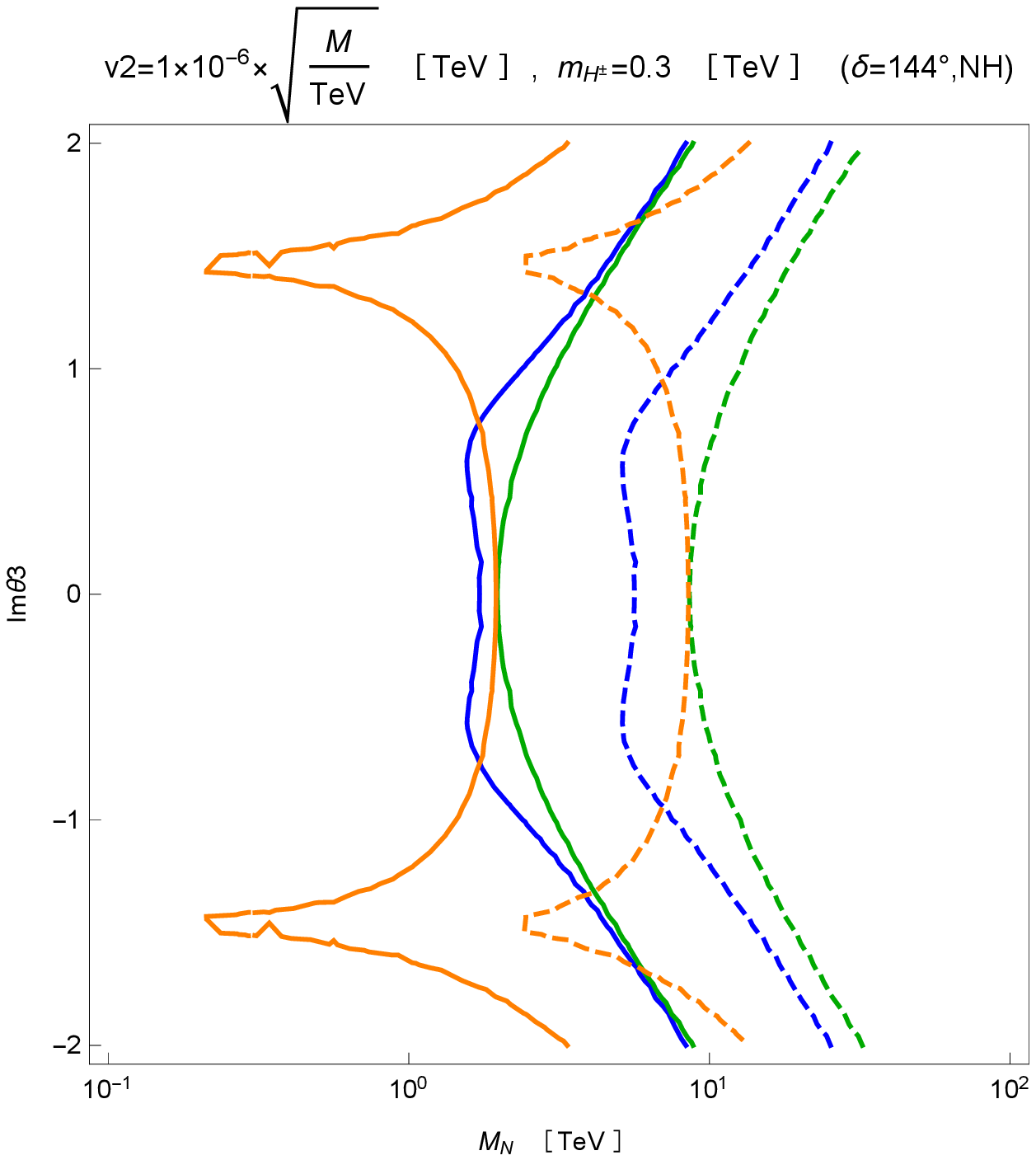}
      \end{minipage} \\
      \\
 
      \begin{minipage}{0.33\hsize}
        \centering
          \includegraphics[keepaspectratio, scale=0.44, angle=0]
                          {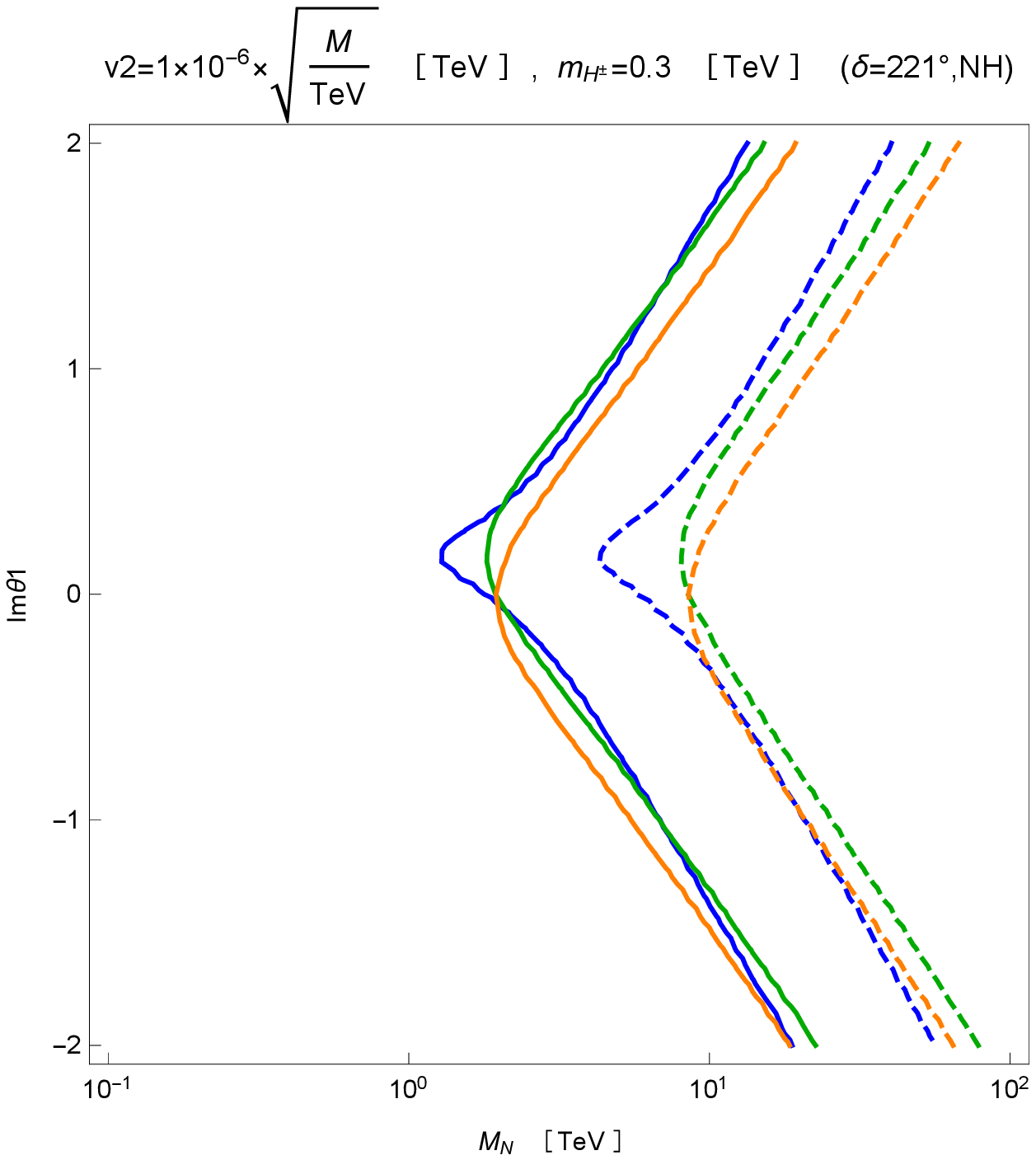}
      \end{minipage}

 
      \begin{minipage}{0.33\hsize}
        \centering
          \includegraphics[keepaspectratio, scale=0.44, angle=0]
                          {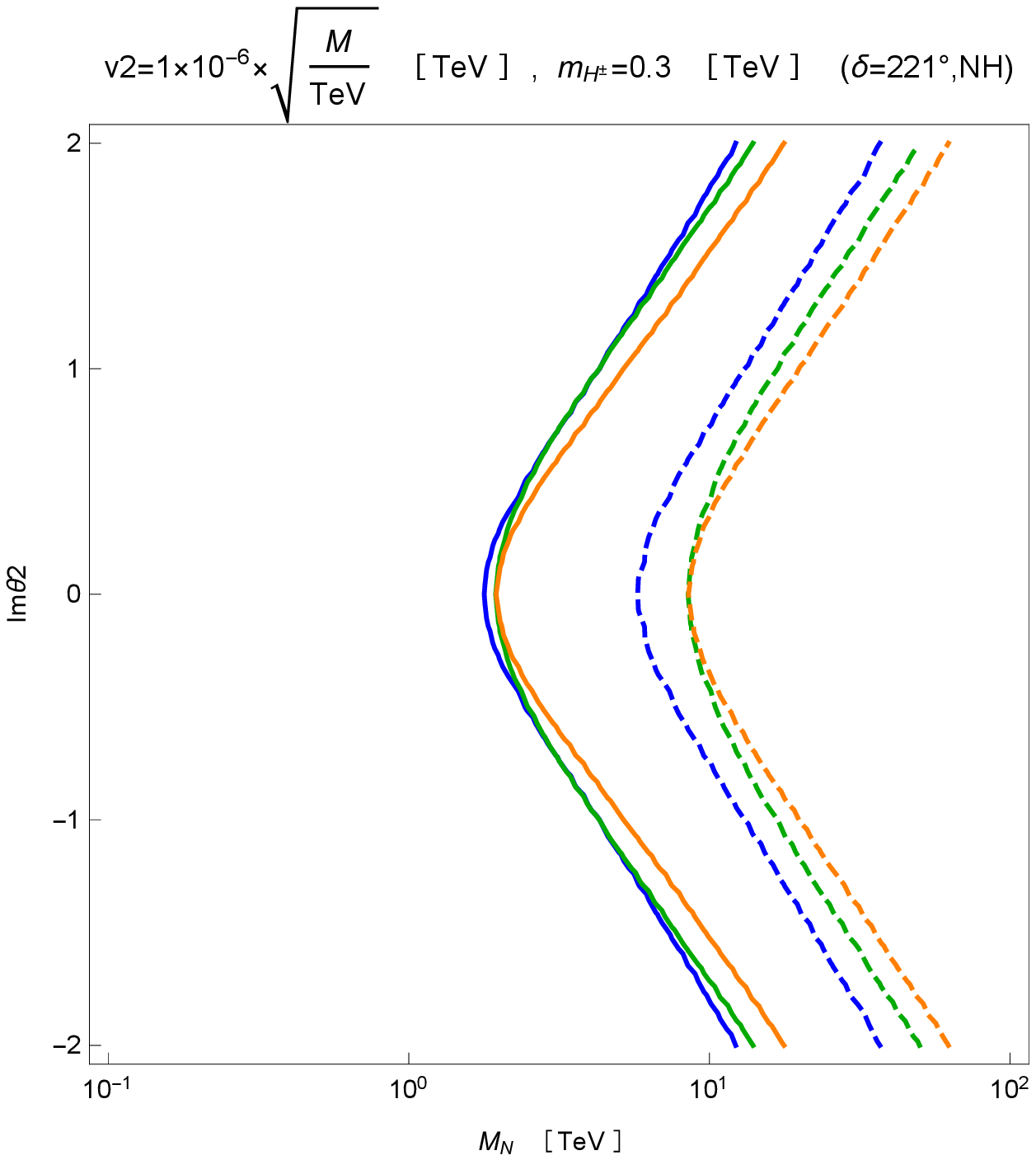}
      \end{minipage}
 
 
      \begin{minipage}{0.33\hsize}
        \centering
          \includegraphics[keepaspectratio, scale=0.44, angle=0]
                          {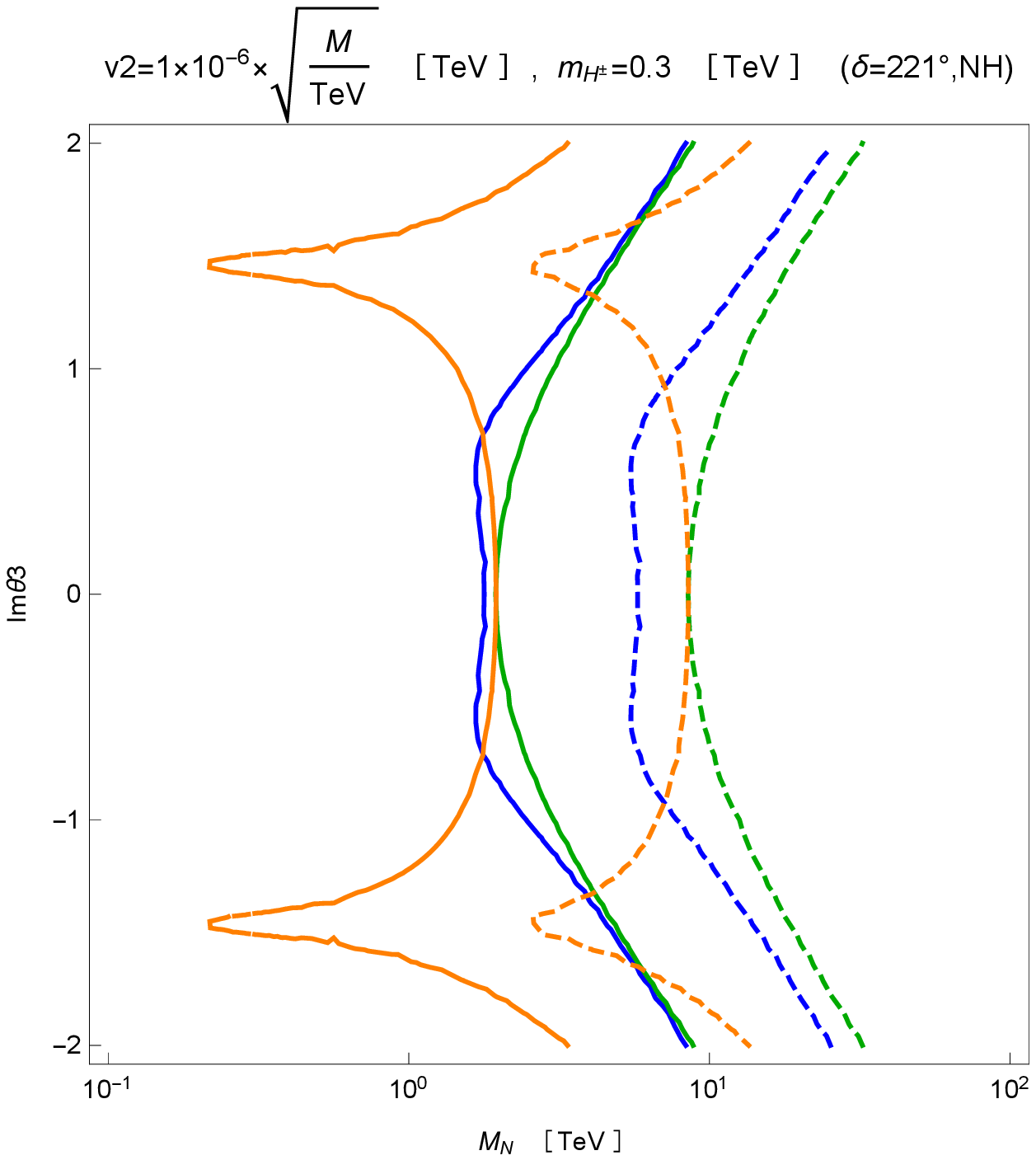}
      \end{minipage} \\ 
      \\
 
      \begin{minipage}{0.33\hsize}
        \centering
          \includegraphics[keepaspectratio, scale=0.44, angle=0]
                          {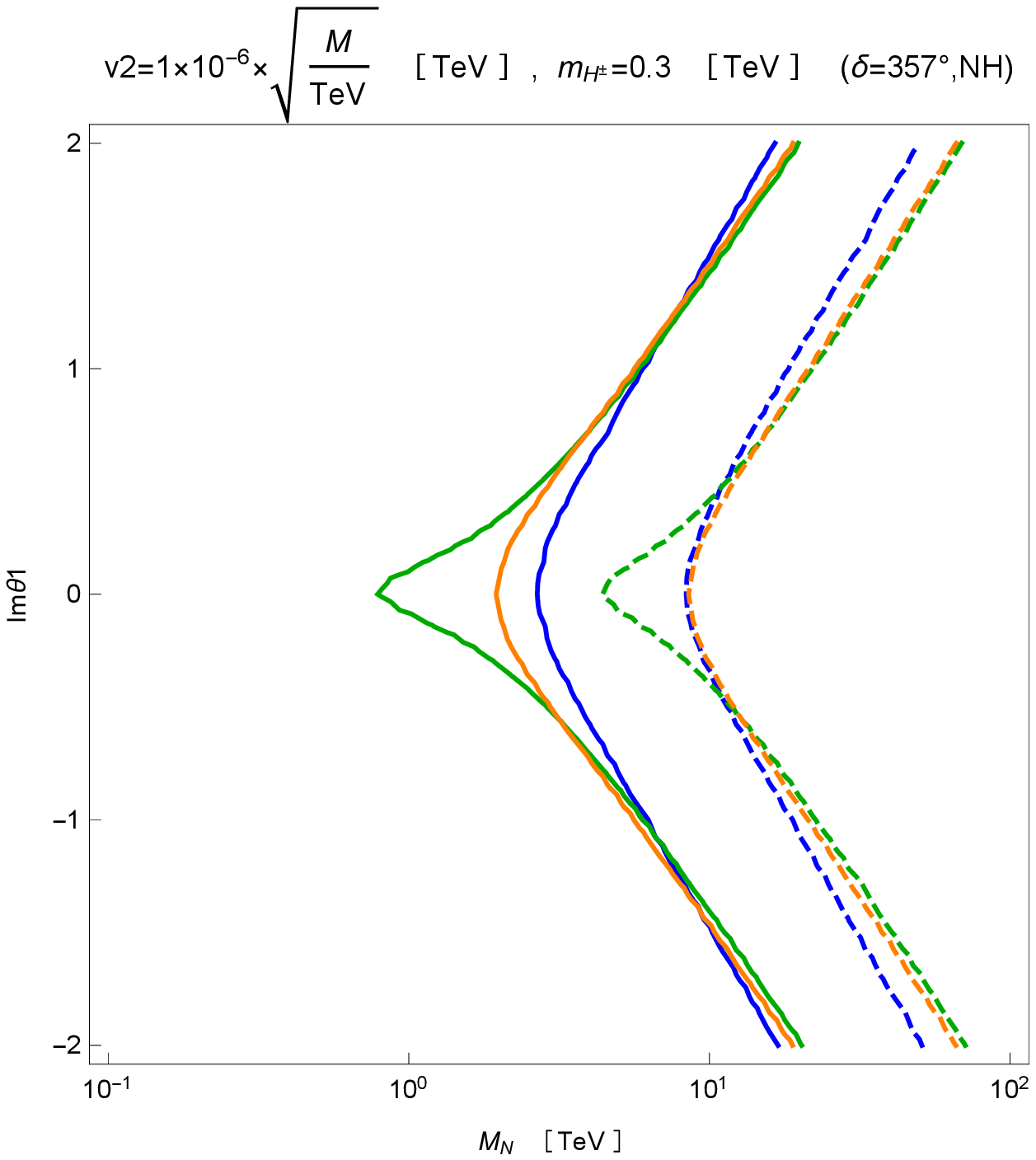}
      \end{minipage}

 
      \begin{minipage}{0.33\hsize}
        \centering
          \includegraphics[keepaspectratio, scale=0.44, angle=0]
                          {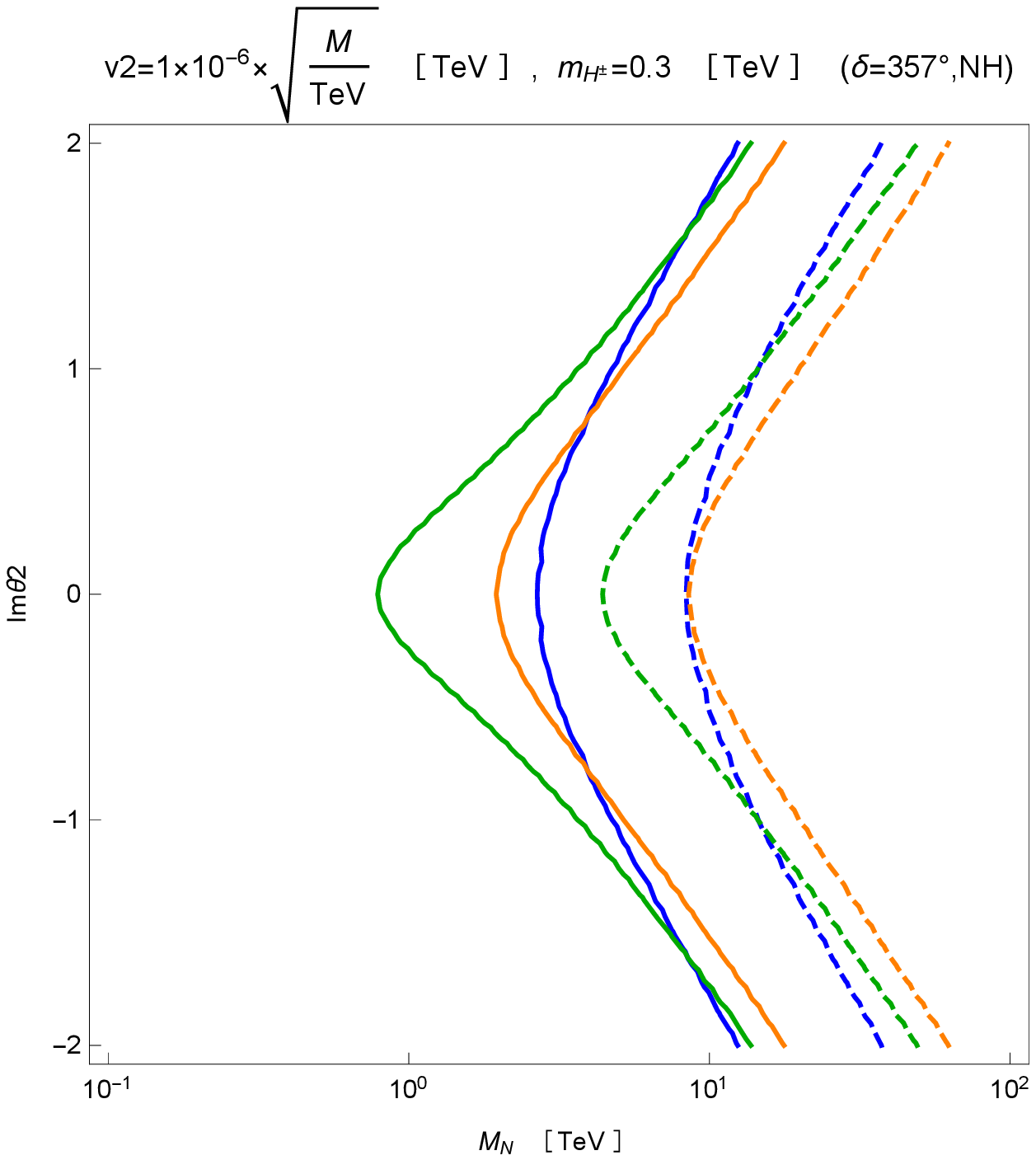}
      \end{minipage}
 
 
      \begin{minipage}{0.33\hsize}
        \centering
          \includegraphics[keepaspectratio, scale=0.44, angle=0]
                          {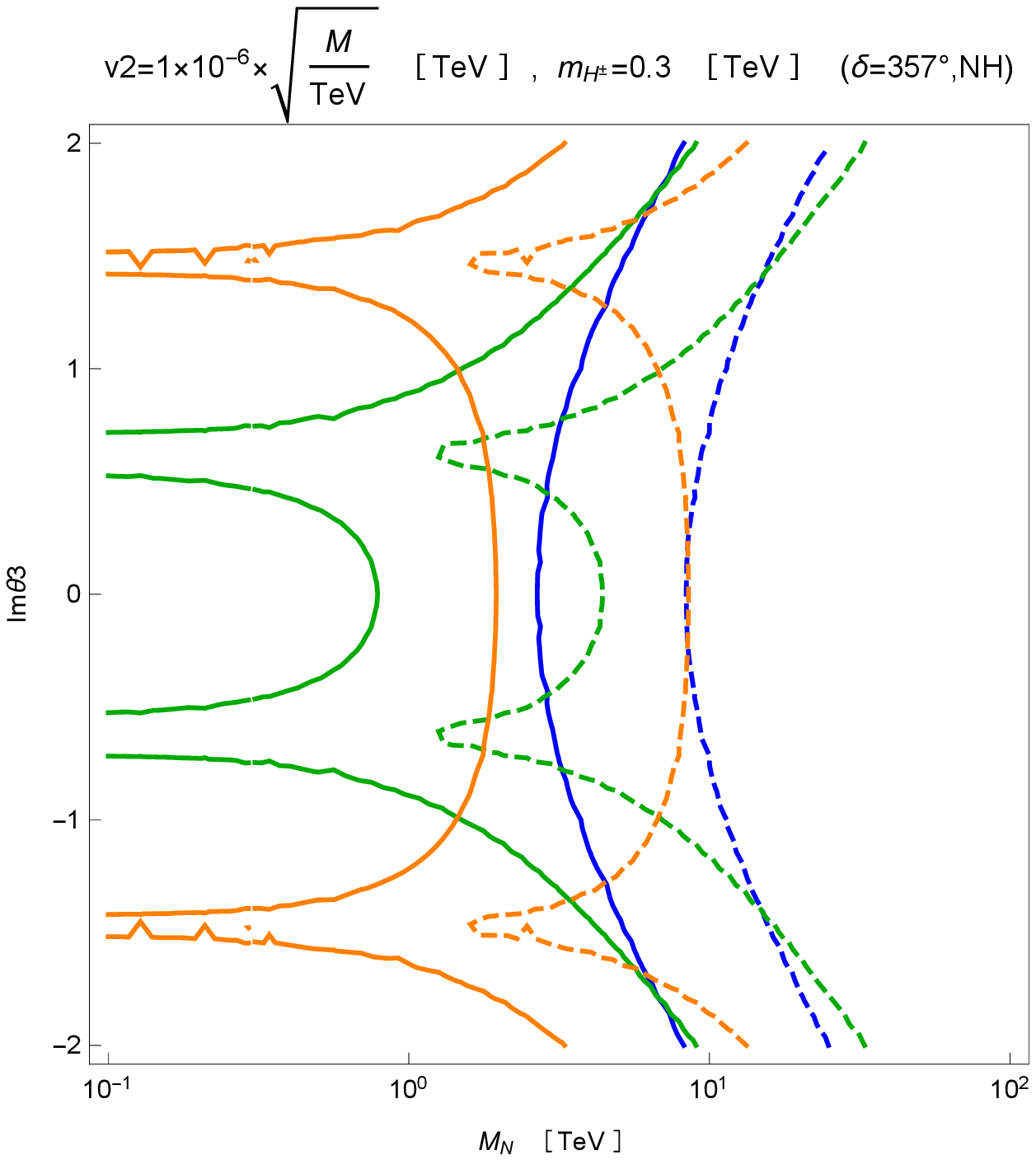}
      \end{minipage} 

    \end{tabular}
     \caption{\footnotesize 
     Prediction for $Br(h\to e\tau)$ and $Br(h\to \mu\tau)$, along with the values of $Br(\mu\to e\gamma)$. 
     The neutrino mass hierarchy is Normal Hierarchy, and we fix $m_{H^\pm}=0.3$ TeV. We take $\delta=144^\circ,~221^\circ~{\rm and}~357^\circ$ in the first, second and third rows. In the first column, we vary Im$\theta_1\neq0$ while fixing Im$\theta_2$=Im$\theta_3=0$. 
In the second column, we vary Im$\theta_2\neq0$ while fixing Im$\theta_1$=Im$\theta_3=0$. In the third column, we vary Im$\theta_3\neq0$ while fixing Im$\theta_1$=Im$\theta_2=0$.
The solid blue line corresponds to $Br(\mu\to e\gamma)=4.2\times10^{-13}$ for $v_2$ in Eq.~(\ref{v2-nh}), 
 and the region on the left of the solid blue line is excluded by the search for $Br(\mu\to e\gamma)$.
The solid green and orange lines correspond to the contours of $Br(h\to e\tau)/Br(h\to \tau\tau)=10^{-12}$ and $Br(h\to \mu\tau)/Br(h\to \tau\tau)=10^{-11}$, respectively, for $v_2$ in Eq.~(\ref{v2-nh}).
The dashed blue line corresponds to $Br(\mu\to e\gamma)=4.2\times10^{-13}$ and the dashed green and orange lines correspond to the contours of
 $Br(h\to e\tau)/Br(h\to \tau\tau)=10^{-12}$ and $Br(h\to \mu\tau)/Br(h\to \tau\tau)=10^{-11}$, respectively, when $v_2$ is multiplied by $1/3$.}
 \label{figprehN}
\end{figure}

In Fig.~\ref{figprehN}, the solid green and orange lines correspond to the contours of $Br(h\to e\tau)/Br(h\to \tau\tau)=10^{-12}$ and 
 $Br(h\to \mu\tau)/Br(h\to \tau\tau)=10^{-11}$, respectively, for NH and $v_2$ in Eq.~(\ref{v2-nh}).
In the same figure, the dashed green and orange lines correspond to the contours of $Br(h\to e\tau)/Br(h\to \tau\tau)=10^{-12}$ and 
 $Br(h\to \mu\tau)/Br(h\to \tau\tau)=10^{-11}$, respectively, when $v_2$ is multiplied by $1/3$.
$Br(\mu\to e\gamma)$ and $Br(Z \to \bar{e}_\alpha e_\beta)$ $(\alpha\neq\beta)$
 both scale with $1/v_2^4$, and so the relative location of their contours does not depend on $v_2$.

\begin{figure}[H]
  \centering
    \begin{tabular}{c}
 
 
      \begin{minipage}{0.33\hsize}
        \centering
          \includegraphics[keepaspectratio, scale=0.35, angle=0]
                          {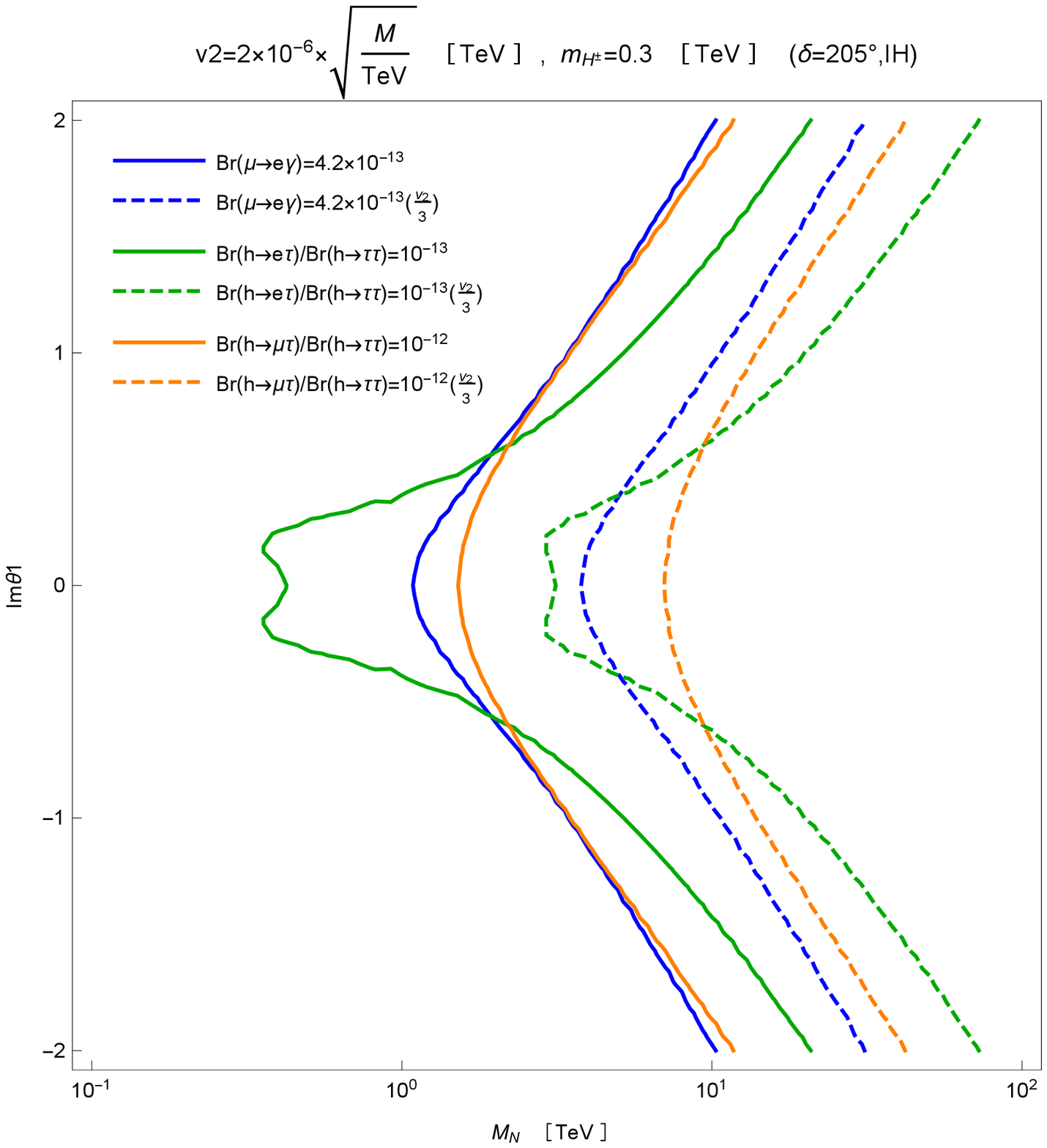}
      \end{minipage}

 
      \begin{minipage}{0.33\hsize}
        \centering
          \includegraphics[keepaspectratio, scale=0.44, angle=0]
                          {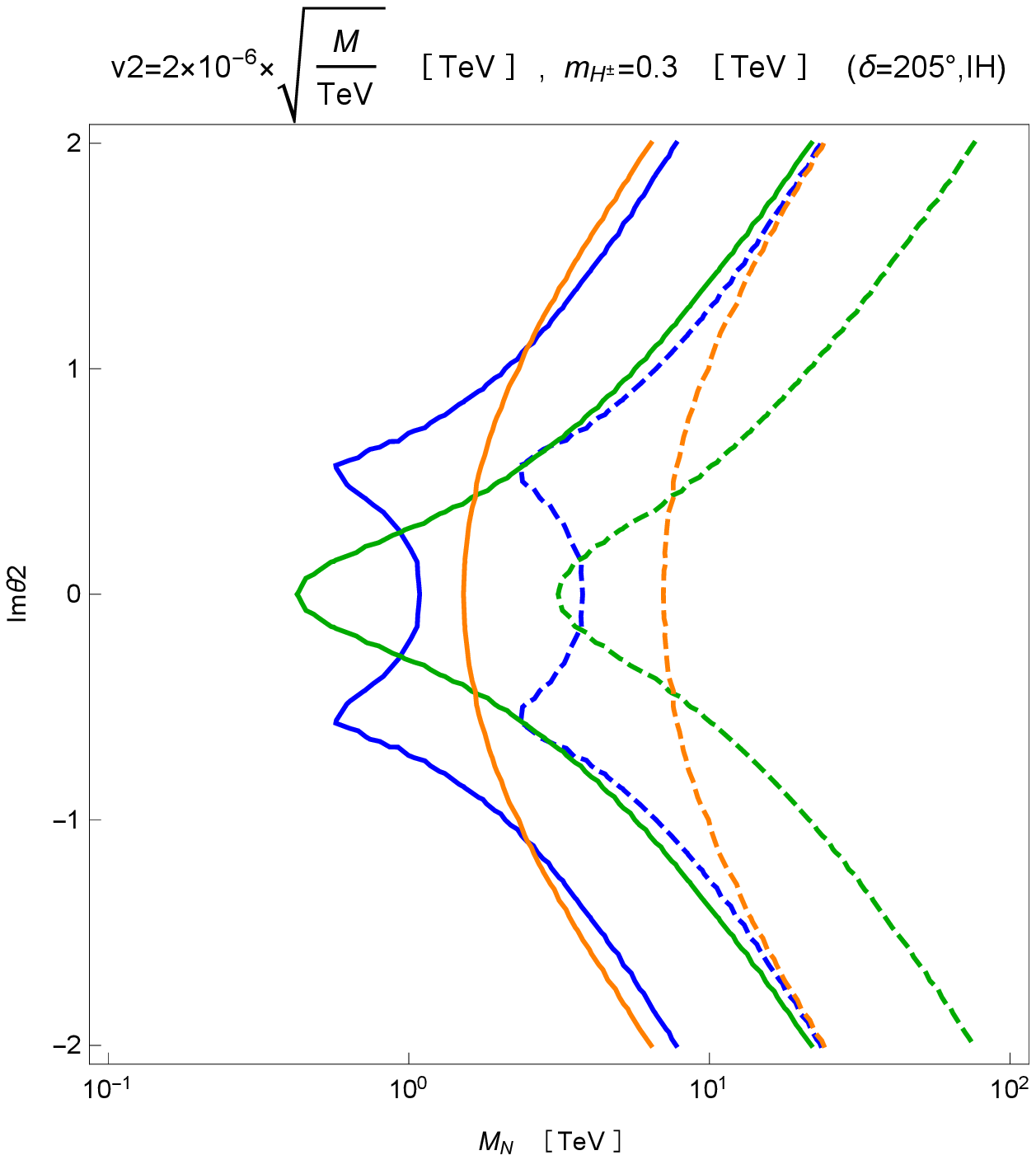}

      \end{minipage}
 
 
      \begin{minipage}{0.33\hsize}
        \centering
          \includegraphics[keepaspectratio, scale=0.44, angle=0]
                          {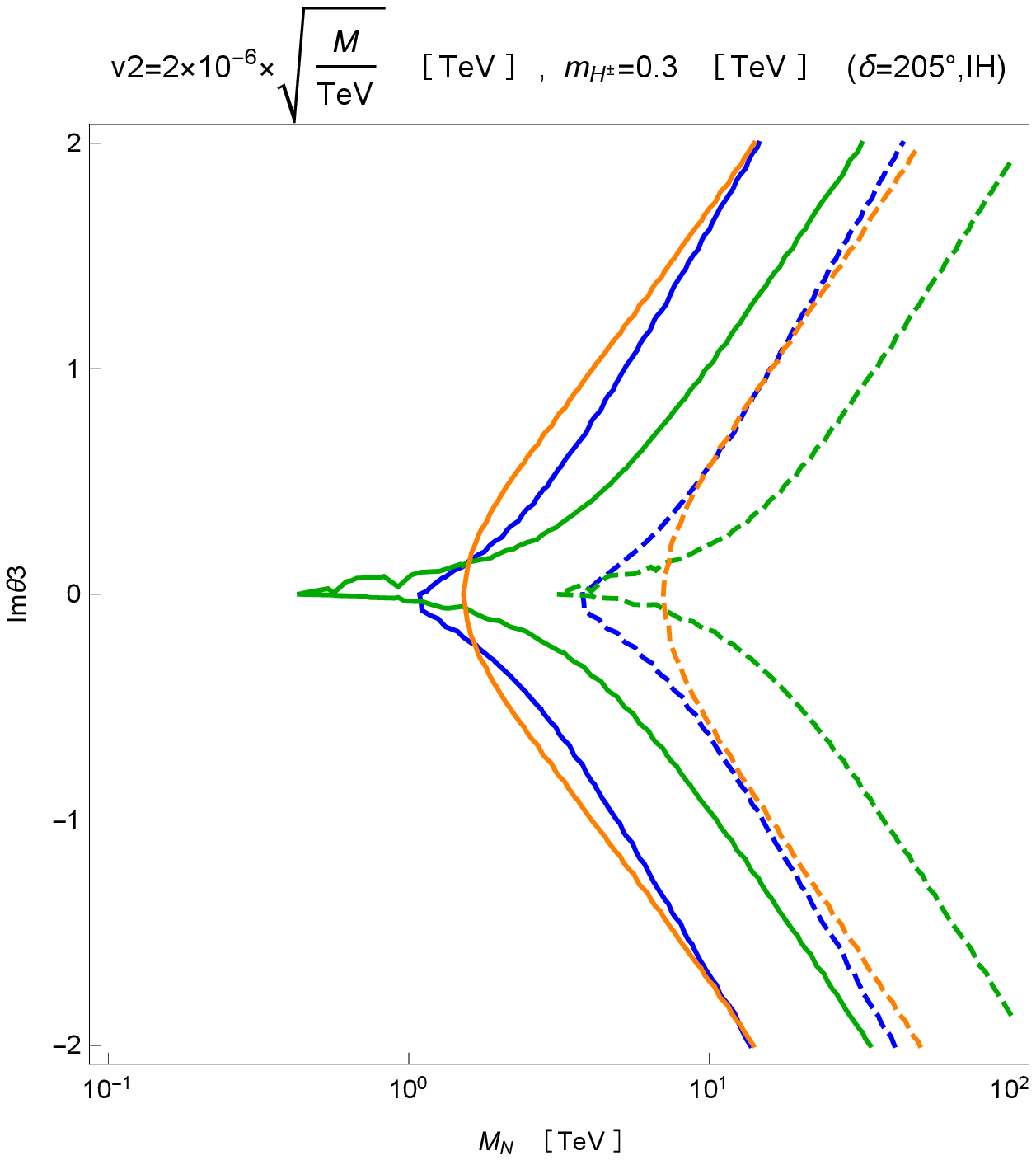}
      \end{minipage} \\
      \\
 
      \begin{minipage}{0.33\hsize}
        \centering
          \includegraphics[keepaspectratio, scale=0.44, angle=0]
                          {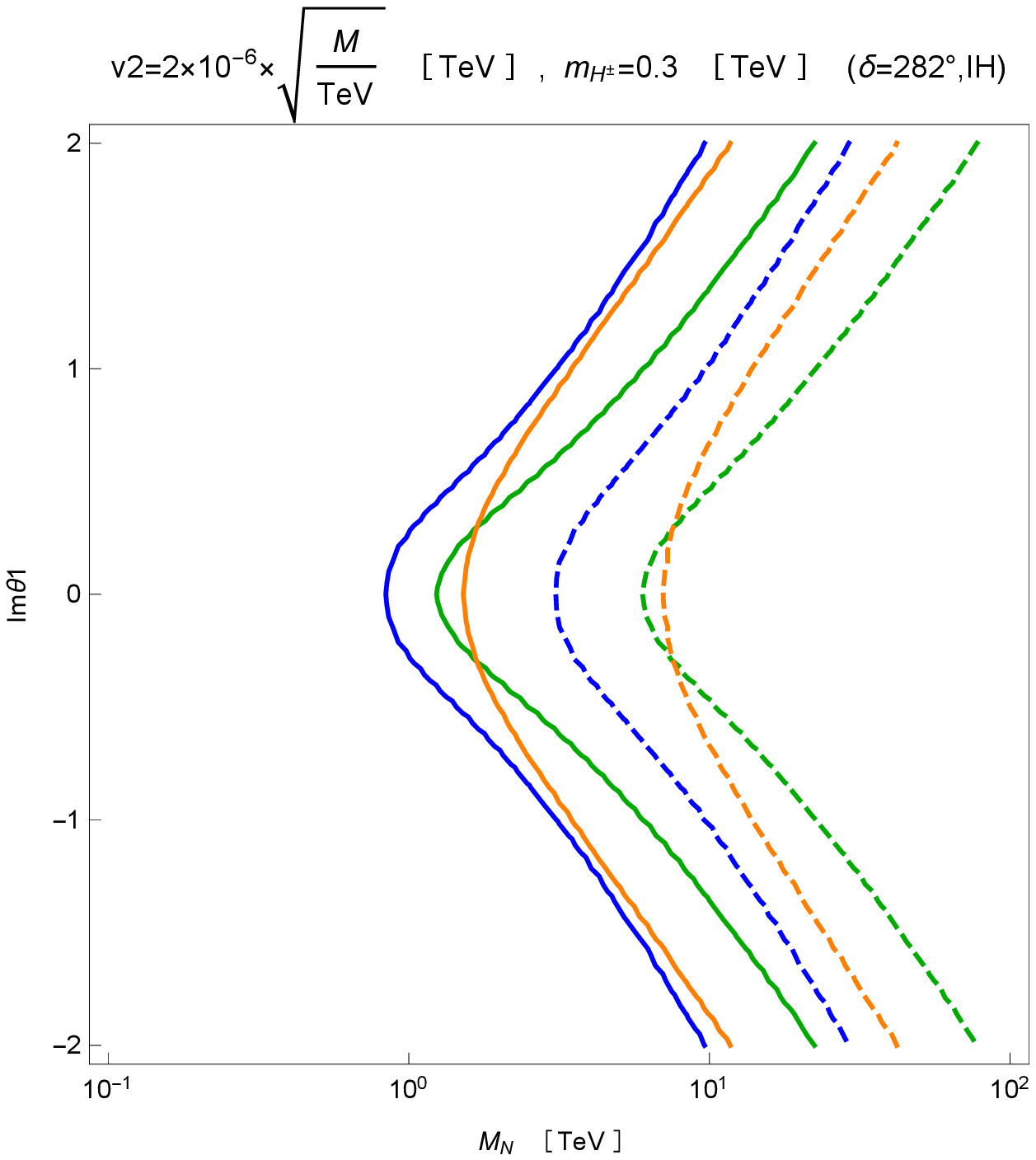}
      \end{minipage}

 
      \begin{minipage}{0.33\hsize}
        \centering
          \includegraphics[keepaspectratio, scale=0.44, angle=0]
                          {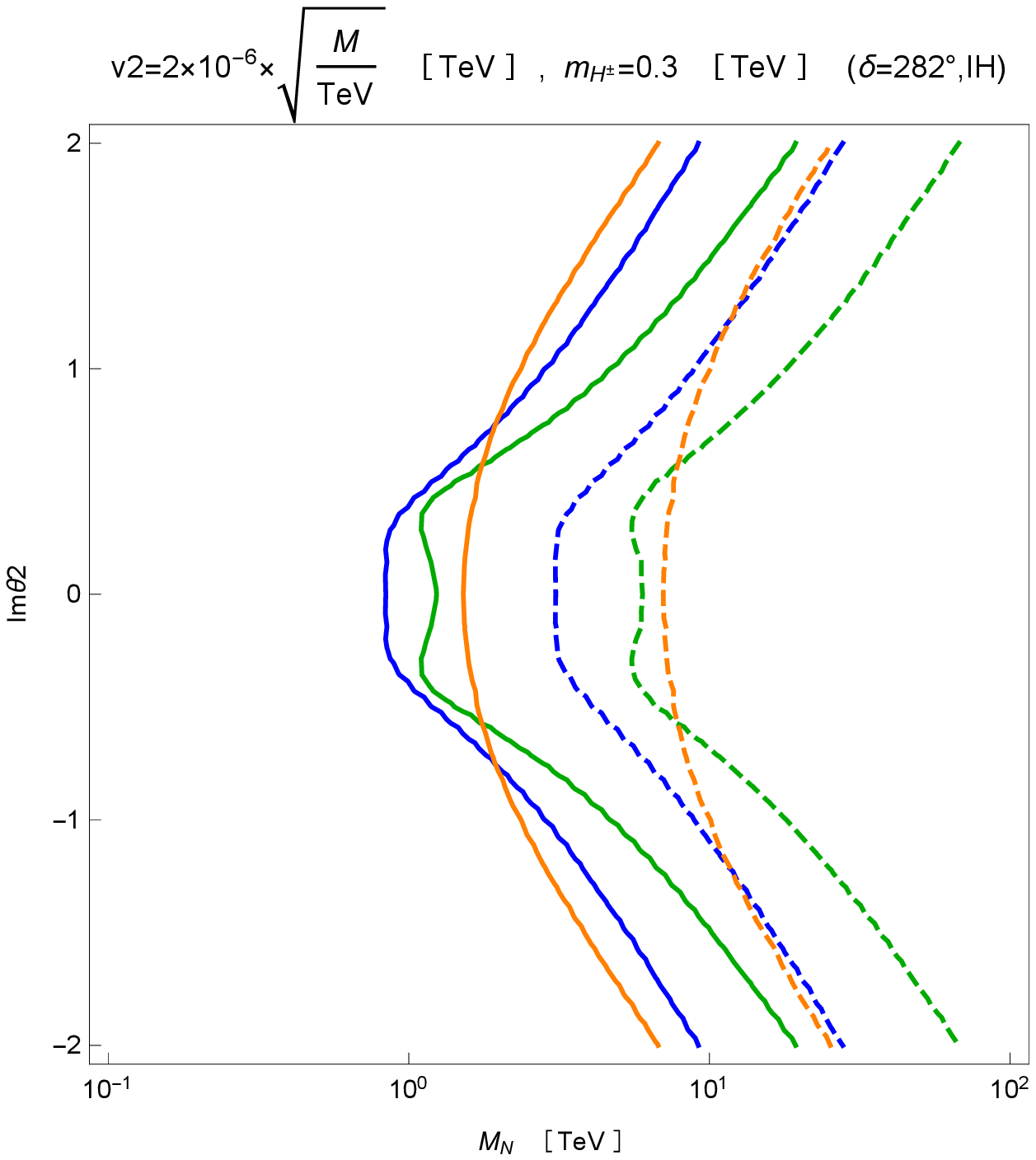}
      \end{minipage}
 
 
      \begin{minipage}{0.33\hsize}
        \centering
          \includegraphics[keepaspectratio, scale=0.44, angle=0]
                          {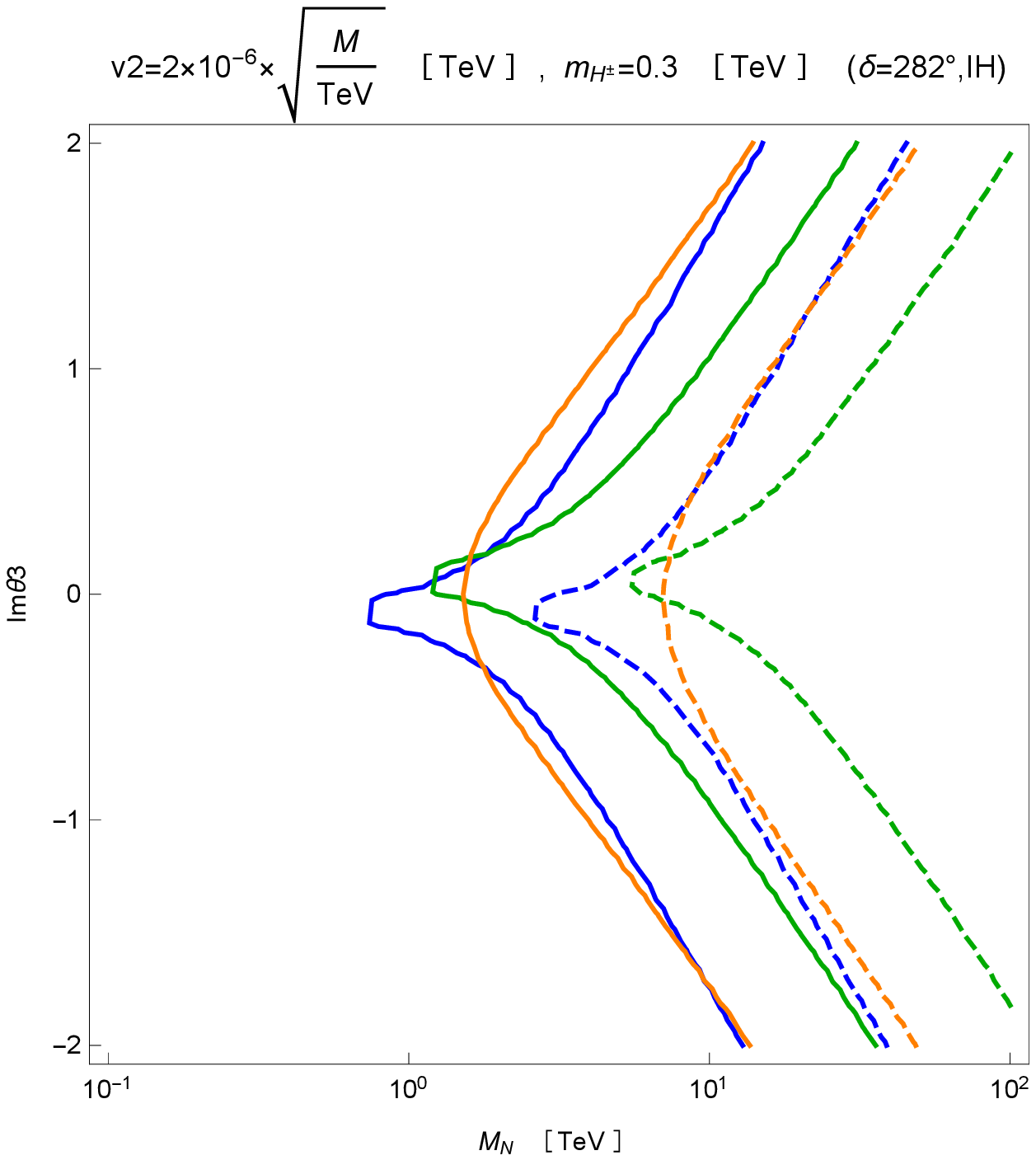}
      \end{minipage} \\ 
      \\
 
      \begin{minipage}{0.33\hsize}
        \centering
          \includegraphics[keepaspectratio, scale=0.44, angle=0]
                          {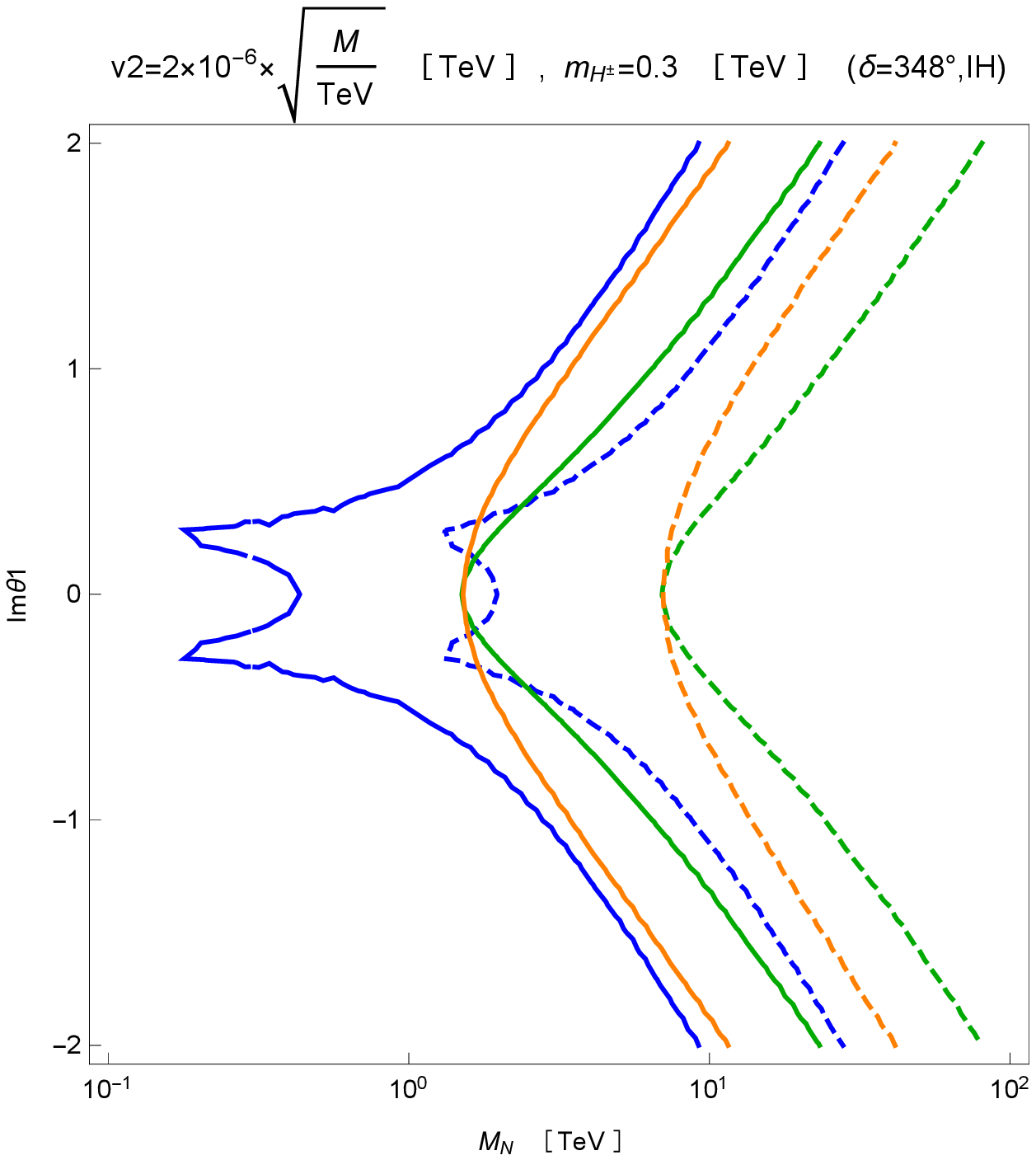}
      \end{minipage}

 
      \begin{minipage}{0.33\hsize}
        \centering
          \includegraphics[keepaspectratio, scale=0.44, angle=0]
                          {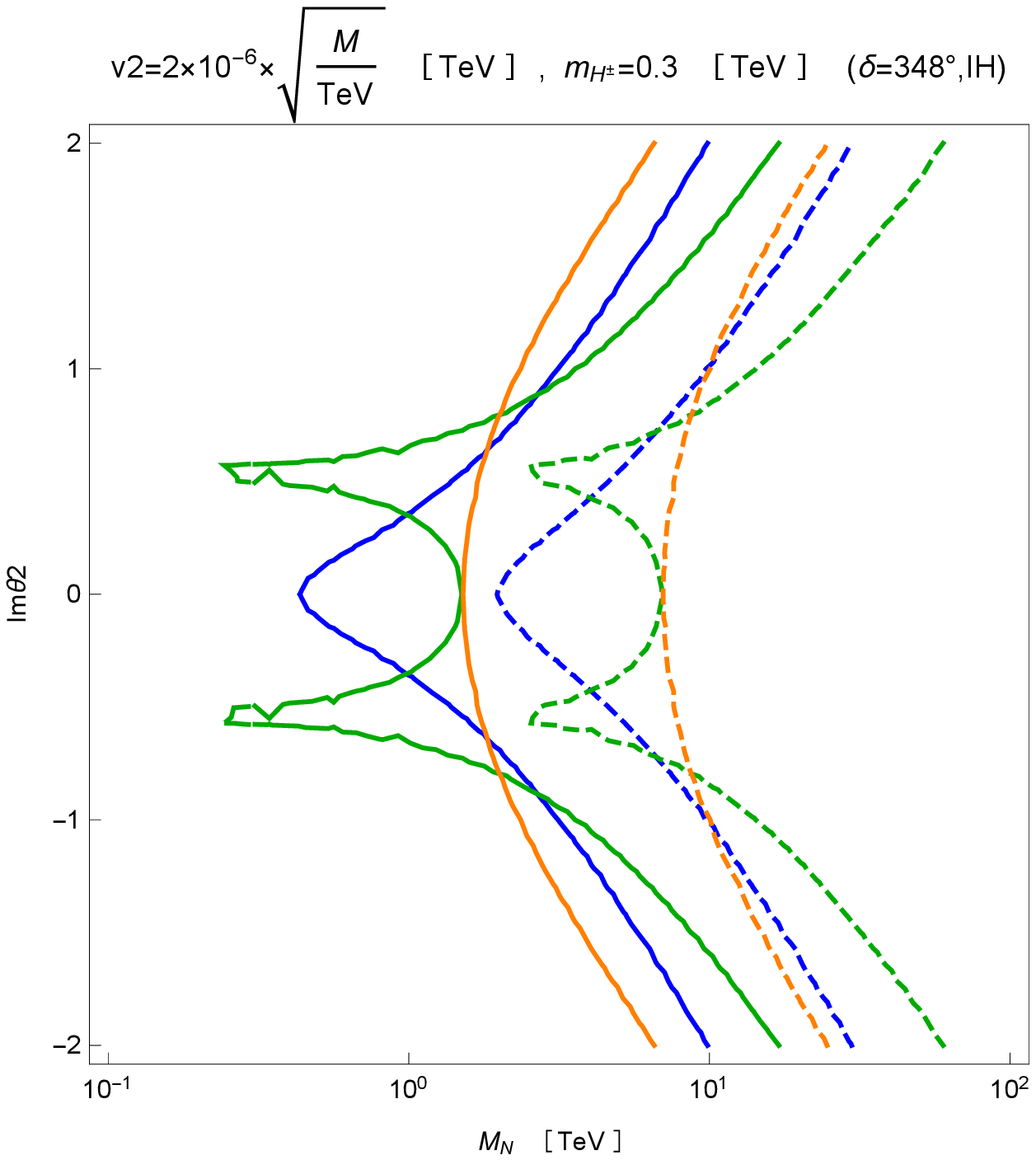}
      \end{minipage}
 
 
      \begin{minipage}{0.33\hsize}
        \centering
          \includegraphics[keepaspectratio, scale=0.44, angle=0]
                          {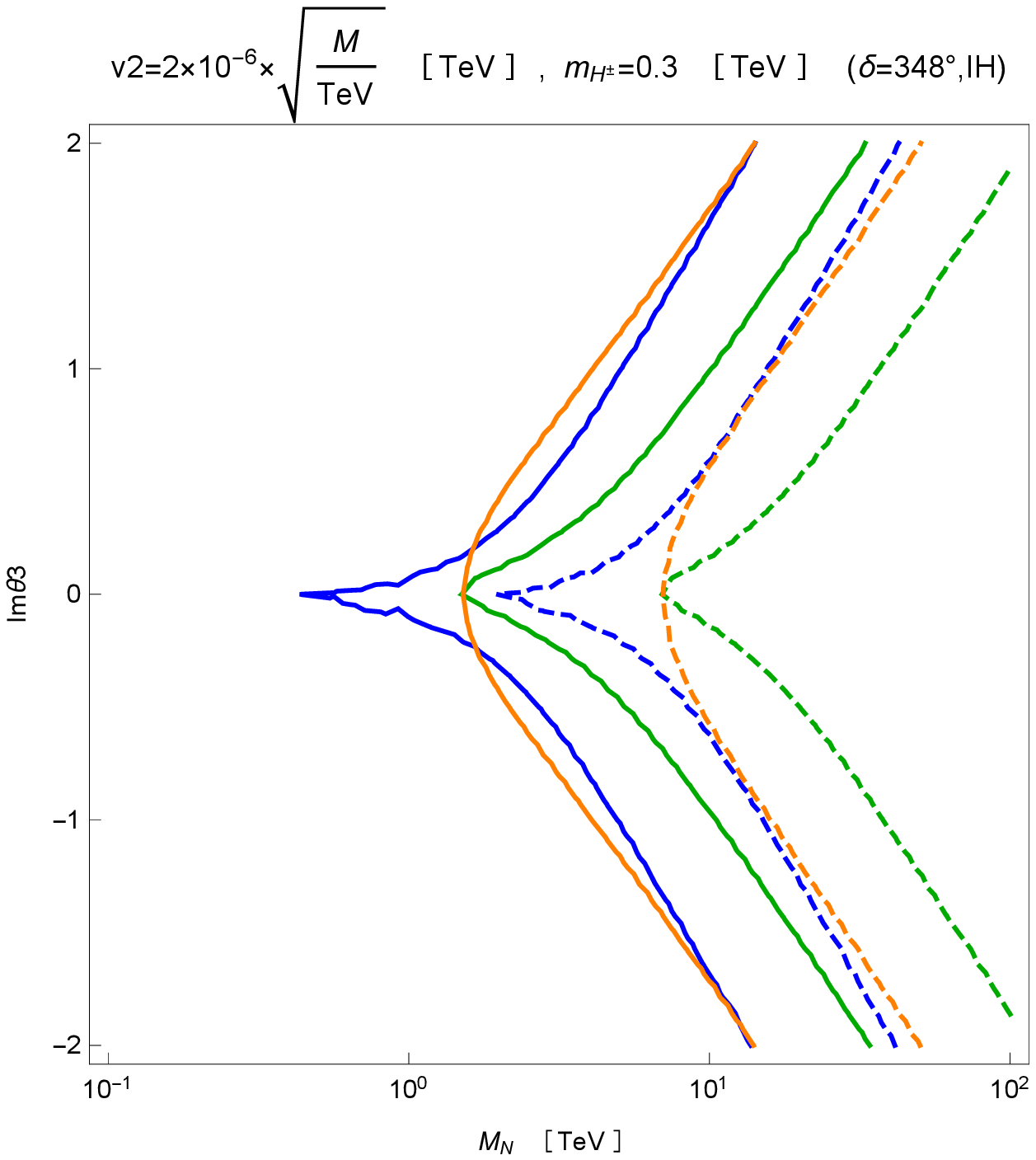}
      \end{minipage} 

    \end{tabular}
     \caption{\footnotesize Same as figure~\ref{figprehN} except that the neutrino mass hierarchy is Inverted Hierarchy and $v_2$ is given in Eq.~(\ref{v2-ih})
     and that the solid green and orange lines correspond to the contours of $Br(h\to e\tau)/Br(h\to \tau\tau)=10^{-13}$ and $Br(h\to \mu\tau)/Br(h\to \tau\tau)=10^{-12}$, respectively,
     and the dashed green and orange lines correspond to the contours of $Br(h\to e\tau)/Br(h\to \tau\tau)=10^{-13}$ and $Br(h\to \mu\tau)/Br(h\to \tau\tau)=10^{-12}$, respectively, when $v_2$ is multiplied by $1/3$.}
 \label{figprehI}
\end{figure}

Figure~\ref{figprehI} is the corresponding figure for IH and $v_2$ in Eq.~(\ref{v2-ih}). Here, the green lines correspond to $Br(h\to e\tau)/Br(h\to \tau\tau)=10^{-13}$
 and the orange lines correspond to $Br(h\to \mu\tau)/Br(h\to \tau\tau)=10^{-12}$.

We observe that for NH we can hope that the $h\to e \tau$ decay is detected at a rate $Br(h\to e\tau)/Br(h\to \tau\tau)\sim10^{-12}$
 and that the $h\to \mu \tau$ decay is detected at a rate $Br(h\to \mu\tau)/Br(h\to \tau\tau)\sim10^{-11}$
 even when the model satisfies the current experimental bound on $Br(\mu \to e\gamma)$.
If IH is the correct mass hierarchy, both $Br(h\to e\tau)$ and $Br(h\to \mu\tau)$ roughly decrease by $1/10$.
Unfortunately, the predicted rate is too small to explain the hint of $h\to \mu \tau$ decay reported by CMS~\cite{Khachatryan:2015kon}.

 \newpage
\newpage

 \newpage
\newpage

\section{Summary}
We have investigated the neutrinophilic Higgs+seesaw model, in which right-handed neutrinos couple only with an extra Higgs field that develops a tiny VEV
 and have Majorana mass, and which realizes the low-scale seesaw naturally.
We have concentrated on CLFV processes induced by loop diagrams of the charged scalar and heavy neutrinos. 
First, we have studied the current constraint on the model's parameter space from the search for $\mu\to e\gamma$. 
Secondly, we have predicted the branching ratios of other CLFV processes ($\mu\to3e$, $\mu+{\rm Al}\to e+{\rm Al}$, $\mu+{\rm Ti}\to e+{\rm Ti}$, $Z\to e\mu$,
$Z\to e\tau$, $Z\to \mu\tau$,  $h\to e\tau$ and $h\to\mu\tau$), and discussed whether these processes can be detected in the future.
An important finding is that, considering the future sensitivities, the $\mu\to3e$, $\mu+{\rm Al}\to e+{\rm Al}$ and $\mu+{\rm Ti}\to e+{\rm Ti}$ processes
 can be detected in a wide parameter region in the future, even when the model satisfies the current stringent bound on the $\mu\to e\gamma$ branching ratio.
\\

\section*{Acknowledgment}
This work is partially supported by Scientific Grants by the Ministry of Education, Culture, Sports, Science and Technology of Japan,
Nos.~17K05415, 18H04590 and 19H051061 (NH), and No.~19K147101 (TY).
\\

\section*{Funding}
Open Access funding: SCOAP.

\end{document}